\documentclass[aps,nofootinbib,showpacs,preprintnumbers,amsmath,amssymb]{revtex4}
\def\be{\begin{equation}}
\def\ee{\end{equation}}
\def\bea{\begin{eqnarray}}
\def\eea{\end{eqnarray}}
\def\bear{\begin{array}}
\def\eear{\end{array}}
\newcommand{\MSbar}{\overline{\rm MS}}  
\newcommand{\m}{{\overline m}}
\newcommand{\A}{{\mathcal{A}}}
\newcommand{\tA}{{\widetilde {\mathcal{A}}}}
\newcommand{\ta}{{\widetilde a}}
\newcommand{\tlA}{{\mathfrak A}}
\newcommand{\tk}{{\widetilde k}}
\newcommand{\B}{{\mathcal{B}}}
\usepackage{graphics}
\usepackage{graphicx}
\usepackage{dcolumn}
\usepackage{bm}
\usepackage{epsfig}
\usepackage{graphicx}
\usepackage{multirow}
\usepackage{dcolumn}
\usepackage{graphicx,epsfig}%

\begin{document}

\preprint{USM-TH-309; arXiv 1210.6117v2; published in: Phys.Rev.D87,054008 (2013)}

\title{Calculation of binding energies and masses of quarkonia in analytic QCD models\footnote{
In comparison with v1: improved presentation; (F)APT calculation performed in the $\beta_2=\beta_3=\cdots =0$ scheme; comparison with the results of other methods in the literature included (cf.~Table IV); part of the text moved into the new Appendix A; the basic conclusions unchanged; extended acknowledgments;
Refs.~[12,38-39,41,44,80-83,89-99,110-11] are new.}} 
\author{C\'esar Ayala$^1$}
 \email{c.ayala86@gmail.com}
\author{Gorazd Cveti\v{c}$^1$$^2$}
 \email{gorazd.cvetic@usm.cl}

\affiliation{$^1$Department of Physics, Universidad T{\'e}cnica Federico
Santa Mar{\'\i}a (UTFSM), Valpara{\'\i}so, Chile\\
$^2$Centro Cient\'{\i}fico-Tecnol\'ogico de Valpara\'{\i}so, UTFSM, Valpara{\'\i}so, Chile}

\date{\today}

\begin{abstract}

We extract quark masses $\m_q$ ($q=b,c$) from the evaluation
of the masses of quarkonia $\Upsilon(1S)$ and $J/\psi(1S)$,
performed in two analytic QCD models, 
and in perturbative QCD in two renormalization schemes.
In analytic QCD the running coupling has no unphysical
singularities in the low-momentum regime.
We apply the analytic model of Shirkov {\it et al.\/}~[Analytic
Perturbation Theory (APT)],
extended by Bakulev {\it et al.\/}~[Fractional Analytic Perturbation 
Theory (FAPT)], and the two-delta analytic
model (2$\delta$anQCD). The latter, in contrast to (F)APT, at higher energies
basically coincides with the perturbative QCD (in the same scheme).
We use the renormalon-free mass $\m_q$ as input. 
The separation of the soft and ultrasoft parts of
the binding energy $E_{q \bar q}$ is performed by the requirement of
the cancellation of the leading infrared renormalon.
The analysis in the 2$\delta$anQCD model
indicates that the low-momentum ultrasoft regime 
is important for the extraction of the masses $\m_q$, especially $\m_b$.
The 2$\delta$anQCD model gives us clues on how to estimate the influence 
of the ultrasoft sector on $\m_q$ in general.
These effects lead to relatively large values $\m_b \approx 4.35 \pm 0.08$ GeV 
in the 2$\delta$anQCD model, which, however, 
are compatible with recent lattice calculations.
In perturbative QCD in $\MSbar$ scheme these effects are
even stronger and give larger uncertainties in $\m_b$. The (F)APT model gives
small ultrasoft effects and the extracted values of $\m_b$ agree
with those in most of the literature ($\m_b \approx 4.2$ GeV).
The extracted values of $\m_c$ in all four models
are about $1.26$-$1.27$ GeV and agree well with those in the literature. 

\end{abstract}

\maketitle

\section{Introduction}
\label{sec:intr}

Most of the calculations of the masses of heavy quarkonia are
based on perturbative expansions. These expansions come from
the knowledge of the static quark-antiquark potential $V(r)$ which has
been calculated up to ${\rm N}^2{\rm LO}$ ($\sim \alpha_s^3/r$) in 
Refs.~\cite{Peter:1996ig,Schroder:1998vy}. At
${\rm N}^3{\rm LO}$ ($\sim \alpha_s^4/r$) level, ultrasoft gluons contribute
\cite{Brambilla:1999qa,Kniehl:1999ud} to $V(r)$ and the
calculation of these terms has been completed in 
Refs.~\cite{Smirnov:2009fh,Anzai:2009tm}. The expansion coefficients
for the mass of the (heavy) $q {\bar q}$ quarkonium
vector ($S=1$) or scalar ($S=0$) ground state (i.e., with:
$n=1$ and $\ell=0$) are given, for example, in 
Ref.~\cite{Penin:2002zv}, for the terms including 
$\sim m_q \alpha_s^5$. The latter term is known
because the potential $V(r)$ is known at $\sim \alpha_s^4/r$ level.
For a review of the topic, see, e.g.~Ref.~\cite{Pineda:2011dg}.

Most of the radiative contributions to the quarkonium mass expansion are 
from the so called soft sector of gluon momenta, 
$Q \sim m_q \alpha_s$.\footnote{
The momentum transfer $q$ of the gluon is spacelike, i.e.,
$q^2 \equiv - Q^2 < 0$, because the scattering of the quark and antiquark
in the quarkonium is of the $t$-channel type.}
In the cases of $b{\bar b}$ and $c{\bar c}$, these are around $2$ GeV
and $1$ GeV, respectively. At $Q \approx 2$ GeV scales, the perturbative
QCD (pQCD) coupling $\alpha_s(Q^2)$ is marginally reliable; 
at $Q \approx 1$ GeV it is unreliable.
This is so because the pQCD coupling $\alpha_s(Q^2)$ suffers from
unphysical (Landau) singularities at low spacelike $q^2$ ($ \equiv - Q^2$),
i.e., at $0 < Q^2 < \Lambda^2$ where $\Lambda^2 \sim 10^{-1} \ {\rm GeV}^2$;
for $Q^2 \sim 1 \ {\rm GeV}^2$ it is dangerously close to these singularities, 
and thus unreliable.     

The aforementioned Landau singularities are not physical because
they do not possess the analytic properties of the spacelike
observables ${\cal D}(Q^2)$ (such as the Adler function), the latter
properties being dictated by the general properties of quantum field theories
\cite{BS,Oehme} including causality. 
Namely, ${\cal D}(Q^2)$ must be an analytic function
in the entire complex $Q^2$ plane, with the exception of the
timelike semiaxis $Q^2 < - M_{\rm thr}^2$, where $M_{\rm thr} \sim 10^{-1}$ GeV
is a particle production threshold. Specifically, the evaluation
of spacelike quantities ${\cal D}(Q^2)$ in pQCD, as a (truncated) power series
of the perturbative running coupling (couplant)
$a_{\rm pt}(\kappa Q^2) \equiv \alpha_s(\kappa Q^2)/\pi$ 
(with $\kappa \sim 1$), does not respect these important analytic
properties of ${\cal D}(Q^2)$. These problems, within the context of QCD,
were first addressed in the seminal works of Shirkov, Solovtsov, Milton {\it et al.\/}
\cite{ShS,MS,MSS,Solovtsov:1997at,Sh1,Sh2}. 
There, the perturbative running coupling
$a_{\rm pt}(Q^2) \equiv \alpha_s(Q^2)/\pi$ was made an analytic (in the aforementioned
sense) function of $Q^2$, $a_{\rm pt}(Q^2) \mapsto \A_1^{\rm (APT)}(Q^2)$,
in the following way: 
in the dispersion relation the discontinuity function
$\rho_1^{\rm (pt)}(\sigma) \equiv {\rm Im} \; a_{\rm pt}(Q^2=-\sigma - i \epsilon)$ 
was kept unchanged on the entire negative axis in the complex $Q^2$-plane
(i.e., for $\sigma \geq 0$), and was set equal to zero on the nonphysical cut
$0 < Q^2 < \Lambda^2$. For this reason, this analytic QCD (anQCD) model 
can be referred to as the minimal analytic (MA) model. 
Here we will refer to this
model by its usual name used in the literature:
Analytic Perturbation Theory (APT). 

Various other analytic QCD (anQCD) models for $\A_1(Q^2)$
can be constructed,
and have been presented in the literature, among them 
Refs.~\cite{Nesterenko,Nesterenko2,Alekseev:2005he,Srivastava:2001ts,Webber:1998um,CV1,CV2,Bdecays,Shirkov:2012ux}. These models fulfill certain 
additional constraints at low and/or at high $Q^2$.
For further literature on various analytic QCD models, 
we refer to the review articles in 
Refs.~\cite{Prosperi:2006hx,Shirkov:2006gv,Cvetic:2008bn,Bakulev}. 
Some newer constructions of anQCD models of $\A_1(Q^2)$
include those based on specific classes of $\beta$ functions with
nonperturbative contributions \cite{Belyakova:2010iw} 
or without such contributions \cite{CKV1,CKV2},
and those based on modifications of the discontinuity function 
$\rho_1^{\rm (pt)} \mapsto \rho_1$
[ $\rho_1(\sigma)  \equiv {\rm Im} \; \A_{1}(Q^2=-\sigma - i \epsilon)$ ]
at low (positive) $\sigma$ where $\rho_1$ is parametrized in a specific manner
(with delta functions), cf.~Refs.~\cite{1danQCD,2danQCD}.

The model of Ref.~\cite{2danQCD}, 
an extension of the model of Ref.~\cite{1danQCD},
will be called the two-delta analytic QCD model 
(2$\delta$anQCD), because its discontinuity
function $\rho_1(\sigma)$ is parametrized in the unknown low-$\sigma$ regime
by two delta functions. It has a specific attractive feature that
in the high-scale regime ($|Q^2| > \Lambda^2$) it practically coincides with
the corresponding (in the scheme) pQCD coupling: $\A_1(Q^2) - a_{\rm pt}(Q^2)
\sim (\Lambda^2/Q^2)^5$. This makes it possible for the model to be applied
with the Operator Product Expansion (OPE) \cite{anOPE} including terms of dimension
$D \leq 8$ without the need to modify the ITEP school  (Institute of Theoretical and Experimental Physics) interpretation of the
OPE \cite{Shifman:1978bx,DMW}. The latter basically states that the terms in OPE
of higher dimension ($D>0$) originate from the infrared regime only. 

Once we have an anQCD model, i.e., a model for the analytic analog 
$\A_1(Q^2)$ of the pQCD coupling $a_{\rm pt}(Q^2)$ 
(or, equivalently, for the discontinuity
function $\rho_1(\sigma)$), the analytic analogs $\A_n(Q^2)$ of the higher powers
$a_{\rm pt}(Q^2)^n$ have to be constructed from $\A_1(Q^2)$. In the case of 
APT ($\A_1^{\rm (APT)}(Q^2)$), for integer $n$, 
the couplings $\A_n^{\rm (APT)}(Q^2)$ were 
constructed in Refs.~\cite{MS,MSS,Solovtsov:1997at}.
However, perturbation expansions of some observables, and effects of
evolution of distribution amplitudes at a chosen loop-level,
are represented sometimes by noninteger powers $a_{\rm pt}(Q^2)^{\nu}$
or logarithmic terms  $a_{\rm pt}(Q^2)^{\nu} \ln^k a_{\rm pt}(Q^2)$
($\nu$ noninteger, $k$ integer). Among them are the pion electromagnetic
form factor $F_{\pi}(Q^2)$ and hadronic decay width of Higgs $\Gamma_{(H \to b {\bar b})}$.
The problems of the analytization of such quantities, within 
the spirit of APT, were considered in Ref.~\cite{KS}, where
the dispersion relations were extended from the coupling parameter
to the general QCD amplitudes. The APT approach to evaluation of
the (factorizable part) of the form factor  $F_{\pi}(Q^2)$ was
performed in Ref.~\cite{Baketal}, where the nonperturbative distribution
amplitude was evaluated by performing numerical evolution with
an analytic coupling parameter. The authors Bakulev, Mikhailov and
Stefanis (BMS) then systematically extended APT 
in Ref.~\cite{BMS1} to the evaluation of
noninteger power analogs
$\A_{\nu}(Q^2)$ and $\A_{\nu, k}(Q^2)$ of the mentioned terms $a_{\rm pt}(Q^2)^{\nu}$ and
$a_{\rm pt}(Q^2)^{\nu} \ln^k a_{\rm pt}(Q^2)$. This extension was
performed for the spacelike quantities, with explicit expressions at
the one-loop level, and extension to the higher-loop level via expansions
of the one-loop expressions. This was then applied to
an evaluation of the factorizable part of $F_{\pi}(Q^2)$ in
Ref.~\cite{BKS}. 
In Ref.~\cite{BMS2} the BMS authors 
extended this construction to the timelike quantities and
applied it to the evaluation of the Higgs
decay width $\Gamma_{(H \to b {\bar b})}(s)$ in APT.
In Ref.~\cite{BMS3} this construction was applied to
the evaluation of the $e^+ e^- \to {\rm hadrons}$ ratio $R$,
and a detailed analysis of the evaluation of  $\Gamma_{(H \to b {\bar b})}$.
In the review work of Bakulev, Ref.~\cite{Bakulev},
several variants of this construction are reviewed, among them
the numerical dispersive approach at the two-loop and higher-loop level
(Secs.~III B, C, D of Ref.~\cite{Bakulev});
a mathematical package for such numerical calculation
is given in Ref.~\cite{BK}. 
The construction of BMS is referred to in the 
literature as the Fractional Analytic Perturbation
Theory (FAPT). 

In the general
case of analytic $\A_1(Q^2)$, the quantities $\A_n(Q^2)$ were constructed
in Refs.~\cite{CV1,CV2} for integer $n$, as linear combinations of
logarithmic derivatives $\tA_k(Q^2) \propto d^{k-1} \A_1(Q^2)/d (\ln Q^2)^{k-1}$
($k \geq n$),\footnote{The relations between such functions $\tA_k$'s and 
$\A_n$'s, allowing a recurrent construction of $\A_n$,
for integer $n$ within the context of the APT model,
were given also in Refs.~\cite{Shirkov:2006nc,Shirkov:2006gv}.}
and were extended to noninteger $n$ in Ref.~\cite{GCAK}.

In this work, we apply two anQCD models of 
$(a_{\rm pt})_{\rm an.} \equiv \A_1$, namely APT of 
Refs.~\cite{ShS,MS,MSS,Solovtsov:1997at,Sh1,Sh2},
and the 2$\delta$anQCD model of Refs.~\cite{2danQCD,anOPE},
to evaluations of the perturbation series of the binding energy 
$E_{q \bar q}$ of heavy quarkonia 
($\Upsilon(1S)$ and $J/\psi(1S)$) and of the quark pole mass $m_q$.
In this way, we will evaluate the masses of these quarkonia
$M_{q \bar q} = 2 m_q + E_{q \bar q}$ as functions of the
($\MSbar$) quark mass $\m_q$.
In the APT model of analytic QCD (i.e., the
model for $\A_1^{\rm (APT)}(Q^2)$, Ref.~\cite{ShS}), we will
need to evaluate not just the integer power analogs
$(a_{\rm pt}^n)_{\rm an. APT} \equiv \A_n^{\rm (APT)}$, 
but also the analogs of the logarithmic terms
$(a_{\rm pt}^n \ln^k a_{\rm pt})_{\rm an. APT} \equiv \A_{n,k}^{\rm (APT)}(Q^2)$
whose evaluation uses the approach of FAPT \cite{BMS1,BMS2},
reviewed in \cite{Bakulev}.
Therefore, we will refer to this method as APT
when it involves only integer power analogs, and as (F)APT when
it involves both aforementioned types of terms.

As input parameter we use the renormalon-free quark mass
$\m_q$ ($\equiv \m_q(\mu^2=\m_q^2)$) of the corresponding quark $q=b, c$
(also called the $\MSbar$ quark mass),
and the anQCD coupling of the model.
Since the quarkonia masses
are well measured, we can extract the values of $\m_q$.
We also perform the same analysis in pQCD in the corresponding 
renormalization schemes.

In Sec.~\ref{sec:anQCD} we briefly describe the (F)APT model
(Sec.~\ref{subs:MA}) and the 2$\delta$anQCD model (Sec.~\ref{subs:2d}).
In Sec.~\ref{sec:Eqq} we present the procedures of evaluation of
the binding energy $E_{q {\bar q}}$ and of the quark pole mass $m_q$,
in terms of the mass $\m_q$ and of the couplings. Furthermore, we explain
how the cancellation of the leading infrared renormalon in the sum 
$2 m_q + E_{q \bar q}$ allows us to separate the ultrasoft from the soft part
of the binding energy.
The numerical results and the extractions of the masses $\m_b$
and $\m_c$ are presented in Sec.~\ref{sec:numres} (Secs.~\ref{subs:mb} and
\ref{subs:mc}, respectively). The evaluations are performed 
in the two aforementioned analytic models (F)APT and 2$\delta$anQCD,
and in pQCD in the corresponding two renormalization schemes
($\MSbar$, and in the scheme of 2$\delta$anQCD called here the Lambert scheme).
In Sec.~\ref{sec:summ} we summarize our results and draw certain conclusions.
Appendix \ref{app0} summarizes the construction of the
higher order analytic analogs $\A_{\nu}$ for the powers $a_{\rm pt}^{\nu}$
and for the logarithmic-type terms $a_{\rm pt}^{\nu} \ln^k a_{\rm pt}$, 
for general $\nu$ and integer $k$, in general anQCD models.
Appendix \ref{app1} contains the expressions of the coefficients 
of the perturbation expansion of $E_{q {\bar q}}$, available 
from the literature. Appendix \ref{app2} contains a
renormalon-based estimation of the coefficient at the $a_{\rm pt}^4$ term in
the perturbation expansion of $m_q/\m_q$. Appendix \ref{app3} has some
useful formulas for the scale and scheme dependence of $a_{\rm pt}$.
  
\section{Two analytic QCD models}
\label{sec:anQCD}

\subsection{(Fractional) Analytic Perturbation Theory}
\label{subs:MA}

As already mentioned in the Introduction, the coupling 
$a_{\rm pt}(Q^2) \equiv \alpha_s(Q^2)/\pi$ in pQCD in the usual
renormalization schemes (such as ${\overline {\rm MS}}$, or the
$\beta_2=\beta_3= \ldots = 0$ scheme),
possesses unphysical (Landau) singularities inside the required
analyticity regime $Q^2 \in \mathbb{C} \backslash (-\infty, 0]$.
Specifically, it has a cut on the semiaxis $(-\infty, \Lambda_{\rm L}^2)$
(where $\Lambda_{\rm L}^2 \sim 10^{-1} \ {\rm GeV}^2$ 
is the ``Landau'' branching point), thus offending the
analyticity requirement on the cut sector $Q^2 \in (0,\Lambda_{\rm L}^2)$.
Application of the Cauchy theorem gives us for the power
$a^{\nu}_{\rm pt}(Q^2)$ ($\nu > 0$) of this spacelike coupling
the following dispersion relation:
\begin{equation}
a^{\nu}_{\rm pt}(Q^2) = 
\frac{1}{\pi} 
\int_{\sigma= - \Lambda_{\rm L}^2 - \eta}^{\infty}
\frac{d \sigma \ {\rm Im}  \; 
a^{\nu}_{\rm pt}(-\sigma-i \epsilon)}{(\sigma + Q^2)},
\label{anptdisp}
\end{equation}
and $\eta \to +0$. Eliminating the aforementioned cut sector, and keeping
the rest of the discontinuity function
$\rho_1^{\rm (pt)}(\sigma) \equiv {\rm Im} \; a_{\rm pt}(Q^2=-\sigma - i \epsilon)$ 
unchanged, results in the (F)APT model 
\cite{ShS,MS,MSS,Solovtsov:1997at,Sh1,Sh2,BMS1,BMS2,Bakulev},
i.e., the spacelike coupling
\begin{equation}
\left( a^{\nu}_{\rm pt}(Q^2) \right)_{\rm an}^{\rm ((F)APT)} \equiv
{{\A_{\nu}^{\rm {((F)APT)}}}}(Q^2) = \frac{1}{\pi} \int_{\sigma= 0}^{\infty}
\; d \sigma 
\frac{{\rm Im}\;{a^{\nu}_{\rm pt}(-\sigma-i \epsilon)}}{(\sigma + Q^2)} \ .
\label{MAanAnu}
\end{equation}
In evaluation of general spacelike observables in pQCD, terms of the type
$a_{\rm pt}^{\nu}(Q^2) \ln^k a_{\rm pt}(Q^2) \ln^{\ell}(Q^2/m^2)$ may appear, 
with $\nu > 0$ and (nonnegative) integers $k$ and $\ell$. 
In APT, in principle,
the factor $\ln^\ell(Q^2/m^2)$ may be included in the FAPT-type analytization
(see, for example, Ref.~\cite{Bakulev}). 
However, the factor  $\ln^\ell(Q^2/m^2)$ is analytic
function in the complex $Q^2$ plane with the exclusion
of the nonpositive axis $Q^2 \leq 0$. On the other hand, this is also the
regime of analyticity of $\A_1^{\rm (APT)}(Q^2)$ [note that $Q^2=0$ is a point
of nonanalyticity of  $\A_1^{\rm (APT)}(Q^2)$ because
$d \A_1^{\rm {(APT)}}(Q^2)/d Q^2 =\infty$ there]. Therefore, we will not include
the factor $\ln^\ell(Q^2/m^2)$ in the (FAPT-type)  analytization.
With these considerations, all the other pQCD terms in such
observables, $a_{\rm pt}^{\nu}(Q^2) \ln^k a_{\rm pt}(Q^2)$, 
are made analytic in (F)APT by the same procedure 
as $a^{\nu}_{\rm pt}(Q^2)$ in Eq.~(\ref{MAanAnu})
\be
\left( a_{\rm pt}^{\nu}(Q^2) \ln^k a_{\rm pt}(Q^2)  
\right)_{\rm an}^{\rm ((F)APT)} \equiv 
{{\A_{\nu,k}^{\rm {((F)APT)}}}}(Q^2)
=
\frac{1}{\pi} \int_{\sigma= 0}^{\infty} \; d \sigma
\frac{{\rm Im} \left[ a_{\rm pt}^{\nu}(-\sigma-i \epsilon) 
\ln^k a_{\rm pt}(-\sigma-i \epsilon) 
\right]}
{(\sigma + Q^2)} \ .
\label{MAangen}
\ee
In this context, the authors of Refs.~\cite{BMS1,BMS2,BMS3,Bakulev}
derived explicit simple formulas for these minimal analytic 
expressions at one-loop
level, and the related more involved expressions at higher-loop levels --
expressions from which the analyticity structure in $Q^2$ and in $\nu$
can be more clearly seen. For our calculational purposes, though,
the dispersion formulas (\ref{MAangen}) will be applied, numerically, 
when the (F)APT model is used, in the same spirit as was performed
by Bakulev in Ref.~\cite{Bakulev} (Sec.~III C there),
and in Ref.~\cite{BK}.

APT has a relatively specific and interesting property,
namely, that the couplings differ from the corresponding pQCD
couplings in a nonnegligible way even at high $|Q^2|$
\be
\A_1^{\rm (APT)}(Q^2) - a_{\rm pt}(Q^2) \sim \Lambda^2/Q^2 
\qquad (|Q^2| > \Lambda^2) \ .
\label{MAptdiff}
\ee
As a consequence, $\A_n^{\rm (APT)}(Q^2)$ and $a_{\rm pt}(Q^2)^n$ differ 
from each other in a nonnegligible way even at high $|Q^2|$ \cite{Sh2};
we cannot approximate $\A_1^{\rm (APT)}(Q^2)$ well with
$a_{\rm pt}(Q^2)$ even at as high $|Q^2|$ as $\sim M_Z^2$. 
The (only) free parameter in 
APT is the ${\overline {\rm MS}}$ scale ${\overline \Lambda}$, which appears 
in the usual perturbation expansion of the underlying pQCD coupling 
$a_{\rm pt}(Q^2;{\overline {\rm MS}})$ 
in inverse powers of $\ln(Q^2/{\overline \Lambda}^2)$ 
[i.e., the $a_{\rm pt}$ whose discontinuity functions appear in the relations
 (\ref{MAanAnu})-(\ref{MAangen})]. 

We will consider that in APT this scale varies in the interval 
${\overline \Lambda}_{N_f=5} = 0.260 \pm 0.030$ GeV.
The analysis of high-energy QCD observables
(with $|Q| \stackrel{>}{\sim} 10^1$ GeV) in Ref.~\cite{Sh2} suggests a value of
${\overline \Lambda}_{N_f=5} \approx 0.290$ GeV [corresponding to
$\alpha_s(M_Z^2,{\overline {\rm MS}}) \approx 0.124$]. The works of 
Refs.~\cite{BMS2,BMS3,Bakulev} use the value 
${\overline \Lambda}_{N_f=5} \approx 0.260$ GeV instead [corresponding to 
$\alpha_s(M_Z^2,{\overline {\rm MS}}) \approx 0.122$ and giving the
timelike (Minkowskian) coupling $\tlA_1(M_Z^2) \approx 0.120$].
On the other hand, if the ``average'' $e^+ e^-$ annihilation ratio $R(s)$
value at $\sqrt{s}=M_Z$ is taken to be $\approx 1.03904$ as used, e.g., in 
Ref.~\cite{Baikov:2008jh}, (F)APT approach of evaluating
$R(M_Z^2)$ gives ${\overline \Lambda}_{N_f=5} \approx 0.225$ GeV \cite{Bakpriv}.
 
In comparison, pQCD analyses give for $\alpha_s(M_Z^2)$ the
world average $0.1184 \pm 0.0007$ \cite{PDG2010}, corresponding to
${\overline {\Lambda}}_{N_f=5} = 0.213 \pm 0.008$ GeV, i.e., significantly lower
scales than for APT.\footnote{Also the anQCD models of 
Refs.~\cite{Nesterenko,Nesterenko2} differ nonnegligibly from pQCD, 
even at high $|Q^2|$.}

Due to the nonnegligible difference, Eq.~(\ref{MAptdiff}), the
threshold effects of heavy quarks have to be implemented in APT.
This is performed by requiring the continuity of the
perturbative coupling [whose discontinuities are used in APT]
\be
a_{\rm pt}(Q^2) \equiv f_{N_f}(Q^2/{\overline {\Lambda}}^2_{N_f}) \ ,
\label{aptf}
\ee 
at positive ``threshold'' values $Q^2 =Q^2_{N_f-1 \mapsto N_f} > 0$
chosen to be for $N_f=4,5,6$ the squares of the 
current heavy quark masses $m_c \equiv m_4$, $m_b \equiv m_5$ 
and $m_t \equiv m_6$, respectively \cite{Sh2}
\be
f_{N-1}(m_N^2/{\overline {\Lambda}}^2_{N-1})=
f_N(m_N^2/{\overline {\Lambda}}^2_N) 
\qquad (N=4,5,6) \ .
\label{contthr}
\ee
Further, it is assumed that, for complex $Q^2$, the values
of $a_{\rm pt}(Q^2)$ involve the scale ${\overline {\Lambda}}_{N_f}$
determined by the absolute value $|Q^2|$
\be
a_{\rm pt}(Q^2) = f_N(Q^2/{\overline {\Lambda}}_N^2) \qquad
{\rm for:} \ m_N^2 < |Q^2| < m_{N+1}^2 \ .
\label{aptglob}
\ee
Thus constructed $a_{\rm pt}(Q^2)$
gives the piecewise discontinuous
global APT discontinuity functions. For example,
for $\rho_{\nu,k}^{\rm (pt)}(\sigma) \equiv
{\rm Im}[ a_{\rm pt}^{\nu} \ln^k a_{\rm pt} (-\sigma-i \epsilon)]$,
this globalization results in
\be
\rho_{\nu,k}^{\rm (pt)}(\sigma) \equiv 
{\rm Im} \left( a_{\rm pt}^{\nu}(-\sigma - i \epsilon) 
\ln^k a_{\rm pt} (-\sigma-i \epsilon) \right) 
= 
{\rm Im} \left[ f_N(z)^{\nu} \ln^k f_N(z) \right] {\bigg |}_
{z= - \sigma/{\overline {\Lambda}}_N^2  - i \epsilon}
\quad {\rm for:} \ (m_N^2 < \sigma < m_{N+1}^2) \ .
\label{rho1ptglob}
\ee
Despite the discontinuity of $\rho_{\nu,k}^{\rm (pt)}(\sigma)$, the
(F)APT analytic analogs (\ref{MAangen}) are analytic functions.  

For the underlying $a_{\rm pt}(Q^2) \equiv \alpha_s(Q^2)/\pi$
we will use the coupling in the $\beta_2=\beta_3= \cdots =0$ 
renormalization scheme \cite{Gardi:1998qr,Magr1,Magr2},
i.e., the coupling which fulfills the two-loop
renormalization group equation (RGE)
\bea
\frac{d a_{\rm pt}(Q^2)}{d \ln Q^2} & = &
- \beta_0 a_{\rm pt}^2 (1 + c_1 a_{\rm pt} ) \ .
\label{RGE2l}
\eea
Here, $c_1=\beta_1/\beta_0$, the constants
$\beta_0= (1/4) (11 - 2 N_f/3)$ and $\beta_1= (102 - 38 N_f/3)/16$
are universal, and the scale $Q^2$ is complex in general, 
$Q^2=|Q^2| \exp(i \phi)$. This equation has an explicit solution
of the form \cite{Gardi:1998qr,Magr1,Magr2}
\bea
a_{\rm pt}(Q^2) = - \frac{1}{c_1} \frac{1}{\left[1 + W_{\mp 1}(z) \right]} \ ,
\label{apt2l}
\eea
$W_{-1}$ and $W_{+1}$ are the branches of the Lambert function
for the case $0 \leq \phi < + \pi$ and $- \pi < \phi < 0$, respectively, and 
\be
z(Q^2) =  - \frac{1}{c_1 e} 
\left( \frac{|Q^2|}{\Lambda^2} \right)^{-\beta_0/c_1} 
\exp \left( - i {\beta_0}\phi/c_1 \right) \ .
\label{zexpr}
\ee 
We call the scale $\Lambda^2$ ($\sim 0.1 \ {\rm GeV}^2$) appearing here 
the Lambert scale. 
The aforementioned values 
${\overline \Lambda}_{N_f=5} = 0.260 \pm 0.030$ GeV, in the
scheme $\beta_2=\beta_3=\cdots=0$,
correspond to $\Lambda_{N_f=5} = 0.322 \pm 0.037$ GeV. 
The solution (\ref{apt2l})
is convenient because it represents an explicit function\footnote{
In Mathematica \cite{Math8} 
the Lambert function $W_n(z)$ is implemented under the name
${\rm  ProductLog}[n,z]$.} and it is thus easy to evaluate it
and the corresponding discontinuity functions $\rho_{\nu,k}^{\rm (pt)}(\sigma)$,
Eq.~(\ref{MAangen}), in practice. 
It has Landau singularities.

Using the aforementioned value 
$\Lambda_5 = 0.322 \pm 0.037$ GeV (${\overline \Lambda}_5 = 0.260 \pm 0.030$ GeV),
the continuity conditions (\ref{contthr}) give us at other values of $N_f$:
$\Lambda_4 = 0.476 \pm 0.050$ GeV (${\overline \Lambda}_4 = 0.366 \pm 0.038$ GeV),
$\Lambda_3 = 0.581 \pm 0.055$ GeV (${\overline \Lambda}_3 = 0.427 \pm 0.040$ GeV),
and
$\Lambda_6 = 0.128 \pm 0.016$ GeV (${\overline \Lambda}_6 =0.110 \pm 0.014$ GeV).
The first three quark flavors are regarded to be massless.

In this context, we mention that the elimination of the
Landau singularities (i.e., analytization) of $a_{\rm pt}(Q^2)$ 
can be performed in various ways, not just in the
``minimal'' way of Eq.~(\ref{MAanAnu}). This analytization,
in general results in an (analytic) running coupling $\A_1(Q^2)$
which differs from the corresponding perturbative one
by a power term
\be
\A_1(Q^2) = a_{\rm pt}(Q^2) + 
{\cal O}\left( \left(\frac{\Lambda^2}{Q^2} \right)^n \right) \ ,
\quad (|Q^2| > \Lambda^2) \ .
\label{diffA1a}
\ee
In the case of APT, the power index is $n=1$, cf.~Eq.~(\ref{MAptdiff}).
This index can
be increased to $n=3$ \cite{Alekseev:2005he,1danQCD}, and even to $n=5$
\cite{2danQCD}, while still maintaining analyticity. 
In Refs.~\cite{CKV1,CKV2} the problem of finding 
perturbative analytic couplings (in specific schemes) was
investigated. It turned out that such couplings always gave
a value of the (strangeless) semihadronic
tau lepton decay ratio $r_{\tau}$ that was significantly too low, 
unless the scheme was changed
in a drastic way which made the perturbation series diverge
strongly after the first four terms. 

Therefore, the appearance 
of the power terms as given in Eq.~(\ref{diffA1a})
seems to be a general feature of procedures which eliminate the
unphysical (Landau) singularities of the coupling 
in the complex $Q^2$-plane. These power terms are of
ultraviolet origin and thus contravene the philosophy of
the ITEP school \cite{Shifman:1978bx},
according to which the power terms in (inclusive)
QCD observables appear in the OPE
and are of infrared origin. Furthermore, in APT we have $n=1$
and the terms $\sim \Lambda^2/Q^2$ appear even in the massless QCD,
i.e., in the case when such terms are not allowed in the usual 
pQCD+OPE approach for observables related with the
vacuum expectation values (such as the Adler function). 
In general, the ITEP
interpretation of the OPE must be abandoned in those anQCD
models in which the index $n$ in Eq.~(\ref{diffA1a}) is low
(e.g., $n \leq 2$), and modified or restricted in others (e.g., when
$n=3,4$).\footnote{Nonetheless, even in APT, OPE can be applied,
and has successfully been applied -- see, for example,
the inclusion of higher-twist terms
in the APT vs pQCD analysis of the Bjorken polarized sum rule
at low $Q^2$, Refs.~\cite{Khandramai:2011zd}. We note that the 
leading-twist ($D=0$) term, in such an APT+OPE approach,
contains implicitly (parts of) the power term contributions
$\sim (\Lambda^2/Q^2)^n$, $n=1,2,\ldots$.}

We evaluate quarkonium masses with the (F)APT model of anQCD,
illustrating the effects of elimination of
Landau singularities on the behavior of the series.
The (F)APT model is relatively simple to apply technically.

We will also apply another anQCD
model to the evaluation of quarkonium masses, 
namely the so called two-delta anQCD (2$\delta$anQCD) \cite{2danQCD},
in which the index in Eq.~(\ref{diffA1a}) is $n=5$. 
Evaluations in such models are technically more demanding, though,
because the evaluation of the analytic analogs of
noninteger powers $a_{\rm pt}^{\nu}$ and of the logarithm-type terms
$a^n \ln^k a$ is more involved \cite{GCAK}.  

\subsection{Two-delta analytic QCD model (2$\delta$anQCD)}
\label{subs:2d}

It is possible to construct such anQCD models which,
at high momenta, practically merge with pQCD; namely, with
the analytic coupling $\A_1(Q^2)$ which fulfills Eq.~(\ref{diffA1a})
with $n >1$. One such model, with $n=3$, was constructed in 
Ref.~\cite{Alekseev:2005he}.\footnote{The higher couplings 
$\A_n(Q^2) = (a_{\rm pt}(Q^2)^n)_{\rm an}$ were not constructed there, though.}
Another such anQCD model, which we can call the one-delta model,
was constructed in Ref.~\cite{1danQCD}, and also satisfies Eq.~(\ref{diffA1a})
with $n=3$. The idea was simple: the perturbative discontinuity function
$\rho_1^{\rm (pt)}(\sigma) \equiv {\rm Im} \; a_{\rm pt}(Q^2=-\sigma - i \epsilon)$, for
$\sigma >0$, was kept unchanged for $\sigma$ down to a ``pQCD-onset-scale''
$M_0^2 \sim 1 \ {\rm GeV}^2$, and the unknown low-energy regime
$0 < \sigma < M_0^2$ was parametrized with one delta function
at a scale $M_1^2$ (such that: $0 < M_1^2 < M_0^2$)\footnote{ 
A similar idea was applied in Refs.~\cite{DeRafael,MagrDual}
directly to spectral functions of the vector current correlators.
Other approaches of eliminating unphysical singularities directly 
from specific (spacelike) obsevables, were presented in 
Refs.~\cite{Milton:2001mq,mes2,Nest3}.}
\be
\rho_1^{\rm (1 \delta)}(\sigma) = 
\pi f_1^2 \Lambda^2 \delta(\sigma - M_1^2) +
\Theta(\sigma - M_0^2) \times
\rho_1^{\rm (pt)}(\sigma) \ ,
\label{1drho1}
\ee
where for $a_{\rm pt}(Q^2)$ [the latter defining $\rho_1^{\rm (pt)}(\sigma)$] the
renormalization scheme with $c_2 = c_3 = \ldots = 0$ was used.
The (Lambert) scale parameter $\Lambda$ was fixed by the value of
$\alpha_s(M_Z^2, \MSbar)$. The dimensionless parameters $s_j \equiv M_j^2/\Lambda^2$
($j=0,1$) and $f_1^2$ were then fixed by the requirement of $n=3$ in
Eq.~(\ref{diffA1a}) (these are in fact two conditions) and by the
requirement that the central experimental value of the strangeless $(V+A)$
$\tau$-decay ratio $r_{\tau}$, namely $(r_{\tau})_{\rm exp} = 0.203 \pm 0.004$
\cite{ALEPH,DDHMZ},\footnote{
$r_{\tau}$ is the QCD part of the $V$+$A$
decay ratio $R_{\tau}(\Delta S=0)$, with the (small) quark mass effects 
subtracted. It is normalized in the canonical way: $r_{\tau} = a + {\cal O}(a^2)$.}
be reproduced in the model.

As argued at the end of the previous subsection, the ITEP interpretation
of the OPE requires that index $n$ in Eq.~(\ref{diffA1a})
be relatively high,\footnote{
Small-size instanton effects may lead to the conditions Eq.~ (\ref{diffA1a})
to be valid only up to index $n$ such that $2 n$ is
the largest dimension of condensates not affected by the instantons.
Instanton-antiinstanton gas scenarios lead to $n < 4 \beta_0$
($=9$ for $N_f=3$), cf.~Ref.~\cite{DMW}.}
$4 < n < 9$. If, instead of one delta, we use two deltas to parametrize
the otherwise unknown low-$\sigma$ regime behavior of the
spectral function $\rho_1(\sigma)$, we obtain an analytic QCD
model with $\A_1(Q^2)$ fulfilling the relation (\ref{diffA1a})
with $n=5$, Ref.~\cite{2danQCD}, 
\bea
\rho_1^{\rm (2 \delta)}(\sigma; c_2) &=&
\pi \sum_{j=1}^2 f_j^2 \Lambda^2 \; 
\delta(\sigma - M_j^2) +  \Theta(\sigma-M_0^2) \times 
\rho_1^{\rm (pt)}(\sigma; c_2) 
\label{rho1o1}
\\
& = & \pi \sum_{j=1}^2 f_j^2 \; \delta(s - s_j) +  
\Theta(s-s_0) \times r_1^{\rm (pt)}(s; c_2) \ .
\label{rho1o2}
\eea
Here, $s=\sigma/\Lambda^2$, and five dimensionless parameters are: 
$s_j = M_j^2/\Lambda^2$ ($j=0,1,2$) and $f_k^2$ ($k=1,2$).
Further, $c_2 = \beta_2/\beta_0$ is the scheme parameter,
and $r_1^{\rm (pt)}(s; c_2) =  \rho_1^{\rm (pt)}(\sigma; c_2)
= {\rm Im} \ a_{\rm pt}(Q^2=-\sigma - i \epsilon; c_2)$ is the perturbative
spectral function of $a_{\rm pt}(Q^2)$ in terms of $s=\sigma/\Lambda^2$. 
Here $ a_{\rm pt}(Q^2)$ is given by
\bea
a_{\rm pt}(Q^2) = - \frac{1}{c_1} \frac{1}{\left[1 - c_2/c_1^2 + W_{\mp 1}(z) \right]} \ ,
\label{aptexact}
\eea
where: $c_j=\beta_j/\beta_0$ ($j=1,2$); $Q^2 = |Q^2| \exp(i \phi)$;
and $W_{-1}$ and $W_{+1}$ are the branches of the Lambert function
for the case $0 \leq \phi < + \pi$ and $- \pi < \phi < 0$, respectively.
The argument $z = z(Q^2)$ in terms of $Q^2$ is given in Eq.~(\ref{zexpr}),
with $\Lambda$ there being the Lambert scale.
The solution (\ref{aptexact}) is
the solution to the RGE where
the beta function has the Pad\'e form $\beta(a_{\rm pt}) \propto
a_{\rm pt}^2 \times {\rm [1/1]}(a_{\rm pt})$
\bea
\frac{d a_{\rm pt}(Q^2)}{d \ln Q^2} & = &
- \beta_0 a_{\rm pt}^2 \frac{\left[ 1 + (c_1 - (c_2/c_1)) a_{\rm pt}
\right]}{\left[ 1 - (c_2/c_1) a_{\rm pt} \right]} \ .
\label{RGE}
\eea
The expansion of beta function $\beta(a_{\rm pt})$ 
on the right-hand side gives
\be
\beta(a_{\rm pt}) = 
- \beta_0 a_{\rm pt}^2 
\left(
1 + c_1 a_{\rm pt} + c_2 a_{\rm pt}^2 + c_3 a_{\rm pt}^3 + 
\ldots \right) \ ,
\label{betapt}
\ee
where the higher renormalization scheme parameters $c_j$ ($j \geq 3$)
are fixed by the leading scheme parameter $c_2$: $c_j = c_2^{j-1}/c_1^{j-2}$.
In the case of (F)APT of the previous subsection, $c_2=0$ was taken
(effectively as two-loop running). 
In the case of the model in this subsection,
the parameter $c_2$ will be varied in an interval, as specified below.
The spacelike
coupling $\A_1(Q^2)$ is then obtained by the dispersion relation
\bea
{{\A_1^{\rm {(2 \delta)}}}}(Q^2) &=& \frac{1}{\pi} \int_{\sigma= 0}^{\infty}
\; d \sigma \frac{\rho_1^{\rm (2 \delta)}(\sigma)}{(\sigma + Q^2)} 
= \sum_{j=1}^2 \frac{f_j^2}{(u +s_j)} +
\frac{1}{\pi} \int_{s_0}^{\infty} ds \; 
\frac{r_1^{\rm (pt)}(s;c_2)}{(s+u)} \ ,
\label{2dA1}
\eea
where we denoted $u = Q^2/\Lambda^2$. 
We call this model the two-delta analytic QCD model (2$\delta$anQCD).
It is to be applied
in the low-momentum regime, i.e., where $N_f=3$ (considering the
current masses of $u$, $d$ and $s$ zero): $|Q^2| < (2 m_c)^2$
($\approx 6.45 \ {\rm GeV}^2$). At $|Q^2| > (2 m_c)^2$ (where $N_f \geq 4$)
it is replaced by the underlying pQCD with the coupling
$a_{\rm pt}(Q^2)$, Eq.~(\ref{aptexact}).
The Lambert scale $\Lambda^2 = \Lambda^2_{N_f=3}$ ($\sim 0.1 \ {\rm GeV}^2$), 
which is the only dimensional parameter of the model, 
is determined by the world average value
$a_{\rm pt}(M_Z^2;\MSbar) = (0.1184 \pm 0.0007)/\pi$, \cite{PDG2010}.
For example, when $c_2=-4.76$, we obtain
$\Lambda \approx (0.260 \pm 0.008)$ GeV. On the other hand,
the five mentioned dimensionless parameters $s_j$ and $f_k^2$ 
are fixed, independently of the value of $\Lambda$, by the
condition Eq.~(\ref{diffA1a}) with $n=5$ (these are four conditions)
\be
\A_1^{(2 \delta)}(Q^2) - a_{\rm pt}(Q^2) \sim 
\left( \frac{\Lambda^2}{Q^2} \right)^5 \ , \qquad (|Q^2| > \Lambda^2)
\ ,
\label{diffAa2d}
\ee
and by the condition that the mentioned central experimental value 
of the strangeless
$(V+A)$ $\tau$-decay ratio $r_{\tau}$ be reproduced by the model, 
namely $r_{\tau} = 0.203$. More specifically, the (four) conditions
Eq.~(\ref{diffAa2d}) determine the four parameters
$s_j$ and $f_j^2$ ($j=1,2$) as a function of $s_0$;
the value of $s_0$ ($\equiv M_0^2/\Lambda^2$) is then
determined (if $\Lambda$ is already fixed) 
by the condition $r_{\tau} = 0.203$.
The (scheme) parameter $c_2$ remains the only free parameter of the model.

\begin{table}
\caption{Values of the parameters of the considered 2$\delta$anQCD model.
We consider $c_2=-4.76$ ($M_0=1.25$ GeV) as the central representative case. 
The Lambert scale values in the corresponding cases are for
the QCD coupling parameter value $\alpha_s^{({\overline {\rm MS}})}(M_Z^2) = 0.1184$.}
\label{t1}  
\begin{ruledtabular}
\begin{tabular}{l|llllllll}
$c_2=\beta_2/\beta_0$ & $s_0$ & $s_1$ & $f_1^2$ & $s_2$ & $f_2^2$  & $\Lambda$ [GeV] & $M_0$ & $\A_1(0)$
\\ 
\hline
-2.10 & 17.09 & 12.523 & 0.1815 & 0.7796 & 0.3462 & 0.363 & 1.50 & 0.544
\\
-4.76 & 23.06 & 16.837 & 0.2713 & 0.8077 & 0.5409  & 0.260 & 1.25 & 0.776
\\
-5.73 & 25.01 & 18.220 & 0.3091 & 0.7082 & 0.6312 & 0.231 & 1.15 & 1.00 
\end{tabular}
\end{ruledtabular}
\end{table}
In Table \ref{t1} we present the results of the model for three
representative values of the parameter $c_2$. The Lambert scale
parameter was fixed by using the central value $0.1184/\pi$ of the
world average 
$a_{\rm pt}(M_Z^2;\MSbar) = (0.1184 \pm 0.0007)/\pi$, \cite{PDG2010}.
It turns out that increasing $c_2$ increases the pQCD-onset scale
$M_0$ ($=s_0 \Lambda^2$), but decreases the coupling
$\A_1(0)$ at $Q^2=0$.  For phenomenological reasons, we prefer
to have the scale $M_0$ ($\sim 1$ GeV)  below the $\tau$ lepton mass
$m_{\tau}$, e.g., $1 \ {\rm GeV} \leq M_0 \leq 1.5$ GeV; and the value of $\A_1(0)$
below unity to avoid instabilities in the infrared. These two
restrictions give us the range of the free parameter $c_2$ 
between $-5.73$ and $-2.10$, as seen in Table \ref{t1}.\footnote{
In Ref.~\cite{2danQCD} we included the case $c_2=-7.15$, for
which $M_0=1.0$ GeV. However, in that case, $\A_1(0)=2.29$, and this
indicates that the model is unstable in the infrared at such $c_2$ values.}
Our central preferred value for $c_2$ is $c_2=-4.76$, for which the 
pQCD-onset scale becomes $M_0=1.25$ GeV.

The (scheme) parameter $c_2$ in Table \ref{t1} was varied
at fixed value $a_{\rm pt}(M_Z^2;\MSbar) = 0.1184/\pi$
so that the relations (\ref{diffAa2d}) and $r_{\tau} = 0.203$
were fulfilled. On the other hand, if we vary the 
coupling parameter within the world average interval,  
$a_{\rm pt}(M_Z^2;\MSbar) = (0.1184 \pm 0.0007)/\pi$,
and keep the dimensionless 
parameters of the model fixed (Table \ref{t1}, the line with $c_2= -4.76$),
the relation (\ref{diffAa2d}) remains fulfilled and
only the Lambert scale $\Lambda$ varies, this resulting in
$r_{\tau} = 0.203 \pm 0.006$. This is acceptably compatible with the 
measured values  $(r_{\tau})_{\rm exp} = 0.203 \pm 0.004$ \cite{ALEPH,DDHMZ}.

For further details on the 2$\delta$anQCD model we refer to Ref.~\cite{2danQCD}. 

The higher order couplings $A_{\nu} 
\equiv \left( a_{\rm pt}^{\nu} \right)_{\rm an}$
in this model are then 
obtained according to the construction for general anQCD models,
explained in Refs.~\cite{CV1,CV2} for integer $\nu$, 
and in Ref.~\cite{GCAK} for general (noninteger) $\nu$ and for the
couplings $\A_{\nu,k} \equiv \left( a_{\rm pt}^{\nu} \ln^k a_{\rm pt} \right)_{\rm an}$.
The details are given here in Appendix \ref{app0}. 

The analytization of expansions 
now consists simply in the replacements
\begin{subequations}
\label{analyt}
\bea
a^{\nu}_{\rm pt}(\mu^2) & \mapsto & 
\left( a^{\nu}_{\rm pt}(\mu^2) \right)_{\rm an} = \A_{\nu}(\mu^2) \ ,
\label{analyt1}
\\
a^{\nu}_{\rm pt}(\mu^2) \ln^k a_{\rm pt}(\mu^2) & \mapsto & 
\left( a^{\nu}_{\rm pt}(\mu^2) \ln^k a_{\rm pt}(\mu^2) \right)_{\rm an} 
= \A_{\nu,k}(\mu^2) \ .
\label{analyt2}
\eea
\end{subequations}
In the special case of APT, i.e., $\A_1^{\rm (APT)}(Q^2)$ of Ref.~\cite{ShS}, 
the analytization procedure
of Appendix \ref{app0} is, in principle,
equivalent to that of Eqs.~(\ref{MAanAnu})-(\ref{MAangen}), 
as argued in Refs.~\cite{CV2,GCAK}.
The (small) differences between the two analytizations
in APT arise due to somewhat 
different truncations applied in the two approaches. 
In the FAPT procedure of Eqs.~(\ref{MAanAnu})-(\ref{MAangen}),
applicable only in APT, 
usually the truncation in the loop expansion $(\leq a_{\rm pt}^{\ell+1})$
is applied to the running coupling 
$a_{\rm pt}(\mu^2)$, which is then reflected in the spectral functions 
${\rm Im}  \; a_{\rm pt}^{\nu}(-\sigma-i \epsilon)$ and
${\rm Im} [ a_{\rm pt}^{\nu}(-\sigma-i \epsilon) \ln^k a_{\rm pt}(-\sigma-i \epsilon) ]$
in Eqs.~(\ref{MAanAnu})-(\ref{MAangen}). In the general approach,
Eqs.~(\ref{tAnu1})-(\ref{Anulnk}) of Appendix \ref{app0}, 
which can be applied also to the APT analytic model
of $\A_1(Q^2)^{\rm (APT)}$ of Ref.~\cite{ShS},
the same kind of truncation for $a_{\rm pt}(\mu^2)$ can be applied in 
$\rho_1^{\rm (pt)}(\sigma) = {\rm Im} \; a_{\rm pt}(Q^2=-\sigma - i \epsilon)$,
in the integral (\ref{tAnu1}); in addition,
the corresponding loop-truncation
in the summations (\ref{AnutAnu}) and (\ref{Anulnk}) is applied --
usually $m \leq \ell - \nu$. 

\begin{figure}[htb]
\includegraphics[width=100mm]{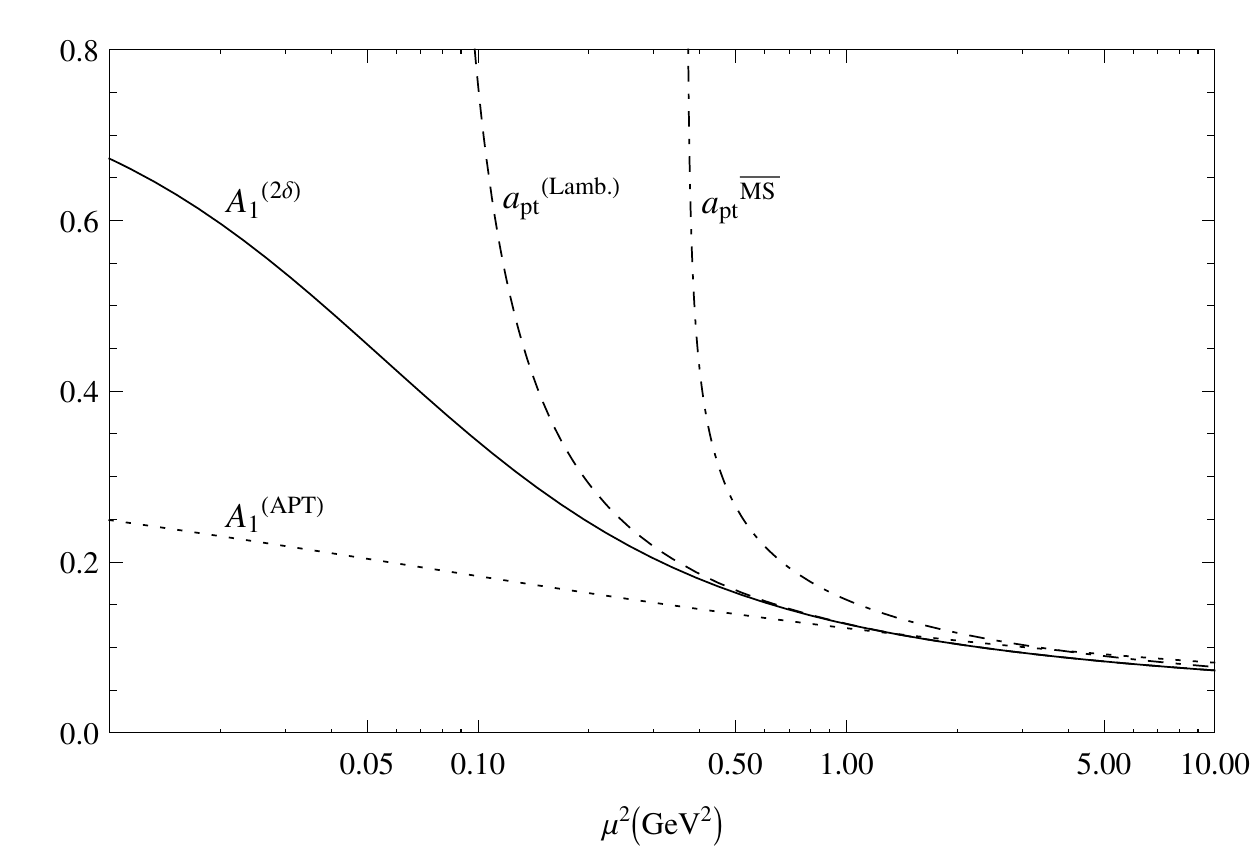}
\vspace{-0.4cm}
\caption{\footnotesize
Analytic couplings $\A_1(\mu^2)$ at low positive $\mu^2$
($\equiv Q^2$),
for APT and the 2$\delta$anQCD model, for the central input values. 
Also included are pQCD couplings $a_{\rm pt}(\mu^2)$
in the $\MSbar$ and Lambert schemes.}
\label{plaA1}
\end{figure}
\begin{figure}[htb] 
\begin{minipage}[b]{.49\linewidth}
\centering\includegraphics[width=80mm]{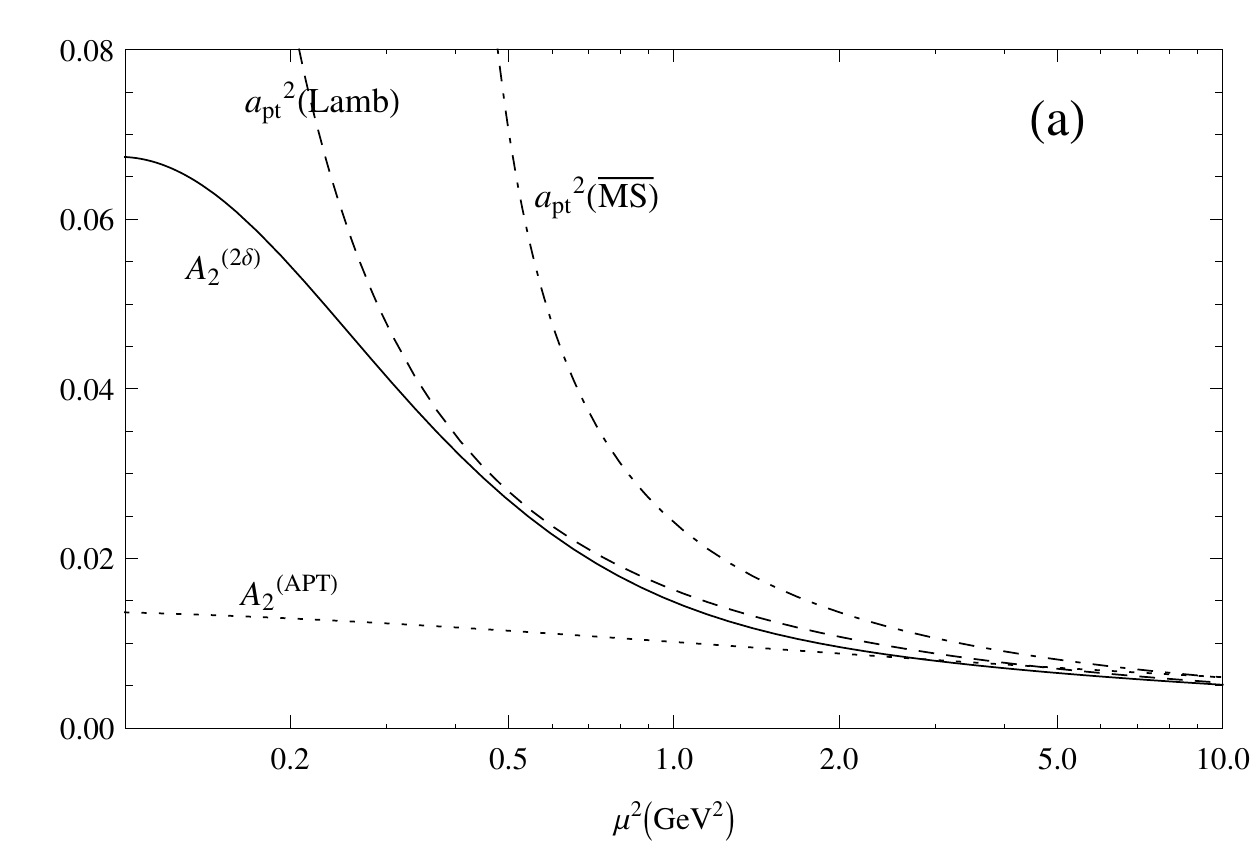}
\end{minipage}
\begin{minipage}[b]{.49\linewidth}
\centering\includegraphics[width=80mm]{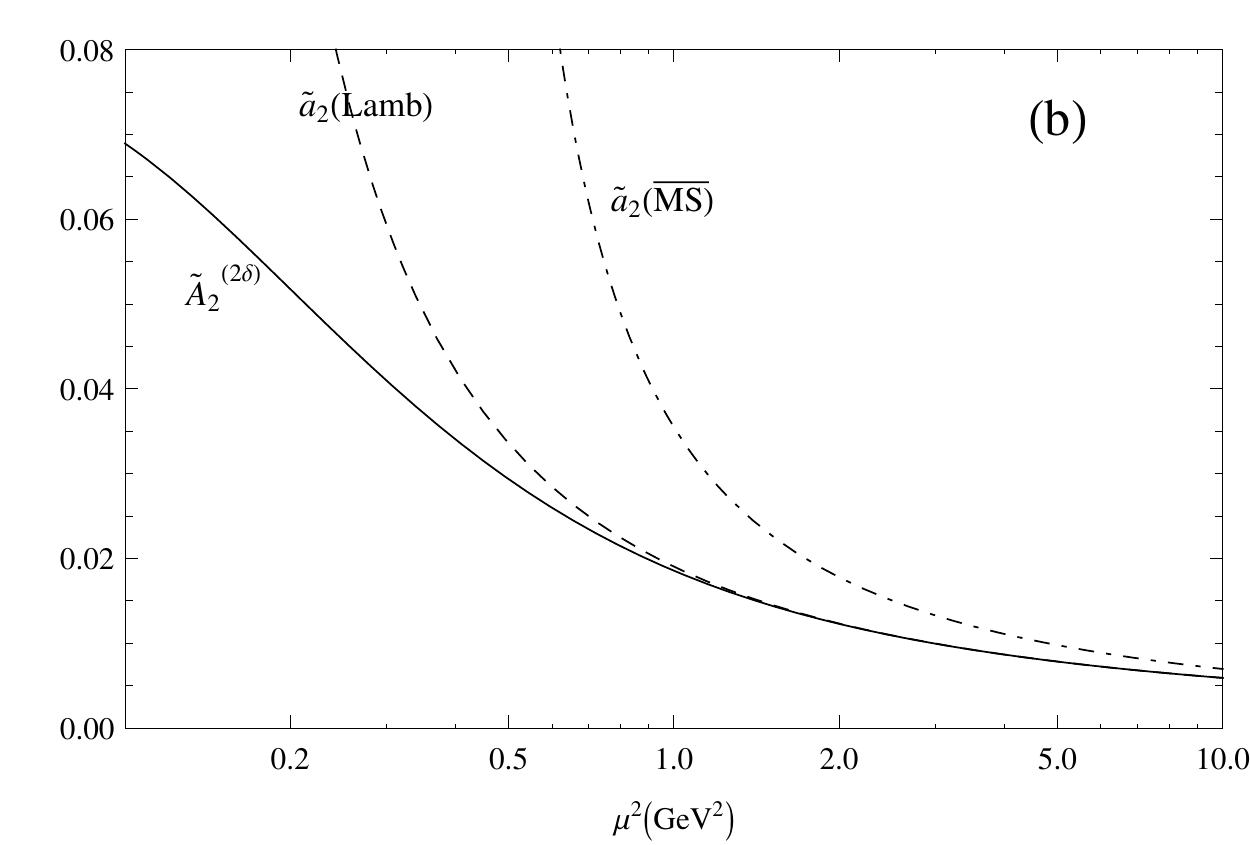}
\end{minipage}
\vspace{-0.4cm}
 \caption{(a) Squares of the pQCD couplings,
$a^2_{\rm pt}$, and their analytic analogs $\A_2$, at low positive $\mu^2$.
(b) Logarithmic derivatives $\ta_{\rm pt,2}$ and $\tA_2$
at low positive $\mu^2$. (See Appendix \ref{app0} for these definitions.)}
\label{plaA2}
 \end{figure}
Figure~\ref{plaA1} shows the analytic running couplings $\A_1(\mu^2)$
of the 2$\delta$anQCD model and APT, 
at positive low values of $\mu^2$ ($\equiv Q^2$), 
for the central input values. 
For comparison, the pQCD couplings 
in the Lambert renormalization scheme (the scheme
of 2$\delta$anQCD, $c_2=-4.76$) 
and in the four-loop $\MSbar$ scheme (at $N_f=3$)
are included.
Figure~\ref{plaA2}(a) shows the
higher couplings $\A_2$ and $a_{\rm pt}^2$,
and Fig.~\ref{plaA2}(b) the logarithmic derivatives 
$\ta_{{\rm pt},2}(\mu^2)$ and $\tA_2(\mu^2)$, which are defined
in Eqs.~(\ref{tan}) and (\ref{tAn}) in Appendix \ref{app0}. In APT
the coupling $\A_2$ was constructed according to Eq.~(\ref{MAanAnu}),
and in the 2$\delta$anQCD model
according to Eq.~(\ref{A2}) truncated at the $\tA_5$ term. 

These figures show that the analytic couplings in the infrared regime,
in comparison with the pQCD couplings,
are very suppressed in (F)APT, and less so in the 2$\delta$anQCD model.\footnote{
Figure \ref{plaA2}(b) shows that the logarithmic derivatives
$\tA_2(\mu^2)$ and $\ta_{{\rm pt},2}(\mu^2)$ in the Lambert scheme
are almost indistinguishable from each other at 
$\mu^2 > 1 \ {\rm GeV}^2$, this being the consequence of
the construction of the 2$\delta$anQCD model: $\A_1(\mu^2) - a_{\rm pt}(\mu^2)
\sim (\Lambda^2/\mu^2)^5$ for $|\mu^2| > \Lambda^2$. Namely, application of
$d/d \ln \mu^2$ to this relation implies $\tA_2(\mu^2) - \ta_{{\rm pt},2}(\mu^2)
\sim (\Lambda^2/\mu^2)^5$. On the other hand, Fig.~\ref{plaA2}(a) shows that
the difference between $\A_2(\mu^2)$ and $a_{\rm pt}^2(\mu^2)$ is significant
even up to $\mu^2 \approx 5 \ {\rm GeV}^2$, this being the consequence of 
the aforementioned truncation at $\tA_5$ in the construction of $\A_2$
(Appendix \ref{app0}).}

The 2$\delta$anQCD model has been successfully applied \cite{anOPE}
in the analysis of $(V+A)$
$\tau$-decay data via Borel sum rules with OPE.

 \section{Perturbation expansion for heavy $q {\bar q}$ ground state energy}
\label{sec:Eqq}

\subsection{General formulas}
\label{subsec:GF}

The analysis of nonrelativistic potential is the starting point 
for the determination of the ground state energy
of $\bar{q}q$ and  thus of the mass of such systems.
The main input in these calculations is the 
mass $\m_q(\m_q^2)$ (here denoted simply as $\m_q$), also
called $\MSbar$ quark mass.
The masses of the heavy quarkonia $\bar{q}q$ 
[$\Upsilon(1S) $ when $q=b$, $J/ \psi(1S)$ when $q=c$] are
well measured, and this allows us to extract the corresponding
mass $\m_q$. By evaluating an observable, 
such as the quark-antiquark binding energy here, 
within anQCD models, at least part of the 
(chirality-conserving) nonperturbative effects get included in
the leading-twist term via the analytization, such as  
Eqs.~(\ref{MAanAnu})-(\ref{MAangen}) in (F)APT, and
(\ref{analyt}) in general analytic QCD models.

The coefficients in the (leading-twist) perturbation expansion of the
ground state binding energy $E_{q {\bar q}}$ (i.e., with:
$n=1$ and $\ell=0$) of heavy quarkonium $q {\bar q}$
in powers of $a_{\rm pt}$ were obtained
up to all terms ${\cal O}( a_{\rm pt}^4)$ in Ref.~\cite{PinYnd}, 
the terms  ${\cal O}( a_{\rm pt}^5 \ln a_{\rm pt})$ in Ref.~\cite{BPSV},
and all terms up to ${\cal O}(a_{\rm pt}^5)$ (including logarithmic)
are given in Ref.~\cite{Penin:2002zv}. 
The last term ($\sim a_{\rm pt}^5$) is now completely known
since the parameter $a_3$ from the static potential
is now known, Refs.~\cite{Smirnov:2009fh,Anzai:2009tm}.
The general structure of the (leading-twist term of the)
ground state binding energy in pQCD is 
\begin{eqnarray}
E_{q \bar q}^{\rm (pt)} & = & - \frac{4}{9} m_q \pi^2 a_{\rm pt}^2(\mu^2) 
{\Big \{} 1 + a_{\rm pt}(\mu^2) 
\left[ K_{1,0} + K_{1,1} L_{\rm pt}(\mu^2) \right] +
a_{\rm pt}^2(\mu^2) 
\left[ K_{2,0} + K_{2,1} L_{\rm pt}(\mu^2) + K_{2,2} L_{\rm pt}^2(\mu^2) \right]
\nonumber\\
&& + a_{\rm pt}^3(\mu^2) 
\left[ K_{3,0,0} + K_{3,0,1} \ln a_{\rm pt}(\mu^2)+ K_{3,1} L_{\rm pt}(\mu^2) + 
K_{3,2} L_{\rm pt}^2(\mu^2) 
+ K_{3,3} L_{\rm pt}^3(\mu^2) \right] + {\cal O}(a_{\rm pt}^4) {\Big \}} \ ,
\label{Eqqpt}
\end{eqnarray}
where 
\begin{eqnarray}
L_{\rm pt}(\mu^2; m_q^2) &=& \frac{1}{2}
\ln \left( \frac{\mu^2}{((4 \pi/3) m_q)^2 a_{\rm pt}^2(\mu^2) } \right) \ ,
\label{Lp}
\end{eqnarray}
$\mu^2$ is the (square of the) renormalization scale,
$m_q$ is the pole mass of the quark, 
and the coefficients $K_{j,k}$ can be obtained by combining the results of
the mentioned literature; cf.~Appendix \ref{app1}.
The typical scale of the process is a soft reference scale
$Q_s^2$ ($\equiv - q^2$), which is 
a typical quark-antiquark momentum transfer 
inside the quarkonium ($Q_s^2 \sim m_q^2 \alpha_s^2$) and
can be fixed by convention. The soft
renormalization scale $\mu^2 \equiv \mu_s^2$ can then be varied around
$Q_s^2$
\be
\mu^2 \equiv \mu_s^2 = \kappa Q_s^2 \quad
(\kappa \sim 1; Q_s^2 \sim m_q^2 \alpha_s^2) \ .
\label{mus1}
\ee
The quarkonium mass is then
\be
M_{q {\bar q}} = 2 m_q + E_{q {\bar q}}(m_q) \ .
\label{Mqq1}
\ee
In principle, the input quantity here could be the quark pole mass $m_q$. 
However, this mass $m_q$, in contrast to the mass $\m_q$, 
suffers from the strong infrared renormalon ambiguity (at Borel parameter
value $b=1/2$, $\Rightarrow \delta m_q \sim \Lambda_{\rm QCD}$). This ambiguity must
cancel in the physical sum [Eq.~(\ref{Mqq1})]. 

It is more convenient
to use as input a renormalon ambiguity-free input mass, such as $\m_q$,
and we will do this.

In the case of the bottom quark, before we relate the pole mass
$m_b$ with the mass $\m_b$, at renormalization
energies $\mu = \m_b$ where $N_f=4$ (and where the charm quark mass 
is considered zero, i.e., decoupled), the effects 
$\delta m_b$ of the nonzero massa $m_c \not=0$ have to be subtracted,
and they are \cite{Brambilla:2001qk}
\be
(\delta m_b)_{m_c} = (\delta m_b)_{m_c}^{(1)} + 
(\delta m_b)_{m_c}^{(2)}
\approx 0.025 \pm 0.005 \ {\rm GeV} \ .
\label{dmbmc}
\ee
These effects were calculated in Ref.~\cite{Brambilla:2001qk} in pQCD,
at the hard  renormalization scale $\mu^2 = \m_b^2$.
We checked that they do not get significantly modified in APT
and in the 2$\delta$anQCD model.

The pole mass $m_q$ and the mass $\m_q$
are then related via the relation
\begin{equation}
\frac{m_q - \delta m_q}{\m_q} =
1 + \frac{4}{3} \left( a_{\rm pt}(\m_q^2) + r_1 a^2_{\rm pt}(\m_q^2) + 
r_2 a_{\rm pt}^3(\m_q^2)
+ r_3 a_{\rm pt}^4(\m_q^2) \right) + {\cal O}(a_{\rm pt}^5) \ ,
\label{mqbmq}
\end{equation}
where $\delta m_q$ is zero when $q=c$, and is given by Eq.~(\ref{dmbmc})
when $q=b$
\be
\delta m_q \equiv
\left\{
\bear{lc}
(\delta m_b)_{m_c} & (q=b) \\
0 & (q=c)
\eear
\right\}
\ .
\label{dmq}
\ee
The coefficient $R_0 = 4/3$ was obtained in Ref.~\cite{Tarrach:1980up}.
Also the coefficients $r_1(\m_q^2)$, and $r_2(\m_q^2)$ are known 
(Refs.~\cite{Gray:1990yh},
\cite{Chetyrkin:1999ys, Chetyrkin:1999qi, Melnikov:2000qh}, respectively)
they are given in Eqs.~(\ref{Sm}) in Appendix \ref{app2},
and $N_f$ in these coefficients is the number of
flavors of quarks lighter than $q$. Specifically, the values are:
$r_1=7.739$ and $r_2=87.224$ for $N_f=3$; and $r_1=6.958$ and $r_2=70.659$
for $N_f=4$. On the other hand, in Appendix \ref{app2}
we estimate the values of $r_3$ to a reasonably 
high level of precision (with less than 4 \% uncertainty) by
a method which uses the structure of the leading infrared renormalon
(at $b=1/2$) of the quantity $m_q/\m_q$:
$r_3(N_f=3) = 1339.4$, $r_3(N_f=4) = 987.3$.

While the relation (\ref{mqbmq}) is written here at the
``hard'' renormalization scale $\mu^2 = \m_q^2$, it is straightforward
to reexpress the sum on the right-hand side of Eq.~(\ref{mqbmq})
at a different, lower, scale $\mu^2$.
In Appendix \ref{app2} this reexpression is presented explicitly,
under the assumption that during the lowering of the scale, $\m_q^2 \to \mu^2$,
we do not cross the quark threshold. The goal is to express in the 
perturbation expansion of the binding 
energy (\ref{Eqqpt}), where the renormalization
scale is soft, the pole mass $m_q$ via the mass $\m_q$, and for this
we need the relation (\ref{mqbmq}) at the soft renormalization scale.
It turns out that for the $b {\bar b}$ system the hard
scale is $\m_b \approx 4$ GeV, i.e., the scale where $N_f=4$; and the
soft scale, Eq.~(\ref{mus1}), is $\mu_s \approx 2$ GeV, i.e., the scale
where it is more reasonable to expect $N_f=3$.\footnote{
Usually, the quark thresholds are taken at $2 m_q$. In the case of
$N_f=4 \mapsto 3$ transition, this is about $2.5$ GeV, above the
soft scale of the $b {\bar b}$ system.} In this case, we take into account 
also the (three-loop) quark threshold transition 
$N_f=4 \mapsto 3$ at $\mu^2 = (2 m_c)^2$, Ref.~\cite{CKS}. We thus
obtain the relation (\ref{mqbmq}) reexpressed
at the soft scale $\mu_s^2$ of Eq.~(\ref{mus1})
\be
m_q(\m_b^2; \mu_s^2) = \delta m_q + \m_q \left\{ 
1 + \frac{4}{3} a_{\rm pt}(\mu_s^2)
\left[ 1 + a_{\rm pt}(\mu_s^2) r_1(\mu_s^2) 
+ a_{\rm pt}^2(\mu_s^2) r_2(\mu_s^2) 
+ a_{\rm pt}^3(\mu_s^2) r_3(\mu_s^2) \right]  
+ {\cal O}(a_{\rm pt}^5) 
\right\} 
\ .
\label{Smmus}
\ee
Further, the renormalization scheme can also be varied in this relation
and in the relations (\ref{Eqqpt})-(\ref{Lp}),
i.e., the changes of the scheme parameters $c_j = \beta_j/\beta_0$ ($j=2,3$)
from the usual $\MSbar$ scheme to other schemes 
affect correspondingly the values of the coupling $a_{\rm pt}(\mu_s^2)$ and
of the coefficients. We recall that in (F)APT the 
chosen scheme here is $c_2=c_3 = \cdots =0$; 
in the 2$\delta$anQCD model and in Lambert pQCD, the scheme
is $c_2=-4.76^{+2.66}_{-0.97}$ and $c_j= c_2^{j-1}/c_1^{j-2}$ for $j \geq 3$.
The relationship between $a_{\rm pt}$'s at two different scales and in
two different renormalization schemes is summarized in
Appendix \ref{app3}, where we also summarize the (three-loop) connection
of $a_{\rm pt}$'s across the quark threshold.

After performing all these transformations, we can
rewrite the original expansion (\ref{Eqqpt}) for $E_{q \bar q}$
in terms of the $\m_q \equiv \m_q(\m_q^2)$ mass, 
with the coupling $a_{\rm pt}$ at any soft renormalization scale $\mu_s$
and in any chosen renormalization scheme ($c_2, c_3, \ldots$)
\begin{eqnarray} 
\lefteqn{
E_{q \bar q}^{\rm (pt)}(Q_s^2; \m_q^2; N_f=3) =  
- \frac{4}{9} (\m_q + \delta m_q) \pi^2 a_{\rm pt}^2(\mu_s^2) 
{\Big \{} 1 + a_{\rm pt}(\mu_s^2) 
\left[ {\cal K}_{1,0} + {\cal K}_{1,1} {\cal L}_{\rm pt}(\mu_s^2) \right]
}
\nonumber\\
&&+ a_{\rm pt}^2(\mu_s^2) 
\left[ {\cal K}_{2,0} + {\cal K}_{2,1} {\cal L}_{\rm pt}(\mu_s^2) + {\cal K}_{2,2} {\cal L}_{\rm pt}^2(\mu_s^2) \right]
\nonumber\\
&& + a_{\rm pt}^3(\mu_s^2) 
\left[ {\cal K}_{3,0,0} + {\cal K}_{3,0,1} \ln a_{\rm pt}(\mu_s^2)+ {\cal K}_{3,1} {\cal L}_{\rm pt}(\mu_s^2) + 
{\cal K}_{3,2} {\cal L}_{\rm pt}^2(\mu_s^2) 
+ {\cal K}_{3,3} {\cal L}_{\rm pt}^3(\mu_s^2) \right] + {\cal O}(a_{\rm pt}^4) {\Big \}} \ ,
\label{calEqqpt}
\end{eqnarray}
where $\mu_s^2 = \kappa Q_s^2$ ($\kappa \sim 1$ being the soft renormalization scale
parameter),
and the logarithm contains now $\m_q$ mass
\begin{eqnarray}
{\cal L}_{\rm pt}(\mu_s^2; \m_q^2) &=& \frac{1}{2}
\ln \left( 
\frac{\mu_s^2}
{\left[ (4 \pi/3) (\m_q+\delta m_q) \right]^2 a_{\rm pt}^2(\mu_s^2) } 
\right) \ .
\label{calLp}
\end{eqnarray}
We note that the new (renormalon-ambiguity-free) mass
which appears naturally in this expansion is not exactly
the mass $\m_q$, but rather
\be
{\widetilde m}_q \equiv \m_q + \delta m_q 
=
\left\{
\bear{lc}
\m_b + (\delta m_b)_{m_c} & (q=b) \\
\m_c & (q=c)
\eear
\right\}
\ ,
\label{tmq}
\ee
where $(\delta m_b)_{m_c}$ is given by Eq.~(\ref{dmbmc}) when $q=b$.

The mentioned soft ``process scale'' $Q_s^2$ ($\sim \m_q^2 \alpha_s^2$)
can be regarded, at least formally, to be a variable complex scale. 
Therefore, the binding energy $E_{q \bar q} (Q_s^2; \m_q^2)$ is, formally, 
a spacelike observable analytic in $Q_s^2$; and the dependence on the
renormalization scale parameter $\kappa$ disappears when the number of terms
in the expansion is infinite.
The analytization of $E_{q \bar q}^{\rm (pt)}(Q_s^2; \m_q^2)$ of
Eq.~(\ref{calEqqpt}), according to Eqs.~(\ref{analyt}) 
[or, in (F)APT: Eqs.~(\ref{MAanAnu})-(\ref{MAangen})], then leads to
\begin{eqnarray}
\lefteqn{
E_{q \bar q}^{\rm (an)}(Q_s^2;\m_q^2) =  - \frac{4}{9} {\widetilde m}_q \pi^2  
{\Big \{} \A_{2}(\kappa Q_s^2) + 
\left[ {\cal K}_{1,0}  \A_{3}(\kappa Q_s^2)  + {\cal K}_{1,1} \B_{3,1}(\kappa Q_s^2)  \right]
} 
\nonumber\\
&&+ \left[ {\cal K}_{2,0} \A_{4}(\kappa Q_s^2)  + {\cal K}_{2,1} \B_{4,1}(\kappa Q_s^2)  + 
{\cal K}_{2,2} \B_{4,2}(\kappa Q_s^2)  \right]
\nonumber\\
&&
+ \left[  {\cal K}_{3,0,0} \A_{5}(\kappa Q_s^2)  +  {\cal K}_{3,0,1} \A_{5,1}(\kappa Q_s^2) +
{\cal K}_{3,1} \B_{5,1}(\kappa Q_s^2)  + {\cal K}_{3,2} \B_{5,2}(\kappa Q_s^2)  + 
{\cal K}_{3,3} \B_{5,3}(\kappa Q_s^2)  \right]
+ {\cal O}(\A_{6,4}) {\Big \}} \ ,
\label{calEqqan}
\end{eqnarray}
where we use, for simplicity, the notation
of Eq.~(\ref{AnutAnu}) with Eq.~(\ref{tAn}) for $\A_{\nu}$'s ($\nu=k$ integer now)
[in (F)APT: Eq.~(\ref{MAanAnu})], and denote by $\B_{n+2,j}$ the following:
\be
\B_{n+2,j}(\kappa Q_s^2) = 
\left( 
a^{n+2}_{\rm pt}(\kappa Q_s^2) \frac{1}{2^j}
\ln^j 
\left( \frac{\kappa Q_s^2}{((4 \pi/3) \m_q)^2 a_{\rm pt}^2(\kappa Q_s^2) } \right) 
\right)_{\rm an}
= 
\frac{1}{2^j} \sum_{s=0}^j 
\left(
\begin{array}{c}
j \\
s
\end{array}
\right)
 (-2)^s \; {\bar f}^{j-s}(\kappa Q_s^2) \A_{n+2,s}(\kappa Q_s^2) \ ,
\label{Bn2j}
\ee
where ${\bar f} \equiv  \ln [ \kappa Q_s^2/(4 \pi /3)^2 /\m_q^2 ]$,
and $\A_{n+2,s}$ were defined in Eq.~(\ref{Anulnk}) [in (F)APT: Eq.~(\ref{MAangen})].

Here, $j=1,\ldots,n$, therefore $\B_{n+2,j}(\kappa Q_s^2) \to 0$
faster than $a_{\rm pt}^2(\kappa Q_s^2)$ when $|Q_s^2| \to \infty$, 
by asymptotic freedom.

The relation (\ref{Smmus}) in its analytized form is
\begin{equation}
m_q(\m_b^2; \mu_s^2) = \delta m_q + \m_q 
\left\{ 
1 + \frac{4}{3} \left[ \A_1(\mu_s^2) + r_1(\mu_s^2) \A_2(\mu_s^2) + 
r_2(\mu_s^2) \A_3(\mu_s^2)
+ r_3(\mu_s^2) \A_4(\mu_s^2) \right] 
+ {\cal O}(\A_5) 
\right\} \ .
\label{Smmusan}
\end{equation}
Since this relation includes at its highest order the term $\sim \A_4$
($\sim a_{\rm pt}^4$), in general analytic QCD it is consistent to use in Eq.~(\ref{Smmusan}) 
for the expressions $\A_n$ ($n=1,2,3,4$) those given in Eq.~(\ref{AnutAnu}) 
[or: Eqs.~(\ref{Atot})] with the 
sum there truncated at (and including) $\tA_4$ ($\sim \A_4 \sim a_{\rm pt}^4$). 
On the other hand, the expression (\ref{calEqqan}) for the binding energy 
includes at its highest order the terms $\sim \A_5$ (more precisely, $\sim \A_{5,3}$), 
therefore it is consistent to use there for the expressions 
$\A_n$ ($2 \leq n \leq 5$) those of Eq.~(\ref{AnutAnu}) [or Eqs.~(\ref{Atot})]
with the sum truncated at 
(and including) $\tA_5$, and expressions $\tA_{n,k}$ of Eq.~(\ref{Anulnk}) 
truncated at  (and including)  $\partial^k \tA_{\nu+5}/\partial \nu^k|_{\nu=0}$ 
($\sim a_{\rm pt}^n \ln^k a_{\rm pt}$), where $k=1,2, 3$.

The mentioned soft reference scale $Q_s^2$ ($\sim \m_q^2 \alpha_s^2$)
can be fixed in pQCD, by convention, by the condition that all the logarithmic
terms in the expansion (\ref{calEqqpt}) disappear
\be
Q^2_{s,{\rm pt}} = (4 \pi/3)^2  {\widetilde m}_q^2 a_{\rm pt}^2(Q_{s, {\rm pt}}^2 ) \ ,
\label{Qspt}
\ee 
where ${\widetilde m}_q$ is defined in Eq.~(\ref{tmq}).
This type of condition, in analytic QCD, would correspond to fixing 
the soft reference 
scale $Q_s$ by requiring ${\cal B}_{n+1,j}(Q_s^2)=0$, for various $n$'s and 
$j=1,\ldots,n$. This fixing is not unique since it depends on $n$ and $j$.
Our convention will be that the leading logarithmic term  
in Eq.~(\ref{calEqqan}) is zero at such scale
\be
{\cal B}_{3,1}(Q_s^2) = 0 \ .
\label{Qsa}
\ee
It will turn out that this condition has a solution in the case of $b {\bar b}$
($\Upsilon(1S)$) in the 2$\delta$anQCD model, 
but not in $J/\psi(1S)$ in that model, and not in
any case of the (F)APT model. In such respective cases, we will use simply the
following simpler analogs of the pQCD condition (\ref{Qspt}):
\bea
Q^2_{s} &=& 
(4 \pi/3)^2  {\widetilde m}_q^2 \tA_2(Q_s^2 ) \quad (J/\psi(1S) \ {\rm in \ 2\delta anQCD}) \ ,
\label{Qsb}
\\
Q_{s,{\rm ((F)APT)}}^2 &=& 
(4 \pi /3)^2  {\widetilde m}_q^2 \A_2^{\rm ((F)APT)}(Q_{s,{\rm ((F)APT)}}^2)
\quad ({\rm (F)APT}) \ .
\label{QsMA}
\end{eqnarray}
These measures of the typical momentum scale of the (nonrelativistic)
quark inside the quarkonium are rather low, $\approx 2$ GeV 
in $\Upsilon(1S)$, and $\alt 1$ GeV in $J/\psi(1S)$.
In pQCD such scales are problematic, because they are not far away
from the unphysical (Landau) singularities
of $a_{\rm pt}(Q^2)$; in analytic QCD models, no such problems appear in principle.

\subsection{Separation of the soft and ultrasoft contributions}
\label{subsec:sus}

The pole mass $m_q$ and the static potential
$V(r)$ both contain the leading
infrared renormalon ($b=1/2$) singularities, and
cancellation of these singularities takes place 
in the sum $2 m_q + V(r)$ \cite{Hoang:1998nz,Brambilla:1999xf,Beneke:1998rk}.
As a consequence,
this cancellation must take place also in the quarkonium mass
$2 m_q + E_{q {\bar q}}$ 
\cite{Kiyo:2000fr,Brambilla:2001fw,Brambilla:2001qk,Pineda:2001zq,Lee:2003hh,Contreras:2003zb}, 
more specifically, in the sum $2 m_q + E_{q {\bar q}}(s)$ where  
$E_{q {\bar q}}(s) = \langle 1 | V(r) | 1 \rangle$ [$\sim V(r_s)$] is the soft part
of the binding energy $E_{q {\bar q}}$, and $|1 \rangle$ denotes
the (ground) state of the quarkonium. The typical soft distance
$r_s$ in the quarkonium is $\sim 1/\sqrt{Q_s^2} \sim 1/(\m_q \pi a)$,
where $a = a_{\rm pt}$ or $\A_1$.
Since $V(r_s) \propto 1/r_s$, we have 
$E_{q {\bar q}}(s)/\m_q \sim V(r_s)/\m_q \propto 1/(r_s \m_q) \sim (\pi a)$.
This leads to the so called ``power mismatch'' in
the renormalon cancellation in the pQCD expansion of the sum
$(2 m_q + E_{q {\bar q}}(s))/\m_q$ \cite{Hoang:1998ng} 
(see also: \cite{Kiyo:2000fr}): the terms $\sim a^n$
in $2 m_q/\m_q$ tend to cancel numerically the terms
$\sim a^{n+1}$ in $E_{q {\bar q}}(s)/\m_q$. Therefore, since the
binding energy $E_{q {\bar q}}$ is now known up to $\sim a_{\rm pt}^5$,
it is very convenient to have the relation $m_q/\m_q$
up to $\sim a^4$, i.e., to have a good estimate for the coefficient $r_3$
and to use it, so that the effects of renormalon cancellation 
in $2 m_q + E_{q {\bar q}}(s)$ can be seen numerically more clearly.
This was the main reason for performing the analysis in 
Appendix \ref{app2} resulting in estimates of $r_3$,
Eq.~(\ref{r3}).
This cancellation, term by term, should be numerically more precise 
(at sufficiently high orders) if the renormalization scales $\mu_s$
used in $m_q/\m_q$ and in $E_{q {\bar q}}(s)/\m_q$ 
[Eqs.~(\ref{Smmus}) and (\ref{calEqqpt})] are taken to be equal.
This renormalon cancellation will be our guiding principle for the
separation of the soft ($s$) and the ultrasoft ($us$) part in
the binding energy
\be
E_{q {\bar q}} = E_{q {\bar q}}(s) + E_{q {\bar q}}(us) \ .
\label{sus1}
\ee
Typical $us$ scales are $\mu_{us} \sim \m_q  \alpha_s^2$, and the $us$
part of the binding energy is $\sim \m_q a^5 \ln a$. 
We can parametrize the $s$-$us$ separation by a dimensionless
parameter $k_{s/us}$ such that the $s$-$us$ factorization
scale $\mu_f$ is written as 
\be
\mu_f  = k_{s/us} \m_q 
\alpha_s(Q_s)^{3/2} \left[ \approx k_{s/us} (Q_s Q_{us})^{1/2} \right] \ ,
\label{muf}
\ee 
where $Q_{us} \sim \m_q \alpha_s^2$ is a (chosen) $us$ reference scale.
It is expected that usually $k_{s/us} \sim 1$, but it does not have to be so always.
The $us$ part can be rewritten, in terms of $k_{s/us}$ as 
(cf.~Ref.~\cite{Contreras:2003zb})
\bea
E^{\rm (pt)}_{q {\bar q}}(us) &=& - \frac{4}{9} {\widetilde m}_q
\pi^2 \left[
{\cal K}_{3,0,0}(us) a_{\rm pt}^5(\mu_{us}^2) + {\cal K}_{3,0,1}(us) 
 a_{\rm pt}^5(\mu_{us}^2) \ln  a_{\rm pt}(\mu_{us}^2) + {\cal O}(a_{\rm pt}^6)
\right] \ ,
\label{us1}
\eea
and in analytic QCD correspondingly
\bea
E^{\rm (an)}_{q {\bar q}}(us) &=& - \frac{4}{9} {\widetilde m}_q
\pi^2 \left[
{\cal K}_{3,0,0}(us) \A_5(\mu_{us}^2) + {\cal K}_{3,0,1}(us) 
 \A_{5,1}(\mu_{us}^2) + {\cal O}(\A_6)
\right] \ ,
\label{us2}
\eea
where $\mu_{us} = \kappa_{us} Q_{us}$ is a $us$ renormalization scale 
($\kappa_{us} \sim 1$), and the two $us$ coefficients are
\be
{\cal K}_{3,0,1}(us) =  7.098 \pi^3 \ , \quad
{\cal K}_{3,0,0}(us) =  \left[ 27.512 + 7.098 \ln \pi - 14.196 \ln(k_{s/us}) 
\right] \pi^3 \ .
\label{Kus}
\ee
The expansion of the soft ($s$) part $E_{q {\bar q}}(s)$
of the binding energy is then, according to Eq.~(\ref{sus1}),
the same as the expansions (\ref{calEqqpt}) and (\ref{calEqqan}), 
with the exception of the
replacements\footnote{
The coefficients ${\cal K}_{3,0,0}(us)$ and ${\cal K}_{3,0,1}(us)$,
representing the leading part of the (quasi)observable $E_{q \bar q}(us)$,
are renormalization scale ($\mu_{us}$) and scheme independent.} 
of two coefficients ${\cal K}_{3,0,0}$ and ${\cal K}_{3,0,1}$
\bea
E_{q \bar q}^{\rm (pt,an)}(s) & = &  E_{q \bar q}^{\rm (pt,an)}
\left(
{\cal K}_{3,0,0} \mapsto  {\cal K}_{3,0,0}-{\cal K}_{3,0,0}(us);
{\cal K}_{3,0,1} \mapsto  {\cal K}_{3,0,1}-{\cal K}_{3,0,1}(us)
\right) \ .
\label{calEqqspt}
\eea
The $s$-$us$ factorization, i.e., the parameter $k_{s/us}$, will then be determined,
in each model, by requiring that the leading infrared renormalon cancellation
in $2 m_q + E_{q \bar q}(s)$ be exact at the last available order, i.e.,
that the ${\cal O}(a^4)$ term in $2 m_q(Q_s^2,\m_q^2)$ and the term
${\cal O}(a^5)$\footnote{
The latter term includes all ${\cal O}(a^5 \ln^k a)$ terms ($k=0,1,2,3$).}
in $E_{q \bar q}(s; Q_s^2;\m_q^2)$ cancel exactly.

The $us$ part of the quarkonium mass, $E_{q \bar q}(us; Q_{us}^2;\m_q^2)$,
will be evaluated in each case according to a procedure which takes into
account those problems of low-scale evaluations which appear in the
considered model (2$\delta$anQCD, pQCD, (F)APT).

We recall that the binding energy $E_{q \bar q}(s)$ is a 
Euclidean quantity because it depends on spacelike quark-antiquark 
momentum transfer $q$ ($q^2 = - {\vec q}^2 \equiv - Q^2 < 0$).
Analytization of such quantities must follow the procedure
(\ref{analyt}).  
On the other hand, the quark pole mass $m_q$ is a Minkowskian quantity 
because it depends on the timelike pole momentum ($q^2 \equiv -Q^2 = m_q^2 > 0$).  
We note that our analytization procedure for the quark pole mass
is again the procedure (\ref{analyt}):
$a_{\rm pt}^n(\m_q^2) \mapsto \A_n(\m_q^2)$ in the relation (\ref{mqbmq})
[and then reexpressing $\A_n(\m_q^2)$ via $\A_k(\mu_s^2)$'s at a lower
soft scale $\mu_s^2$, for renormalon cancellation]. 
This procedure, for the Minkowskian quantities, is analogous 
to the fixed order perturbation theory (FOPT) in pQCD, Ref.~\cite{BeJa},
where the couplings in the corresponding contour integral, 
on the contour $Q^2 = \m_q^2 \exp(i \phi)$,
are Taylor-expanded around the spacelike point $Q^2 = \m_q^2 > 0$.
As a result, the kinematic $\pi^2$-terms appear in the
expansion coefficients $r_j$. 

Another analytization
of the pole mass expansion would involve contour integration
of the corresponding Euclidean quantity
with (exact) RGE-running couplings along the contour, cf.~Ref.~\cite{CKSir}.
This procedure is analogous to the Contour Improved Perturbation Theory
(CIPT) in pQCD; 
in such a case, the aformentioned $\pi^2$ terms are effectively
resummed, Refs.~\cite{RKP,KK}. 
We decided not to pursue this CIPT type of analytization, because
it is technically more demanding due to the additional running of the
mass factor $\m_q(\mu^2)$; and because in this approach the
renormalon cancellation mechanism, due to the mentioned resummations,
probably changes its practical form. This problem remains to be
addressed in the future.

\section{Numerical results}
\label{sec:numres}  
\subsection{Bottom mass extraction}
\label{subs:mb}
In this section, we extract from the mass of $b \bar b$ quarkonium the
mass $\m_b \equiv \m_b(\m_b^2)$ in (F)APT, 
2$\delta$anQCD model, and in pQCD in two renormalization schemes 
($\MSbar$ and in the Lambert scheme of 2$\delta$anQCD).  
For this, we use the relation between $\m_b$ and the 
well-measured mass of the $b \bar b$ quarkonium $\Upsilon(1S)$
 \cite{PDG2010}
\begin{eqnarray}
2 m_b(\m_b^2;\mu_s^2)+
E_{b \bar b}(s; Q_s^2;\m_b^2;\mu_s^2, N_f=3) 
+ E_{b \bar b}(us; Q_{us}^2; \m_b^2; \mu_{us}^2) =
M_{\Upsilon(1S)}^{\rm (exp)} (= 9.460 \ {\rm GeV}) \ .
\label{mass}
\end{eqnarray}
The dependence on the (soft) renormalization scale
$\mu_s^2 = \kappa_s Q_s^2$ 
in the pole mass $m_b$ and in the soft binding energy $E_{b \bar b}(s)$
occurs due to the truncation of the series for these two quantities.
For the same reason, the ultrasoft binding energy $E_{b \bar b}(us)$
has strong dependence on the ultrasoft renormalization scale
$\mu_{us}^2$ ($=\kappa_{us} Q_{us}^2$)
due to the drastic truncation of this quantity at its 
leading order ($\sim a^5 \ln a$). As mentioned in the previous Section,
the separation of the $s$ and $us$ parts of the binding
energy will be performed here by determining the $s$-$us$
separation parameter $k_{s/us}$, Eqs.~(\ref{sus1})-(\ref{calEqqspt}), 
by the requirement of 
cancellation of the leading renormalon in $2 m_b + E_{b \bar b}(s)$.

In contrast to the other three models, (F)APT gives 
a very small central value for the $s$-$us$ separation parameter 
$k_{s/us} \approx 6 \times 10^{-10}$.
This reflects the difficulty in the (F)APT scenario to exactly enforce
the leading renormalon cancellation of the $\sim \A_4$
term of $2 m_b$ with the corresponding $\sim \A_5$ term of $E_{b \bar b}(s)$.
If, on the other hand, we impose in (F)APT the condition
$k_{s/us} \sim 1$, more specifically the central value 
$k_{s/us} = 1$ and variation in the interval
$(0.1, 10.)$, the results change somewhat,
the central extracted value of $\m_b$ increases by about $0.050$ GeV,
and the absolute values of the $us$ part of the binding energy and 
of various other uncertainties of $\m_b$ get reduced.
We will consider in (F)APT only the 
natural range $\sim 1$ of the $s$-$us$ separation parameter,
$k_{s/us} = 1.0^{+9.0}_{-0.9}$, rather than the exceedingly small
values of $k_{s/us}$ required by the exact renormalon cancellation.
In the other three models (2$\delta$anQCD, and in Lambert and $\MSbar$ pQCD), the
renormalon cancellation is imposed without any problems, resulting
in the values of $k_{s/us}$ within the interval between $10^{-1}$ and $10^1$.

In the previous section we mentioned that we take
$N_f=3$ for the number of active flavors in the binding energy,
i.e., in this case the
$m_c$ mass is considered to be infinite (decoupled).
It turns out that, while the effects of the finiteness of $m_c$ 
cannot be neglected in the relation between $m_b$ and $\m_b$,
Eqs.~(\ref{dmbmc})-(\ref{mqbmq}),
these effects can be safely neglected in the binding energy
$E_{b \bar b}(N_f=3)$;
cf.~Ref.~\cite{Brambilla:2001qk} based on
Refs.~\cite{Gray:1990yh,Eiras:2000rh,Hoang:2000fm}.

Application of the formalism described in Secs.~\ref{subs:MA}
and \ref{subs:2d} (with Appendix \ref{app0})
for the calculation of the couplings of the analytic QCD models (F)APT
and 2$\delta$anQCD, and in Sec.~\ref{sec:Eqq} for the calculation of $2 m_b$,
$E_{b \bar b}(s)$ and $E_{\bar b b}(us)$ in terms of these couplings, 
then gives us the following results:
\begin{subequations}
\label{mbMAtot}
\bea
\m_b ( {\rm (F)APT}) &=& {\bigg \{}
4.155 \pm 0.002_{(us)} 
+ \left( \bear{c} +0.005 \\ -0.004 \eear \right)_{(s/us)} 
\!\!+ \left( \bear{c} -0.019 \\ +0.020 \eear \right)_{(\Lambda)}
\nonumber\\
&&+ \left( \bear{c} -0.004 \\ +0.002 \eear \right)_{(\mu_s)} 
\!\!\mp 0.005_{(m_c)}  {\bigg \}} \; {\rm GeV} \,
\label{mbMA}
\\
& = & 4.155 \pm 0.022  \; {\rm GeV} \ , 
\quad {\rm with:} \ 
Q_{s,\rm ((F)APT)} = 1.60  \; {\rm GeV} \ , \
k_{s/us} = 1.0 \ . 
\label{mbMAquadr}
\eea 
\end{subequations}
\begin{subequations}
\label{mb2dtot}
\bea
\m_b({\rm 2 \delta anQCD}) &=& {\bigg \{}
4.353 
+ \left( \bear{c} - 0.068 \\ +0.071 \eear \right)_{(us)} 
\!\!+ \left( \bear{c} +0.015 \\ -0.016 \eear \right)_{(s/us)} 
\!\!\mp 0.005_{(\alpha_s)}
\nonumber\\
&& + \left( \bear{c} -0.023 \\ +0.034 \eear \right)_{(c_2)} 
\!\!+ \left( \bear{c} +0.017 \\ -0.025 \eear \right)_{(\mu_s)} 
\!\!\mp 0.005_{(m_c)} {\bigg \}}
 \; {\rm GeV} \,
\label{mb2d}
\\
& = & 4.353 \pm 0.084 \; {\rm GeV} \ , 
\quad {\rm with:} \ 
Q_s = 2.08 \ {\rm GeV} \ , \
k_{s/us} = 0.238 \ . 
\label{mb2dquadr}
\eea
\end{subequations}
We will comment on the above uncertainties below. 
For completeness, we give here also the results
of the same kind of analysis in pQCD, first in the Lambert renormalization scheme (i.e., the scheme as used in 2$\delta$anQCD: $c_2=-4.76$,
$c_j = c_2^{j-1}/c_1^{j-2}$ for $j \geq 3$); and in the (four-loop) $\MSbar$ scheme:
\begin{subequations}
\label{mbLamtot}
\bea
\m_b({\rm pQCD Lamb.\/}) &=& {\bigg \{}
4.382 
+ \left( \bear{c} - 0.091 \\ +0.097  \eear \right)_{(us)} 
\!\!+ \left( \bear{c} -0.013 \\ +0.017 \eear \right)_{(s/us)}
\!\!\pm 0.010_{(\alpha_s)}
\nonumber\\
&&+ \left( \bear{c} +0.027 \\ -0.008 \eear \right)_{(c_2)}
\!\!+ \left( \bear{c} -0.002 \\ -0.041  \eear \right)_{(\mu_s)} 
\!\!\mp 0.005_{(m_c)}  {\bigg \}}
 \; {\rm GeV} \,
\label{mbLam}
\\
& = & 4.382 \pm 0.111 \; {\rm GeV}  \ , 
\quad {\rm with:} \ 
Q_{s, \rm pt} = 1.73 \ {\rm GeV} \ , \
k_{s/us} = 0.306 \ . 
\label{mbLamquadr}
\eea
\end{subequations}
\begin{subequations}
\label{mbbMStot}
\bea
\m_b({\rm pQCD} \MSbar) &=&  {\bigg \{}
4.505 
+ \left( \bear{c} - 0.177 \\ +0.200 \eear \right)_{(us)} 
\!\!+ \left( \bear{c}-0.082 \\ +0.084  \eear \right)_{(s/us)}
\!\!+ \left( \bear{c}+0.031 \\ -0.027  \eear \right)_{(\alpha_s)}
\nonumber\\
&&+ \left( \bear{c}-0.004 \\ -0.075   \eear \right)_{(\mu_s)} 
\!\!\mp 0.005_{(m_c)} {\bigg \}}
 \; {\rm GeV} \,
\label{mbbMS}
\\
& = & 4.505 \pm 0.231 \; {\rm GeV}  \ , 
\quad {\rm with:} \ 
Q_{s, \rm pt} = 1.87 \ {\rm GeV} \ , \
k_{s/us} = 0.248 \ . 
\label{mbbMSquadr}
\eea
\end{subequations}
The value of the $s$-$us$ separation parameter $k_{s/us}$ was
determined in all cases by the aforementioned renormalon
cancellation in the sum 
$M_{\Upsilon(1S)}(s) = 2 m_b + E_{b \bar b}(s;k_{s/us})$, except in
the (F)APT case, as discussed above. Below we present
these resulting sums, for the central
choices of the aforementioned four results,
where we combine in each parenthesis the (positive) terms
$\sim a^n$ of $2 m_b$ and the corresponding (negative) terms
$\sim a^{n+1}$ of $E_{b \bar b}(s)$
($n=0,1,2,3,4$), in order to see more clearly the tendency 
of the renormalon cancellation; the $us$ part is given separately
\begin{subequations}
\label{MUMAtot}
\bea
M_{\Upsilon(1S)}(s; {\rm (F)APT}) & = & 8.361 + (1.145-0.152) +
(0.281-0.174) + (0.161-0.154) + (0.081-0.093) \; {\rm GeV} 
\nonumber\\
& = &
 8.361 + 0.993 + 0.107 + 0.006 - 0.013 \; {\rm GeV}
\; ( = 9.454 \; {\rm GeV}) \ ,
\label{MUsMA}
\\
E_{b \bar b}(us; {\rm (F)APT}) & = & 0.006 \mp 0.003 \; {\rm GeV} \ , \qquad 
(\m_b=4.155 \ {\rm GeV}, \; k_{s/us} = 1.0 ) \ ;
\label{MUusMA}
\eea 
\end{subequations}
\begin{subequations}
\label{MU2dtot}
\bea
M_{\Upsilon(1S)}(s; {\rm 2 \delta anQCD}) & = & 8.756 + (0.999-0.131) +
(0.373-0.222) + (0.233-0.193) + (0.568-0.568) \; {\rm GeV} 
\nonumber\\
& = &  8.756 + 0.868 +0.151 +0.040 + 0.000 \; {\rm GeV}
\; ( = 9.815 \; {\rm GeV}) \ ,
\label{MUs2d}
\\
E_{b \bar b}(us; {\rm 2 \delta anQCD}) & = & -0.355 \pm 0.151 \; {\rm GeV} \ ,
\qquad 
(\m_b=4.355 \ {\rm GeV}, \; k_{s/us} = 0.238 ) \ ;
\label{MUus2d}
\eea
\end{subequations}
\begin{subequations}
\label{MULamtot}
\bea
M_{\Upsilon(1S)}(s; {\rm pQCD Lamb.}) & = & 8.814 + (1.095-0.170) +
(0.327-0.220) + (0.346-0.376) + (0.413-0.413) \; {\rm GeV} 
  \nonumber\\
& = &  8.814 + 0.925 +0.107 -0.031 + 0.000 \; {\rm GeV}
\; ( = 9.816 \; {\rm GeV}) \ ,
\label{MUsLam}
\\
E_{b \bar b}(us; {\rm pQCD Lamb.}) & = & -0.355 \pm 0.201 \; {\rm GeV} \ ,
\qquad 
(\m_b=4.382 \ {\rm GeV}, \; k_{s/us} = 0.306 ) \ ;
\label{MUusLam}
\eea
\end{subequations}
\begin{subequations}
\label{MUbMStot}
\bea
M_{\Upsilon(1S)}(s;  {\rm pQCD} {\MSbar}) & = & 9.060 + (1.183-0.193) +
(0.399-0.263) + (0.330-0.439) + (0.475-0.475) \; {\rm GeV} 
\nonumber\\
& = &  9.060 + 0.991 +0.136 -0.109 + 0.000 \; {\rm GeV}
\; ( = 10.078 \; {\rm GeV}) \ ,
\label{MUsbMS}
\\
E_{b \bar b}(us; {\rm pQCD} {\MSbar}) & = & -0.617 \pm 0.394 \; {\rm GeV} \ ,
\qquad 
(\m_b=4.505 \ {\rm GeV}, \; k_{s/us} = 0.248 ) \ ;
\label{MUusbMS}
\eea
\end{subequations}
We can see from Eqs.~(\ref{MUsMA})-(\ref{MUsbMS}) explicitly
that for the chosen corresponding central values of the parameter
$k_{s/us}$ the renormalon cancellation 
is exact in the last term [the fifth term, named $t_5(s)$] 
in the sum for the soft mass $M_{\Upsilon(1S)}(s)$, except in
the case of (F)APT where $k_{s/us}=1.0$ was chosen and the cancellation
in $t_5(s)$ is approximate.

\begin{figure}[htb] 
\begin{minipage}[b]{.49\linewidth}
\centering\includegraphics[width=80mm]{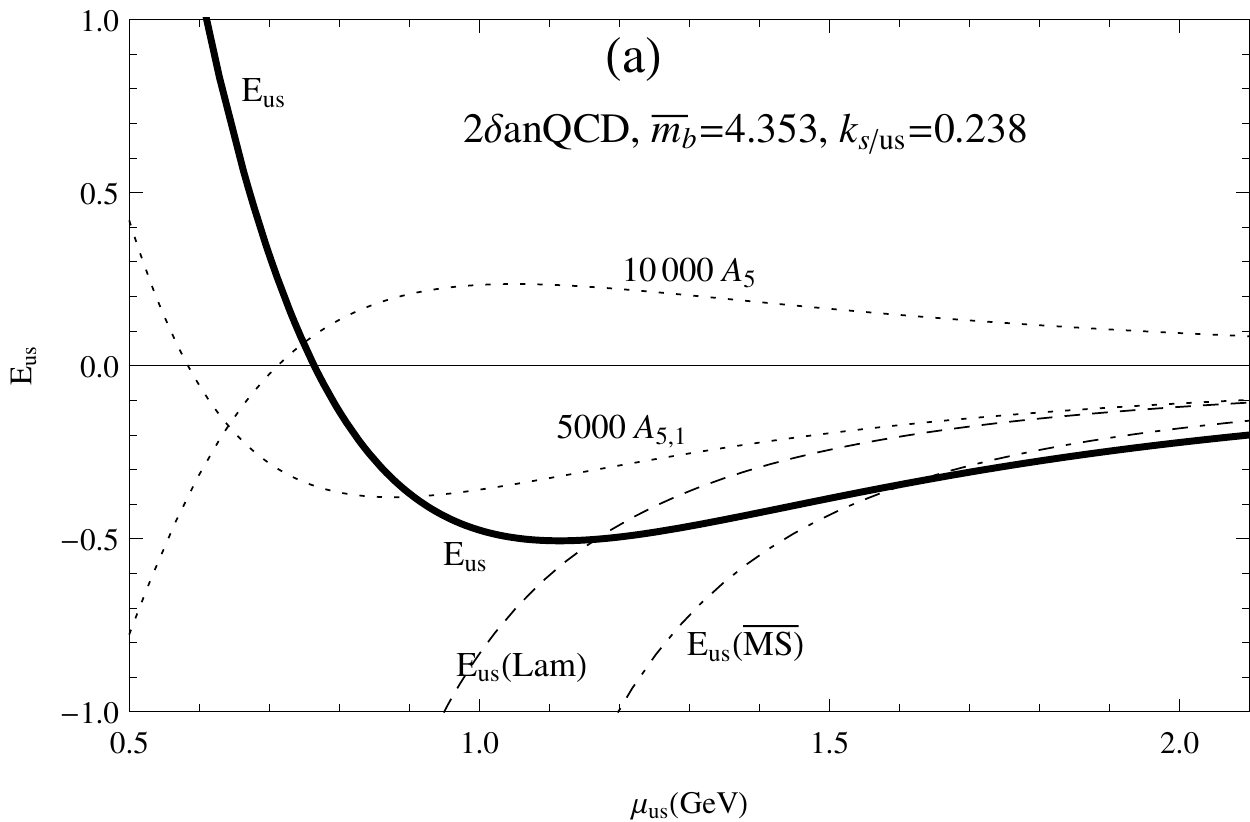}
\end{minipage}
\begin{minipage}[b]{.49\linewidth}
\centering\includegraphics[width=80mm]{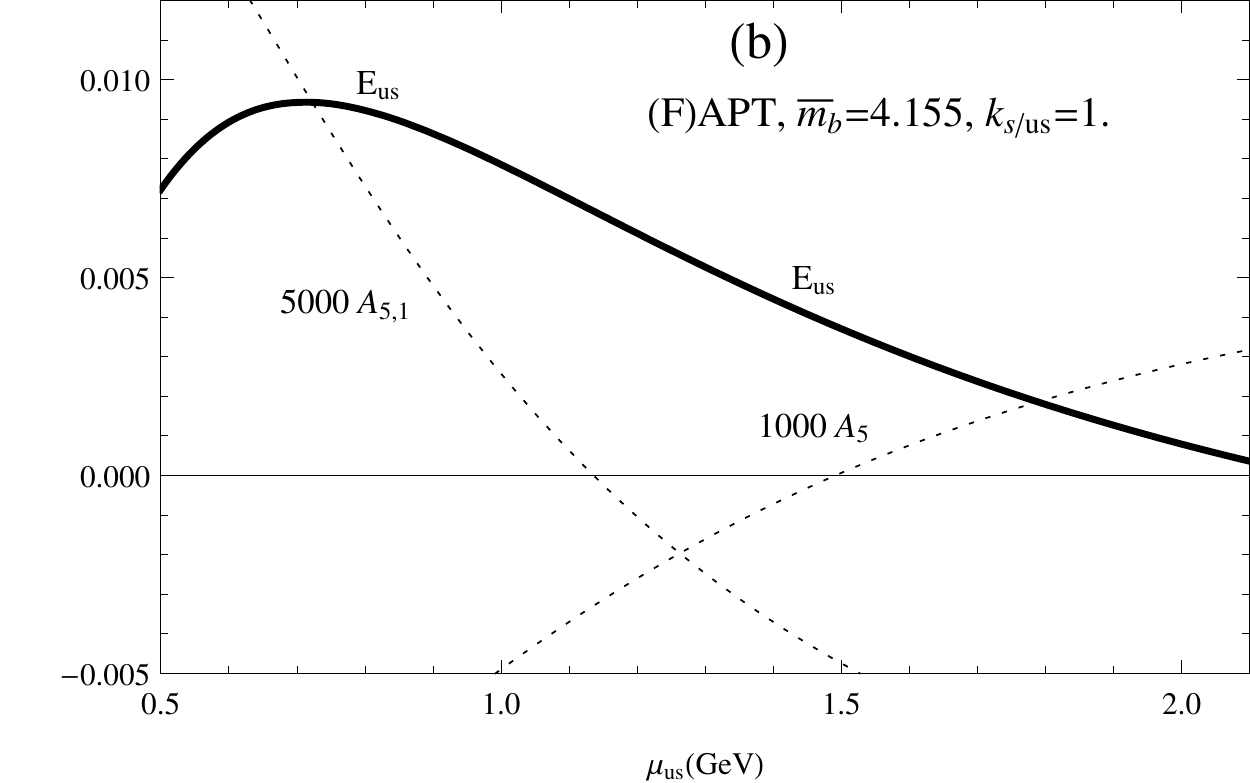}
\end{minipage}
\vspace{-0.4cm}
 \caption{(a) Ultrasoft binding energy $E_{b \bar b}(us)$
(in GeV) as a function of the $us$ renormalization scale $\mu_{us}$, for the
central input values of the 2$\delta$anQCD model for the $\Upsilon(1S)$ system
($\m_b=4.353$ GeV; $k_{s/us} = 0.238$;
$c_2=-4.76$; $s_0=23.06$):
solid line for 2$\delta$anQCD; dashed line for Lambert pQCD ($c_2=-4.76$);
dash-dotted line for the four-loop $\MSbar$ pQCD ($c_2=4.471$). 
For all three curves the same $\m_b$ and  $k_{s/us}$ values are used. 
Also included are the properly rescaled
2$\delta$anQCD couplings $\A_5$ and $\A_{5,1}$ as functions of the scale $\mu$
($=\mu_{us}$). (b) Same as in (a), but for the (F)APT model
(with $c_2 =4.471$), with the corresponding central values of that model
($\m_b=4.155$ GeV and $k_{s/us}=1$). In all the curves $N_f=3$ was taken.}
\label{plEusnf4}
 \end{figure}

The extracted values of $\m_b$, Eqs.~(\ref{mbMAtot})-(\ref{mbbMStot}),
have a strong uncertainty coming from the ultrasoft ($us$) regime 
and from the related $s/us$ separation.
The origin of this uncertainty lies in the strong
dependence of the $us$ binding energy $E_{b \bar b}(us; \mu_{us}^2)$
on the $us$ renormalization scale $\mu_{us}$ and on the
$s$-$us$ separation parameter $k_{s/us}$, cf.~Eqs.~(\ref{muf})-(\ref{Kus}).
The behavior of the $us$ binding energy $E_{b \bar b}(us; \mu_{us}^2)$
in the three models (2$\delta$anQCD, and pQCD in the two schemes), 
as a function of the $us$ renormalization scales $\mu_{us}$ in the
low-momentum regime, is presented in Fig.~\ref{plEusnf4}(a),
and in the case of (F)APT in Fig.~\ref{plEusnf4}(b).
In the 2$\delta$anQCD model and in pQCD, we do not consider the scales
$\mu_{us}$ below $(\mu_{us})_{\rm min} = 1.1$ GeV,
because at $\mu_{us} \approx 1.1$ GeV the coupling $\A_5(\mu_{us})$
($= \tA_5(\mu_{us})$ in our approach)\footnote{
We recall that in the 2$\delta$anQCD model
we calculate in $m_q(\m_q^2;\mu_s^2)$ the couplings
$\A_n(\mu_s^2)$ as sums of $\tA_p(\mu_s^2)$ with $p=n, n+1, \ldots, 4$;
and in $E_{q \bar q}(s)$ [and $E_{q \bar q}(us)$]
the couplings $\A_{n,k}(\mu_s^2)$ [and $\A_{n,k}(\mu_{us}^2)$] as 
(derivatives of the) sums of $\tA_p(\mu_s^2)$ [and  $\tA_p(\mu_{us}^2)$]
with $p=n, n+1, \ldots, 5$; cf.~discussion in Sec.~\ref{subs:2d}, 
Appendix \ref{app0}, and Sec.~\ref{sec:Eqq}.}
in the 2$\delta$anQCD model 
reaches a local maximum, indicating that the model may
not necessarily give reliable values of $\A_5$ and $\A_{5,1}$ below such scales.
Therefore, in general, we will not consider $\mu_{us}$ lower than 1.1 GeV,
in the 2$\delta$anQCD model and in pQCD.
$E_{b \bar b}(us;mu_{us}^2)$ in 2$\delta$anQCD reaches a local minimum at 
$\mu_{us}$ slightly above 1.1 GeV.
Therefore, we estimate the $us$ part of the binding
energy in the 2$\delta$anQCD model in the following way 
[we use the notation $E_{b \bar b}(us; \mu_{us}^2)$]:
\be
E_{b \bar b}(us; {\rm 2 \delta anQCD}) 
= \frac{1}{2} \left[ E_{b \bar b}(us; Q_s^2)
+ (E_{b \bar b}(us))_{\rm min} \right] \pm
 \frac{1}{2} \left[ E_{b \bar b}(us; Q_s^2)
- (E_{b \bar b}(us))_{\rm min} \right] \ ,
\label{Ebbus2d}
\ee
where the soft reference scale $Q_s$ was determined by the condition
(\ref{Qsa}), and gave for the central values of the input
parameters ($s_0$ and $c_2$ of the second line of Table \ref{t1};
$\alpha_s(M_Z^2;\MSbar)=0.1184$; $\m_b=4.353$ GeV) 
the soft reference scale value $Q_s \approx 2.08$ GeV.

In the two pQCD approaches, $E_{b \bar b}(us)$ decreases 
monotonously when $\mu_{us}$ decreases below the
soft reference scale $Q_{s, {\rm pt}}$ of Eq.~(\ref{Qspt}). 
For pQCD in the Lambert scheme ($c_2=-4.76$), 
with central values of the input parameters,
the soft reference scale turns out to be $Q_{s, {\rm pt}} \approx 1.73$ GeV,
and the $us$ binding energy at $(\mu_{us})_{\rm min} =1.1$ GeV reaches the value of
$-0.56$ GeV. In the $\MSbar$ scheme, however, 
$E_{b \bar b}(us; 1.1^2 {\rm GeV}^2) \approx -1.5$ GeV,
which is exceedingly low and
indicates failure of the method already at such scales,
due to vicinity of Landau singularities of the running
coupling.\footnote{\label{cuts}
It turns out that the Landau cut in the perturbative coupling
$a_{\rm pt}(\mu^2)$ starts in the four-loop ($N_f=3$)
$\MSbar$ scheme already at about $\mu = \Lambda_{\rm L} \approx 0.607$ GeV.
In the Lambert scheme [Eq.~(\ref{aptexact}) 
with the central value $c_2=-4.76$] there is one pole at
$\mu  = \Lambda_{\rm p} \approx  0.262$ GeV; the
cut begins at an even lower value $\mu=\Lambda_{\rm L.} \approx 0.208$ GeV. 
The 2$\delta$anQCD coupling
$\A_1(\mu^2)$ has, of course, no Landau singularities 
(poles and cuts), 
although it almost coincides with the Lambert pQCD coupling
$a_{\rm pt}(\mu^2)$ at higher scales $|\mu^2| > 1 \ {\rm GeV}^2$, Ref.~\cite{2danQCD}.}
Therefore, in $\MSbar$ we take as the minimal
acceptable scale $(\mu_{us})_{\rm min}=1.2$ GeV, where 
$E_{b \bar b}(us; 1.2^2 {\rm GeV}^2) \approx -1.0$ GeV 
(and $Q_{s, {\rm pt}} \approx 1.87$ GeV) when the central values of the
input parameters are used. Thus, in pQCD we estimate the $us$ binding energy
as
\be
E_{b \bar b}(us) 
= \frac{1}{2} \left[ 
E_{b \bar b}(us; Q_{s,{\rm pt}}^2)
+ E_{b \bar b}(us; (\mu_{us}^2)_{\rm min}) \right] \pm
 \frac{1}{2} \left[ E_{b \bar b}(us; Q_{s, {\rm pt}}^2)
- E_{b \bar b}(us;  (\mu_{us}^2)_{\rm min}) \right] \ ,
\label{EbbuspQCD}
\ee
with $(\mu_{us})_{\rm min} = 1.1$ GeV and $1.2$ GeV in the Lambert and
$\MSbar$ schemes, respectively. 

In the (F)APT case, on the other hand,  $Q_s$ is fixed according 
to Eq.~(\ref{QsMA}) and gives, for the central input
parameter values, the value $Q_{s, \rm ((F)APT)} \approx 1.60$ GeV. 
In (F)APT no practical problems appear at scales $\mu_{us} < 1.1$ GeV.
$E_{b \bar b}(us,{\rm (F)APT})$ reaches a moderate maximum value of
$0.009$ GeV at $\mu_{us} \approx 0.7$ GeV for the chosen central value $k_{s/us}=1.0$. 
Therefore, we estimate the 
$us$ part of the binding energy in (F)APT in the following way:
\be
E_{b \bar b}(us; {\rm (F)APT}) 
= \frac{1}{2} \left[  E_{b \bar b}(us; Q_{s, \rm ((F)APT)}^2) +
(E_{b \bar b}(us))_{\rm max} \right] \pm
 \frac{1}{2} \left[ E_{b \bar b}(us; Q_{s, \rm ((F)APT)}^2) -
(E_{b \bar b}(us))_{\rm max} \right] \ .
\label{EbbusMA}
\ee

In Eqs.~(\ref{mbMAtot})-(\ref{mbbMStot}), the uncertainties 
in $\m_b$ originating from these determinations of the $us$ binding energy
are denoted by the subscript $(us)$. 

The related uncertainties for the extracted values of $\m_b$
originate from the variation
of the $s$-$us$ separation parameter $k_{s/us}$, and are
denoted by the subscript $(s/us)$ in Eqs.~(\ref{mbMAtot})-(\ref{mbbMStot}).
The parameter $k_{s/us}$ was varied in such a way that the last 
[fifth, $t_5(s)$] term in the
series for the soft mass $M_{\Upsilon(1S)}(s)$ 
[cf.~Eqs.~(\ref{MUsMA})-(\ref{MUsbMS})] varies between the
penultimate term $t_4(s)$ of these series, and its negative
$-t_4(s)$, these two cases correspond to the upper
and the lower entry of $(s/us)$ uncertainty of $\m_b$, respectively.
In the (F)APT case the exact renormalon cancellation
was not achieved and the parameter $k_{s/us}$ was varied between
0.1 and 10, i.e., $k_{s/us} = 1.0^{+9.0}_{-0.9}$.

The other uncertainty in the determination of $\m_b$ comes from
the uncertainty of the $\Lambda$ scale. In (F)APT it comes from
${\overline \Lambda}_5 =0.260 \pm 0.030$ GeV and is denoted 
by the subscript $(\Lambda)$ in Eq.~(\ref{mbMA}). In the 2$\delta$anQCD model
and in the two pQCD approaches (the Lambert scheme and $\MSbar$ scheme),
this uncertainty comes from $\alpha_s(M_Z^2;\MSbar) = 0.1184 \pm 0.007$
\cite{PDG2010} and is denoted by the subscript $(\alpha_s)$
in Eqs.~(\ref{mb2d}), (\ref{mbLam}), and (\ref{mbbMS}).

Yet another uncertainty of $\m_b$, in the 2$\delta$anQCD model
and in Lambert scheme pQCD,
comes from the variation of the (Lambert) renormalization scheme 
parameter $c_2 = -4.76^{+2.66}_{-0.97}$, cf.~Table \ref{t1} and 
Eqs.~(\ref{aptexact})-(\ref{betapt}).
and is denoted in Eqs.~(\ref{mb2d}) and (\ref{mbLam}) by the subscript $(c_2)$.
The scheme in (F)APT was fixed by the underlying pQCD solution, 
Eqs.~(\ref{RGE2l})-(\ref{apt2l}): $c_2=c_3=\cdots=0$, i.e., effectively
the two-loop solution.

The uncertainty due to the variation of the soft renormalization scale $\mu_s$
was denoted in Eqs.~(\ref{mbMAtot})-(\ref{mbbMStot}) by the subscript $(\mu_s)$.
We varied $\mu_s^2$ around the central value $(Q_s^2)_{\rm centr.}$ 
of the soft reference scale, between $2 (Q_s^2)_{\rm centr.}$ and
$(1/2) (Q_s^2)_{\rm centr.}$. The scale  $(Q_s^2)_{\rm centr.}$ is determined by 
Eqs.~(\ref{Qspt}), (\ref{Qsa}) and (\ref{QsMA}) in pQCD,
2$\delta$anQCD and (F)APT, respectively, for
central values of the input parameters $\m_b$, $k_{s/us}$, etc.

Finally, the uncertainty $\delta m_b(m_c \not= 0) = \pm 0.005$ GeV due to nonzero
$c$ mass [Eq.~(\ref{dmbmc}; cf.~also Eqs.~(\ref{Smmus})-(\ref{tmq})]
results in the uncertainties $\mp 0.005$ GeV of $\m_b$, denoted
in Eqs.~(\ref{mbMAtot})-(\ref{mbbMStot}) by the subscript $(m_c)$.

We see in Eqs.~(\ref{mbMAtot})-(\ref{mbbMStot})
that the largest resulting uncertainty in the determination
of $\m_b$ is the one originating from the uncertainty of
the determination of the $us$ binding energy (except in
(F)APT where $|E_{b \bar b}(us)|$ values are small). These uncertainties
are larger in the two pQCD approaches, due to the influence of
the nearby (unphysical) Landau singularities in the running couplings.
The contribution of the $us$ regime to the quarkonium mass,
in the 2$\delta$anQCD model and in pQCD, increases the predicted value of $\m_b$.
This is so because the $us$ binding energies are in these cases 
significant and negative; cf.~also Fig.~\ref{plEusnf4}(a). 
If we had ignored the existence and separation of the $us$ contributions,
i.e., if we had used in the entire binding energy
$E_{b \bar b}$ simply a common soft renormalization scale
$\mu_s \sim Q_s$, the predicted values of $\m_b$ in the 2$\delta$anQCD model and in 
Lambert and $\MSbar$ pQCD would have decreased, by $-0.068$,
$-0.091$, and $-0.177$ GeV, respectively, as can be deduced from
the $us$-origin uncertainties in Eqs.~(\ref{mb2d})-(\ref{mbbMS}).
On the other hand, in (F)APT the choice $\mu_{us} = Q_{s, \rm ((F)APT)}$ would 
only slightly increase the central value of $\m_b$, by $0.002$ GeV
[cf.~Eq.~(\ref{mbMA})], basically because the values of $|E_{b \bar b}(us)|$
in (F)APT are much smaller; cf.~Fig.~\ref{plEusnf4}(b).

For better visibility, we present the results for the central
extracted values of $\m_b$ of the aforementioned four models 
in Table \ref{t2}, and for various uncertainties 
$\delta \m_b$ in Table \ref{t3}.
\begin{table}[htb]
\caption{Extracted central values of $\m_b$ in the four models, for the
central input parameter values (with the total uncertainties $\delta \m_b$).
Included are the corresponding input parameter $k_{s/us}$, and the resulting
scales: soft reference scale $Q_s$ (soft renormalization scale
is taken $\mu_s=Q_s$); soft mass $M_{\Upsilon(1S)}(s)$; averaged
ultrasoft energy ${\bar E}_{b \bar b}(us)$ and its
ambiguity $\delta {\bar E}_{b \bar b}(us)$
[cf.~Eqs.~(\ref{Ebbus2d})-(\ref{EbbusMA})]. 
All scales are given in GeV. Note that $M_{\Upsilon(1S)}(s) +
{\bar E}_{b \bar b}(us) = 9.460$ GeV, i.e., the physical mass $M_{\Upsilon(1S)}$.}
\label{t2}
\begin{ruledtabular}
\begin{tabular}{c|c|c|c|ccc}
 model & $\m_b (\delta \m_b)$  & $k_{s/us}$ & $Q_s$ & $M_{\Upsilon(1S)}(s)$ & ${\bar E}_{b \bar b}(us)$ & $\delta E_{b \bar b}$ \\
\hline
(F)APT &      $4.155(\pm0.022)$ & 1.000 & 1.596 & 9.454 & 0.006 & $\mp 0.003$ \\
2$\delta$anQCD &     $4.353(\pm0.084)$ & 0.238 & 2.084 & 9.815 &-0.355 & $\pm 0.151$ \\
pQCD Lamb.   &$4.382(\pm0.111)$ & 0.306 & 1.729 & 9.816 &-0.355 & $\pm 0.201$ \\
pQCD $\MSbar$ &$4.505(\pm0.231)$ & 0.248 & 1.869 &10.078 &-0.617 & $\pm 0.394$
\end{tabular}
\end{ruledtabular}
\end{table}
\begin{table}[htb]
\caption{Uncertainties $\delta \m_b$ of the extracted 
value of $\m_b$ coming from various sources: (1) from the
evaluation of the $us$ sector; (2) from the variation of
the $s$-$us$ separation parameter $k_{s/us}$; (3) from the 
variation of $\alpha_s$ (or, in (F)APT: variation of $\Lambda$);
(4) from the variation of the $c_2$ parameter (in 2$\delta$anQCD,
and in pQCD in the Lambert scheme); (5) from the
variation of the soft renormalization scale $\mu_s$; (6) from
the uncertainty of $(\delta m_b)_{\rm m_c}$ of Eq.~(\ref{dmbmc}).
See the text for details.}
\label{t3}
\begin{ruledtabular}
\begin{tabular}{c|c|cc|c|cc|c|c}
 model & $\delta \m_b(us)$  & $\delta \m_b(s/us)$ & ($k_{s/us}$) & 
$\delta \m_b(\alpha_s)$ & $\delta \m_b(c_2)$ & ($c_2$) &
$\delta \m_b(\mu_s)$ & $\delta \m_b(m_c)$ 
\\
\hline
\multirow{2}{10mm}{\rm (F)APT} 
&  +0.002 & +0.005 & (1.0+9.0) & -0.019 & -- & (--) & -0.004 & -0.005 
\\
& -0.002& -0.004 & (1.0-0.9) & +0.020 & -- & (--) & +0.002 & +0.005
\\
\hline
\multirow{2}{10mm}{2 $\delta$ an- QCD} 
&  -0.068 & +0.015 & (0.238-0.100) & -0.005 & -0.023 & (-4.76+2.66) & +0.017 & -0.005
\\
 & +0.071 & -0.016 & (0.238+0.173) & +0.005 & +0.034 & (-4.76-0.97) & -0.025 & +0.005
\\
\hline
\multirow{2}{10mm}{\rm pQCD Lamb.}   
&  -0.091 & -0.013 & (0.306+0.194) & +0.010 & +0.027 & (-4.76+2.66) & -0.002 & -0.005
\\
 & +0.097 & +0.017 & (0.306-0.119) & -0.010 & -0.008 & (-4.76-0.97) & -0.041 & +0.005
\\
\hline
\multirow{2}{10mm}{\rm pQCD $\MSbar$} 
&  -0.177 & -0.082 & (0.248+0.642) & +0.031 & -- & (--) & -0.004 & -0.005
\\
 & +0.200 & +0.084 & (0.248-0.181) & -0.027 & -- & (--) & -0.075 & +0.005
\end{tabular}
\end{ruledtabular}
\end{table}

We wish to address here briefly the question of nonperturbative (NP, higher-twist)
contribution to the quarkonium mass.
In the heavy quark system such as $b \bar b$, 
the NP contribution can be estimated,
and in the leading order it comes from the gluon condensate and is 
given by \cite{Voloshin:hc}
\be
E_{b \bar b}(us)^{\rm (NP)} \approx 
\m_b 
\pi^2 \frac{624}{425} 
\left( \frac{4 \pi}{3} \m_b \right)^{-4} \frac{1}{a^4_{\rm pt}(\mu_{us}^2)}
\langle \frac{\alpha_s}{\pi} G_{\mu \nu} G^{\mu \nu} \rangle \ .
\label{EbbusNP1}
\ee
The factor $1/a^4_{\rm pt}(\mu_{us}^2)$ in (four-loop) $\MSbar$
pQCD is unreliable for realistic $us$ scales $\mu_{us} \approx 1$ GeV,
due to the vicinity of the Landau singularities (cf.~footnote \ref{cuts}). 
In the 2$\delta$anQCD model, for purposes of estimation, we 
replace $1/a^4_{\rm pt}(\mu_{us}^2)$ by $1/\A_4(\mu_{us}^2)$ or by $1/\tA_4(\mu_{us}^2)$.
In the interval $(1.1 \ {\rm GeV} < \mu_{us} < 1.3 \ {\rm GeV})$ 
we have $\A_4 \sim \tA_4 \sim 10^{-4}$; the couplings
$\A_4$ and $\tA_4$ cover in this interval the values between
$0.5 \times 10^{-4}$ and $2 \times 10^{-4}$. For these values, and using the
central value of the gluon condensate $\langle (\alpha_s/\pi) G^2 \rangle
= 0.009 \ {\rm GeV}^2$ \cite{Ioffe:2002be} (cf.~also Refs.~\cite{Ioffe} and
\cite{anOPE}), and for $\m_b=4.3$ GeV, we obtain the following estimate:
\be
E_{b \bar b}(us)^{\rm (NP)} = 0.05^{+0.05}_{-0.02} \ {\rm GeV} \ .
\label{EbbusNP2}
\ee
This effect is relatively small and has large uncertainties. If we take it
into account, then the central extracted values of $\m_b$ in this subsection
decrease somewhat, the decrease being 
$(\triangle \m_b)_{\rm (NP)} \approx (-0.025^{-0.025}_{+0.010})$ GeV.

We wish to comment briefly on the following aspect: the results
of this subsection show that 
the extracted values $\m_b$ and various uncertainties
$\delta \m_b$ are similar in the 2$\delta$anQCD model and the 
corresponding Lambert pQCD. The main reason for this lies in the fact that
the scheme parameters ($c_j$, $j \geq 2$) are the same in both 
frameworks, and that the two corresponding running couplings 
practically merge at high momenta $|Q^2| > \Lambda^2$: 
$\A_1(Q^2) - a_{\rm pt}(Q^2) \sim (\Lambda^2/Q^2)^5$. 
Nonetheless, the evaluation methods for these two cases
differ somewhat due to the different types of truncations involved.
In pQCD, the quantities $2 m_b$ and $E_{b \bar b}$ were calculated 
as truncated series of powers $a_{\rm pt}^n$, truncated at
$a_{\rm pt}^4$ and $a_{\rm pt}^5$, respectively.
In the 2$\delta$anQCD model, they were effectively calculated as series in
logarithmic derivatives $\tA_n$, truncated at $\tA_4$ and $\tA_5$, 
respectively; namely, the analytic power analogs
$\A_n$ in $2 m_q$ were evaluated as a series in $\tA_k$'s up to $k=4$,
and in $E_{b \bar b}$ as a series in $\tA_k$'s up to $k=5$
(cf.~Appendix \ref{app0}).

For comparison, we present in Table \ref{bmbcomp} 
a list of extracted values
of $\m_b$ (and $\m_c$) by various methods in the literature:
lattice calculations, sum rules (pQCD+OPE), 
and from meson spectra (pQCD). The latter pQCD calculations
account for the renormalon cancellation,
but most of them either do not consider the ultrasoft contributions,
or they include them unseparated from the soft contributions (using the
same scale in the soft and the ultrasoft).
\begin{table}
\caption{\label{bmbcomp} Values of 
$\m_b$ and $\m_c$ masses extracted from 
lattice calculations, sum rules (pQCD+OPE) or from 
quarkonium spectrum (pQCD).} 
\begin{ruledtabular}
\begin{tabular}{l c ll}
reference (year) & method & $\m_b$ (GeV) & $\m_c$ (GeV)
\\
\hline
ETM (2011) \cite{Dimopoulos:2011gx,Blossier:2010cr} & lattice ($N_f=2$)& $4.29 \pm 0.14$ (\cite{Dimopoulos:2011gx}) & $1.28 \pm 0.04$ (\cite{Blossier:2010cr})
\\
GST (2008) \cite{Guazzini:2007ja} & lattice (quenched) & $4.42 \pm 0.06$ & --
\\
DGPS (2006) \cite{Della Morte:2006cb} & lattice (quenched) & $4.347 \pm 0.048$ & --
\\
DGPTP (2003) \cite{deDivitiis:2003iy} & lattice (quenched) & $4.33 \pm 0.10$ & $1.319 \pm 0.028$
\\
\hline
Narison (2011) \cite{Narison:2011xe} & sum rules & $4.177 \pm 0.011$ & $1.261 \pm 0.016$
\\
HPQCD (2010) \cite{McNeile:2010ji} & sum rules\footnote{
Uses sum rules with lattice input.} & $4.164 \pm 0.023$ & $1.279 \pm 0.006$
\\
CKMM {\it et al.\/} (2009) \cite{Chetyrkin:2009fv} & sum rules & $4.163 \pm 0.016$
& $1.279 \pm 0.013$
\\
PS (2006)  \cite{Pineda:2006gx} & sum rules & $4.19 \pm 0.06$ & --
\\
BCS (2006)  \cite{Boughezal:2006px} & sum rules & $4.205 \pm 0.058$ &
$1.295 \pm 0.015$
\\
\hline
LKW (2011) \cite{Laschka:2011zr} & spectrum & $4.18^{+0.05}_{-0.04}$ & 
$1.28^{+0.07}_{-0.06}$
\\
CCG (2003) \cite{Contreras:2003zb} & spectrum & $4.24 \pm 0.07$ & --
\\
Lee (2003) \cite{Lee:2003hh} & spectrum & $4.20 \pm 0.04$ & --
\\
BSV (2001) \cite{Brambilla:2001qk}  & spectrum & $4.19 \pm 0.02 \pm 0.025$ & --
\\
Pineda 2001 \cite{Pineda:2001zq} & spectrum & $4.210 \pm 0.090 \pm 0.025$ & $1.210 \pm 0.070 \pm 0.079$
\\
\hline
\multirow{4}{20mm}{This work} &
\multirow{4}{20mm}{spectrum} &  $4.16 \pm0.02$ ((F)APT) & $1.257 \pm 0.012$  ((F)APT)
\\
& & $4.35 \pm 0.08$ (2$\delta$anQCD) & $1.266 \pm 0.017$  (2$\delta$anQCD)
\\
& & $4.38 \pm 0.11$ (pQCD Lamb.) & $1.265 \pm 0.027$ (pQCD Lamb.)
\\ 
& & $4.50 \pm 0.23$ (pQCD $\MSbar$) & $1.272 \pm 0.078$  (pQCD $\MSbar$) 
\end{tabular}
\end{ruledtabular}
\end{table} 
We can see that (F)APT results agree with those of the usual
pQCD calculations of the quarkonium spectrum and
those of the sum rules (pQCD+OPE).
The 2$\delta$anQCD results are incompatible with those results,
but are compatible with the results of lattice calculations.
The same can be claimed for the estimates of our pQCD approach
(in the Lambert and $\MSbar$ schemes), but the uncertainties there,
coming principally from the ultrasoft sector, are larger, especially
in $\MSbar$ scheme.

\subsection{Charm mass extraction}
\label{subs:mc}
 
\begin{figure}[htb] 
\begin{minipage}[b]{.49\linewidth}
\centering\includegraphics[width=80mm]{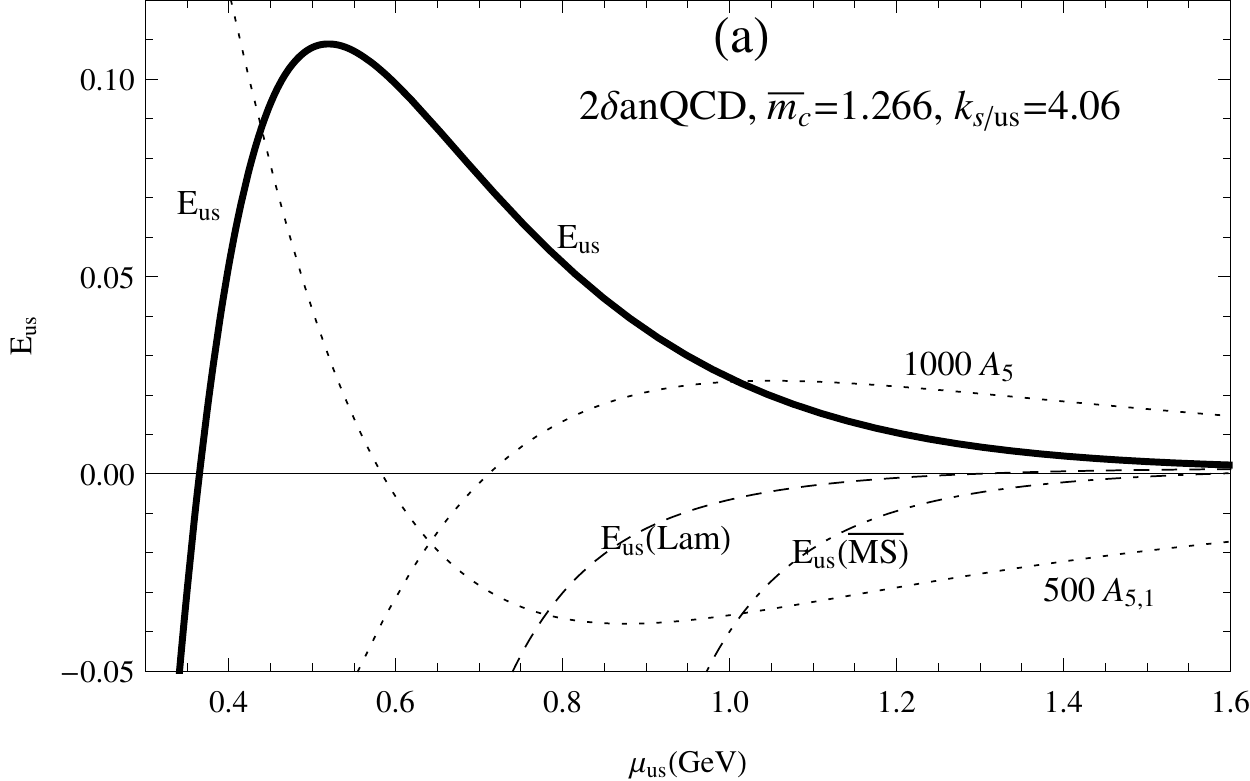}
\end{minipage}
\begin{minipage}[b]{.49\linewidth}
\centering\includegraphics[width=80mm]{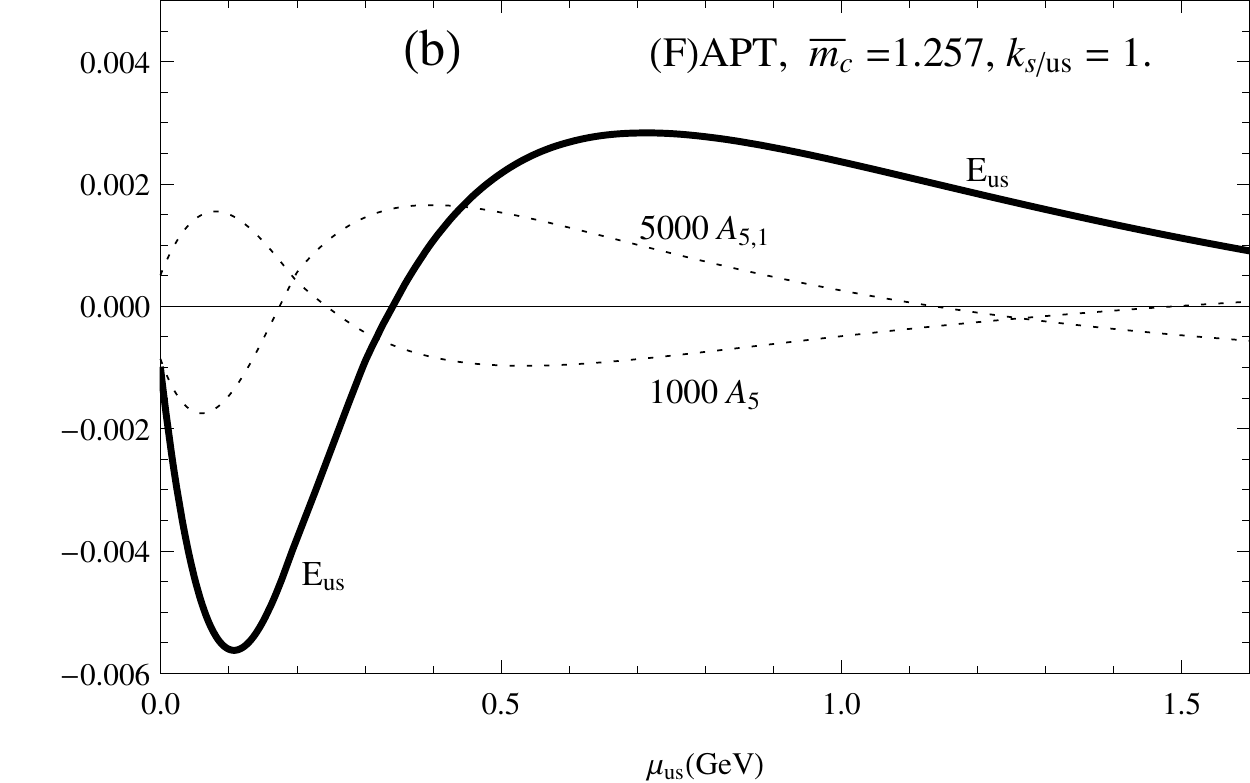}
\end{minipage}
\vspace{-0.4cm}
 \caption{(a) Ultrasoft binding energy $E_{c \bar c}(us)$
(in GeV) as a function of the $us$ renormalization scale $\mu_{us}$, for the
central input values of the 2$\delta$anQCD model for the $J/\psi(1S)$ system
($\m_c=1.266$ GeV; $k_{s/us} = 4.06$; $c_2=-4.76$; $s_0=23.06$):
solid line for 2$\delta$anQCD; dashed line for Lambert pQCD ($c_2=-4.76$);
dash-dotted line for the four-loop $\MSbar$ pQCD ($c_2=4.471$). 
For all three curves the same $\m_c$ and  $k_{s/us}$ values are used. 
Included are the properly rescaled
2$\delta$anQCD couplings $\A_5(\mu_{us}^2)$ and $\A_{5,1}(\mu_{us}^2)$ 
as functions of the scale $\mu_{us}$. 
(b) Same as in (a), but for the (F)APT model
(with $c_2 =4.471$), with the corresponding central values of that model
($\m_c=1.257$ GeV and $k_{s/us}=1$). In all the curves $N_f=3$ was taken.}
\label{plEusnf3}
 \end{figure}

In this case, $q \bar q = c \bar c$, and the quarkonium mass is now
$M_{J/\psi(1S)}=3.0969$ GeV \cite{PDG2010}. 
We basically repeat the analysis as in the case of $b \bar b$.
There are some differences, though:
\begin{itemize}
\item
The relation (\ref{mqbmq})
is now at $N_f=3$; therefore, the transition to the relation at 
the soft renormalization scales, Eq.~(\ref{Smmus}), which is
also at $N_f=3$, has now no threshold transition complication.
\item
The typical soft reference scales 
[Eqs.~(\ref{Qspt}) and (\ref{Qsb})] are now significantly lower:
$Q_s \approx 1$ GeV or even lower
(in the $b \bar b$ case we had: $Q_s \approx 2$ GeV).
This, in conjunction with our suggestion that the considered
models 2$\delta$anQCD and pQCD (in the Lambert and $\MSbar$ schemes)
are not necessarily to be trusted at scales below $1.1$ GeV, implies that
the typical soft renormalization scales $\mu_s$ should be chosen
significantly higher than $Q_s$. We choose
$\mu_s^2 = (3 \pm 1) Q_s^2$ in these three models. 
In this case, the lowest possible soft scale, $(\mu_s)_{\rm min}
= \sqrt{2} Q_s$ is slightly above $1.1$ GeV.
In (F)APT, the soft renormalization scale will also be varied in
this way, giving, however, somewhat lower central value for the
renormalization scale:, 
$\mu_s = \sqrt{3} Q_{s,\rm ((F)APT)} \approx 1.004$ GeV.\footnote{
If we used in (F)APT a lower definition of the
central renormalization scale, $\mu_s = Q_{s,\rm ((F)APT)}$
($\approx 0.58$ GeV), the predicted central value of $\m_c$ would
go up by only $0.005$ GeV.}
The scale variation $\mu_s^2 = (3 \pm 1) Q_s^2$ results in small
uncertainties $(\delta \m_c)_{(\mu_s)}$ of the extracted mass $\m_c$,
but these uncertainties may be underestimated because $\mu_s^2 > Q_s^2$.
\item
In general, the cancellation of the leading renormalon now
implies for the $s$-$us$ separation parameter $k_{s/us}$ such values for which
the absolute values of $E_{c \bar c}(us)$
are significantly smaller than in the $b \bar b$ case, and
consequently, the $us$ ambiguities are smaller.
Since we now use for the central choice of the soft renormalization
scale $\mu_s^2 = 3 Q_s^2$,
$E_{c \bar c}(us)$ is calculated in the 2$\delta$anQCD model and in
pQCD in the following way:
\be
E_{c \bar c}(us) 
= \frac{1}{2} \left[ 
E_{c \bar c}(us; 3 Q_s^2)
+ E_{c \bar c}(us; 1.1^2 {\rm GeV}^2) \right] \pm
 \frac{1}{2} \left[ E_{c \bar c}(us; 3 Q_s^2)
- E_{c \bar c}(us; 1.1^2 {\rm GeV}^2) \right] \ ,
\label{Eccus}
\ee
where the soft reference scale $Q_s$ in the 2$\delta$anQCD model 
is determined by Eq.~(\ref{Qsb})\footnote{
Note that in the 2$\delta$anQCD model in the $c \bar c$ case the condition
(\ref{Qsa}) cannot be fulfilled.} 
and in pQCD by  Eq.~(\ref{Qspt}).
In (F)APT, we do not have practical problems
at low scales $\mu < 1.1$ GeV, and the $us$ energy as a
function of low scale $\mu_{us}$ turns out to have a moderate local
maximum and a moderate local minimum; hence we use
\be
E_{c \bar c}(us; {\rm (F)APT}) 
= \frac{1}{2} \left[  E_{c \bar c}(us)_{\rm max} +
E_{c \bar c}(us)_{\rm min} \right] \pm
 \frac{1}{2} \left[ E_{b \bar b}(us)_{\rm max} -
E_{c \bar c}(us)_{\rm min} \right] \ ,
\label{EccusMA}
\ee
where these values are quite small: in the central case
($\m_c=1.257$, $k_{s/us}=1.0$), 
the local maximum  is $E_{c \bar c}(us)_{\rm max} \approx +0.003$ GeV
and is reached at $\mu_{us} \approx 0.71$ GeV, and the local minimum is
$E_{c \bar c}(us)_{\rm min} \approx -0.006$ GeV and is reached
at very low scale $\mu_{us} \approx 0.11$ GeV [cf.~Fig.~\ref{plEusnf3}(b)].
\item
The exact renormalon cancellation requirement in (F)APT
gives again an exceedingly small value of the $s$-$us$ 
separation parameter, $k_{s/us} \approx 3 \times 10^{-9}$.
In (F)APT we vary the parameter $k_{s/us}$ again around its central
chosen value $1.0$, in the interval between $0.1$ and $10.x$, just as 
it was done in the $b \bar b$ case of (F)APT. 
\end{itemize}

The described behavior of the $E_{c \bar c}(us; \mu_{us}^2)$
in the analytic 2$\delta$anQCD model and in pQCD in the two schemes, 
as a function of the ultrasoft scales $\mu_{us}$, 
is presented in Fig.~\ref{plEusnf3}(a),
and in the case of (F)APT in Fig.~\ref{plEusnf3}(b).
Comparing Figs.~\ref{plEusnf3}(a) and (b), we see that 
$|E_{c \bar c}(us)| \sim 10^{-2}$-$10^{-1}$ GeV in the 2$\delta$anQCD model, 
and $\sim 10^{-3}$ GeV in (F)APT.
Furthermore, comparing with the corresponding curves in 
Figs.~\ref{plEusnf4}(a) and (b) for the $b \bar b$, 
we can see that in the 2$\delta$anQCD model
and in the two pQCD approaches, the absolute values $|E_{c \bar c}(us)|$ are
by almost two orders of magnitude smaller than $|E_{b \bar b}(us)|$, principally
because the renormalon cancellation gives us in the two cases significantly
different $s$-$us$ separation parameter values: $k_{s/us}(c \bar c) =4.06$
and $k_{s/us}(b \bar b) = 0.238$. Further, the values of
$|E_{q \bar q}(us)|$ (for $q=c, b$) in pQCD are larger 
than in the 2$\delta$anQCD model, especially in $\MSbar$ pQCD.
In the (F)APT case, the values of
$|E_{q \bar q}(us)|$  ($q=c, b$) are quite small, being in $c \bar c$
case smaller by almost one order of magnitude. All this is reflected also 
in the numerical results of this and of the previous subsection.

The resulting extracted values of $\m_c$ are
\begin{subequations}
\label{mcMAtot}
\bea
\m_c ( {\rm (F)APT}) &=& {\big \{}
1.257 \mp 0.002_{(us)} 
\mp 0.002_{(s/us)} 
\mp 0.011_{(\Lambda)}
\mp 0.002_{(\mu_s)} 
{\big \}} \; {\rm GeV} \,
\label{mcMA}
\\
& = & 1.257 \pm 0.012  \; {\rm GeV} \ , 
\quad {\rm with:} \ 
\sqrt{3} Q_{s,\rm ((F)APT)} = 1.00  \; {\rm GeV} \ , \
k_{s/us} = 1.0 \ . 
\label{mcMAquadr}
\eea 
\end{subequations}
\begin{subequations}
\label{mc2dtot}
\bea
\m_c({\rm 2 \delta anQCD}) &=& {\bigg \{}
1.266 
\pm 0.003_{(us)}
\!\!\pm 0.007_{(s/us)} 
\!\!\mp 0.005_{(\alpha_s)}
\!\!+ \left( \bear{c} -0.014 \\ +0.003 \eear \right)_{(c_2)} 
\!\!+ \left( \bear{c} -0.002 \\ +0.005 \eear \right)_{(\mu_s)} 
{\bigg \}}
 \; {\rm GeV} \,
\label{mc2d}
\\
& = & 1.266 \pm 0.017 \; {\rm GeV} \ , 
\quad {\rm with:} \ 
\sqrt{3} Q_s = 1.42 \ {\rm GeV} \ , \
k_{s/us} = 4.06 \ .
\label{mc2dquadr}
\eea
\end{subequations}
\begin{subequations}
\label{mcLamtot}
\bea
\m_c({\rm pQCD Lamb.\/}) &=& {\bigg \{}
1.265 
\pm 0.001_{(us)}
\!\!+ \left( \bear{c} +0.021 \\ -0.021 \eear \right)_{(s/us)} 
\!\!\mp 0.004_{(\alpha_s)}
\!\!\pm 0.000_{(c_2)} 
\!\!+ \left( \bear{c} -0.003 \\ +0.015 \eear \right)_{(\mu_s)} 
{\bigg \}}
 \; {\rm GeV} \,
\label{mcLam}
\\
& = & 1.265 \pm 0.027 \; {\rm GeV} \ , 
\quad {\rm with:} \ 
\sqrt{3} Q_s = 1.38 \ {\rm GeV} \ , \
k_{s/us} = 5.59 \ . 
\label{mcLamquadr}
\eea
\end{subequations}
\begin{subequations}
\label{mcbMStot}
\bea
\m_c({\rm pQCD} \MSbar) &=&  {\bigg \{}
1.272 
\mp 0.011_{(us)}
\!\!+ \left( \bear{c} +0.066 \\ -0.075 \eear \right)_{(s/us)} 
\!\!\mp 0.002_{(\alpha_s)}
\!\!+ \left( \bear{c} -0.003 \\ +0.017 \eear \right)_{(\mu_s)} 
{\bigg \}}
 \; {\rm GeV} \,
\label{mcbMS}
\\
& = & 1.272 \pm 0.078 \; {\rm GeV} \ , 
\quad {\rm with:} \ 
\sqrt{3} Q_s = 1.58 \ {\rm GeV} \ , \
k_{s/us} = 3.08 \ . 
\label{mcbMSquadr}
\eea
\end{subequations}
In order to see more clearly the
renormalon cancellation, we present below, as in Sec.~\ref{subs:mb}, 
the sum for the soft mass $M_{J/\psi(1S)}(s) = 2 m_c + E_{c \bar c}(s)$, 
combining in each parenthesis
the (positive) term $\sim a^n$ from $2 m_c$ and the 
(negative) term $\sim a^{n+1}$
from the soft binding energy $E_{c \bar c}(s)$
($n=0,1,\ldots,4$), for the
central input values of parameters $\m_c$, $\mu_s$ ($=\sqrt{3} Q_s$) and $k_{s/us}$;
separately we present below also $E_{c \bar c}(us)$.
\begin{subequations}
\label{MJMAtot}
\bea
M_{J/\psi(1S)}(s; {\rm (F)APT}) & = & 2.513 + (0.409-0.056) +
(0.228-0.082) + (0.133-0.071) + (0.046-0.023) \; {\rm GeV} 
\nonumber\\
& = &
 2.513 + 0.354 + 0.146 + 0.062 + 0.023 \; {\rm GeV}
\; ( = 3.098 \; {\rm GeV}) \ ,
\label{MJsMA}
\\
E_{c \bar c}(us; {\rm (F)APT}) & = & -0.001 \pm 0.004 \; {\rm GeV} \ , \qquad 
(\m_c=1.257 \ {\rm GeV}, \; k_{s/us} = 1.0 ) \ ;
\label{MJusMA}
\eea 
\end{subequations}
\begin{subequations}
\label{MJ2dtot}
\bea
M_{J/\psi(1S)}(s; {\rm 2 \delta anQCD}) & = & 2.531 + (0.349-0.053) +
(0.328-0.171) + (0.253-0.150) + (0.979-0.979) \; {\rm GeV} 
\nonumber\\
& = &  2.531 + 0.296 +0.157 +0.103 + 0.000 \; {\rm GeV}
\; ( = 3.087 \; {\rm GeV}) \ ,
\label{MJs2d}
\\
E_{c \bar c}(us; {\rm 2 \delta anQCD}) & = & 0.010 \mp 0.006 \; {\rm GeV} \ ,
\qquad 
(\m_c=1.266 \ {\rm GeV}, \; k_{s/us} = 4.06 ) \ ;
\label{MJus2d}
\eea
\end{subequations}
\begin{subequations}
\label{MJLamtot}
\bea
M_{J/\psi(1S)}(s; {\rm pQCD Lamb.}) & = &  2.530 + (0.354-0.061) +
(0.303-0.141) + (0.404-0.305) + (0.662-0.662) \; {\rm GeV} 
\nonumber\\
& = &  2.530 + 0.293 +0.161 +0.099 + 0.000 \; {\rm GeV}
\; ( = 3.083 \; {\rm GeV}) \ ,
\label{MJsLam}
\\
E_{c \bar c}(us; {\rm pQCD Lamb.}) & = & 0.013 \mp 0.003 \; {\rm GeV} \ ,
\qquad 
(\m_c=1.265 \ {\rm GeV}, \; k_{s/us} = 5.59 ) \ ;
\label{MJusLam}
\eea
\end{subequations}
\begin{subequations}
\label{MJbMStot}
\bea
M_{J/\psi(1S)}(s; {\rm pQCD} {\MSbar}) & = &  2.544 + (0.368-0.066) +
(0.348-0.163) + (0.455-0.356) + (0.778-0.778) \; {\rm GeV} 
\nonumber\\
& = &  2.544 + 0.302 +0.185 +0.099 + 0.000 \; {\rm GeV}
\; ( = 3.131 \; {\rm GeV}) \ ,
\label{MJsbMS}
\\
E_{c \bar c}(us; {\rm pQCD} {\MSbar}) & = &  -0.034 \pm 0.024 \; {\rm GeV} \ ,
\qquad 
(\m_c=1.272 \ {\rm GeV}, \; k_{s/us} = 3.08 ) \ ;
\label{MJusbMS}
\eea
\end{subequations}

As in the previous subsection in the case of $b \bar b$,
we present now for the case $c \bar c$, for better visibility, 
the results for the central
extracted values of $\m_c$ of the four models in Table \ref{t4}, 
and for various uncertainties $\delta \m_c$ in Table \ref{t5}.
\begin{table}
\caption{Extracted central values of $\m_c$ in the four models, for the
central input parameter values (with the total uncertainties $\delta \m_c$).
Included are the corresponding input parameter $k_{s/us}$, and the resulting
scales: soft renormalization scale $\mu_s = \sqrt{3} Q_s$;
soft mass $M_{J/\psi(1S)}(s)$; averaged
ultrasoft energy ${\bar E}_{c \bar c}(us)$ and its
ambiguity $\delta {\bar E}_{c \bar c}(us)$
[cf.~Eqs.~(\ref{Eccus})-(\ref{EccusMA})]. 
All scales are given in GeV. Note that $M_{J/\psi(1S)}(s) +
{\bar E}_{c \bar c}(us) = 3.097$ GeV, i.e., the physical mass $M_{J/\psi(1S)}$.}
\label{t4}
\begin{ruledtabular}
\begin{tabular}{c|c|c|c|ccc}
 model & $\m_c (\delta \m_c)$  & $k_{s/us}$ & $\sqrt{3} Q_s$ & $M_{J/\psi(1S)}(s)$ & ${\bar E}_{c \bar c}(us)$ & $\delta E_{c \bar c}$ \\
\hline
(F)APT &      $1.257(\pm0.012)$ & 1.00 & 1.004 & 3.098 & -0.001 & $\pm 0.004$ \\
2$\delta$anQCD &     $1.266(\pm0.017)$ & 4.06 & 1.422 & 3.087 & +0.010 & $\mp 0.006$ \\
pQCD Lamb.   &$1.265(\pm0.027)$ & 5.59 & 1.382 & 3.083 & +0.013 & $\mp 0.003$ \\
pQCD $\MSbar$ &$1.272(\pm0.078)$ & 3.08 & 1.585 & 3.131 & -0.034 & $\pm 0.024$
\end{tabular}
\end{ruledtabular}
\end{table}
\begin{table}
\caption{Uncertainties $\delta \m_c$ of the extracted 
value of $\m_c$ coming from various sources: (1) from the
evaluation of the $us$ sector; (2) from the variation of
the $s$-$us$ separation parameter $k_{s/us}$; (3) from the 
variation of $\alpha_s$ (or, in (F)APT: variation of $\Lambda$);
(4) from the variation of the $c_2$ parameter (in 2$\delta$anQCD,
and in pQCD in the Lambert scheme); (5) from the
variation of the soft renormalization scale $\mu_s$. 
See the text for details.}
\label{t5}
\begin{ruledtabular}
\begin{tabular}{c|c|cc|c|cc|c}
 model & $\delta \m_c(us)$  & $\delta \m_c(s/us)$ & ($k_{s/us}$) & 
$\delta \m_c(\alpha_s)$ & $\delta \m_c(c_2)$ & ($c_2$) &
$\delta \m_c(\mu_s)$ 
\\
\hline
\multirow{2}{10mm}{\rm (F)APT} 
&  -0.002 & -0.002 & (1.0+9.0) & -0.011 & -- & (--) & -0.002
\\
&  +0.002 & +0.002 & (1.0-0.9) & +0.011 & -- & (--) & +0.002
\\
\hline
\multirow{2}{10mm}{\rm 2dan- QCD} 
&  +0.003 & +0.007 & (4.06-3.61) & -0.005 & -0.014 & (-4.76+2.66) & -0.002
\\
&  -0.003 & -0.007 & (4.06+39.5) & +0.005 & +0.003 & (-4.76-0.97) & +0.005
\\
\hline
\multirow{2}{10mm}{\rm pQCD Lamb.}   
&  +0.001 & +0.021 & (5.59-5.355) & -0.004 & 0.000 & (-4.76+2.66) & -0.003
\\
&  -0.001 & -0.022 & (5.59+148.4) & +0.004 & 0.000 & (-4.76-0.97) & +0.015
\\
\hline
\multirow{2}{10mm}{\rm pQCD $\MSbar$} 
&  -0.011 & +0.066 & (3.08-2.86) & -0.002 & -- & (--) & -0.003
\\
&  +0.011 & -0.075 & (3.08+66.9) & +0.002 & -- & (--) & +0.017
\\
\end{tabular}
\end{ruledtabular}
\end{table}

The nonperturbative (NP) contribution coming from the gluon condensate, 
cf.~Eq.~(\ref{EbbusNP1}) for the $b \bar b$ system in the previous subsection,
is unreliable for the lighter $c \bar c$ system, 
since the next-to-leading corrections are in this case large
and tend to make the result unreliable, cf.~Ref.~\cite{Pineda:1996uk}.

Comparing the results for $\m_c$ in this subsection with 
those for $\m_b$ in the previous subsection, we see that the
soft-ultrasoft separation parameter $k_{s/us}$ in
the 2$\delta$anQCD model and pQCD is now larger:
$k_{s/us} \approx 3$-$5$, while in $\m_b$ case we had $k_{s/us} \approx
0.2$-$0.3$. This is a consequence of the requirement of
the leading renormalon cancellation.
As a result, the ultrasoft contributions
to the $J/\psi(1S)$ mass are by an order of magnitude
smaller (in absolute value)
than those to the $\Upsilon(1S)$ mass, surprisingly. 
The extracted values of $\m_c$ thus
suffer from less (ultrasoft) uncertainty than the extracted
values of $\m_b$. On the other hand, in (F)APT, the ultrasoft
sector is always suppressed, a consequence of the suppressed 
(F)APT couplings in the infrared.

The extracted values of $\m_c$ obtained in this work, in
all four models, are compatible with all those 
obtained in the literature (from lattice, sum rules, and spectrum
calculations), as can be seen from Table \ref{bmbcomp}.
  
\section{Summary}
\label{sec:summ}
We evaluated, in two analytic QCD models and in perturbative QCD (pQCD, 
in two schemes), the quark-antiquark binding energies 
(up to ${\rm N}^3{\rm LO}$) and masses of ground state
$q \bar q$ quarkonia ($q=b, c$), as functions of the 
quark mass $\m_q$ [$\equiv \m_q(\mu^2=\m_q^2)$], also called the $\MSbar$ quark mass. 
In analytic QCD models the
QCD running coupling $\A_1(Q^2)$ has no unphysical (Landau) singularities
in the $Q^2$ plane. 

The use of the analytic QCD models was motivated by the fact that the 
typical soft ($s$) momentum scales $Q_s$ in the 
ground bound states of quarkonia are 
low ($Q_s \approx 2$ GeV and $1$ GeV, for $b {\bar b}$ and $c {\bar c}$, 
respectively), and that the typical ultrasoft ($us$)
momentum scales $Q_{us}$ are
even lower. This, in conjunction with the fact that
Landau singularities of the pQCD coupling $a_{\rm pt}(Q^2)$ reach
relatively high momenta: 
$Q \approx 0.61$ GeV in the usual (four-loop) $\MSbar$ scheme
(with $c_2 \equiv \beta_2/\beta_0 = 4.471$), 
and $Q \approx 0.26$ GeV in the Lambert scheme ($c_2=-4.76$).
So we can apply in analytic QCD generally more natural
renormalization scales at which the pQCD couplings are
sometimes ``out of control.''

One analytic QCD model applied here was the Analytic Perturbation Theory (APT)
of Shirkov, Solovtsov, Solovtsova, and Milton {\it et al.\/}
(Refs.~\cite{ShS,MSS,Sh1,Sh2}),
which has been extended by Bakulev, Mikhailov and Stefanis to the  
Fractional Analytic Perturbation Theory (FAPT) for calculation of
the fractional power analogs (Refs.~\cite{BMS1,BMS2,BMS3}).
(F)APT can be regarded as a model with minimal analytization of pQCD in the
{\it conceptual\/} sense. Namely, it keeps the perturbative discontinuity
function $\rho_1^{\rm (pt)}(\sigma) \equiv {\rm Im} \; a_{\rm pt}(Q^2=-\sigma - i \epsilon)$
unchanged on the entire positive-$\sigma$ semiaxis, while removing the
(perturbative) discontinuity at $\sigma < 0$ in order to ensure the
analyticity of $\A_1^{\rm (APT)}(Q^2)$. It thus contains no additional regulators
in the positive low-$\sigma$ regime. One of the strengths of (F)APT is that 
it has as a parameter only the pQCD-type $\Lambda$ scale; i.e., 
it contains no new parameters.
As a result, it has finite coupling $\A_1^{\rm (APT)}(Q^2)$ at $|Q^2| \to 0$,
and $\A_1^{\rm (APT)}(Q^2) - a_{\rm pt}(Q^2) \sim (\Lambda^2/Q^2)$ at $|Q^2| > \Lambda^2$.
The latter means that it behaves somewhat differently from the underlying 
pQCD (with the same $\Lambda$) even at high squared momenta $|Q^2|$. 
The value of the scale $\Lambda$ is adjusted so that the high-$|Q^2|$ QCD 
phenomenology is reproduced.

The other analytic QCD model applied here was the two-delta
analytic QCD model (2$\delta$anQCD), Refs.~\cite{2danQCD,anOPE}. This model 
can be regarded as a model with minimal analytization of pQCD in the
{\it numerical\/} sense. Namely, in this model the behavior of 
the discontinuity function $\rho_1^{(2 \delta)} (\sigma) \equiv
{\rm Im} \; \A_1^{(2 \delta)}(-\sigma - i \epsilon)$ in the unknown low-$\sigma$ 
regime ($0 \leq \sigma \alt 1$ GeV) is parametrized (with two deltas)
in such a way that: (a) at $|Q^2| > \Lambda^2$ the model becomes practically
indistinguishable from the (underlying) pQCD,
$\A_1^{\rm (2 \delta)}(Q^2) - a_{\rm pt}(Q^2) \sim (\Lambda^2/Q^2)^5$; and (b) the 
measured value of the semihadronic strangeless $V+A$ decay ratio of the
$\tau$ lepton (the hitherto best measured inclusive low-energy QCD observable),
$r_{\tau} = 0.203$, is reproduced. These conditions fix most of the
mentioned low-$\sigma$ regime parameters. The value of the scale $\Lambda$ 
is the same as in the (underlying) pQCD, so that the high-$|Q^2|$ QCD  
phenomenology is reproduced. In contrast with (F)APT, in the 2$\delta$anQCD model
one relevant parameter remains variable, namely the parameter $c_2$ 
($\equiv \beta_2/\beta_0$), which we vary in the phenomenologically viable
interval, i.e., approximately $-6 < c_2 < -2$.
 
The main conclusions of this work are the following:
analytic QCD approaches which at high energies follow the 
pQCD behavior closely
(such as the 2$\delta$anQCD model) indicate that
the ultrasoft regime in the $\Upsilon(1S)$ quarkonium 
($b \bar b$) is important. 
Our approach, together with the leading renormalon cancellation condition,
gives us clues about how to estimate
the effects of the ultrasoft regimes in pQCD.
In both the 2$\delta$anQCD model and in pQCD we obtain, as a consequence,
extracted values of $\m_b$ which are significantly higher
($\m_b \geq 4.3$ GeV)
than most of those ($\m_b \approx 4.2$ GeV)
obtained in the sum rule approaches 
(which use pQCD+OPE) and in the
usual pQCD calculations of meson spectra. These approaches usually
either do not include the ultrasoft contributions, or
they include them unseparated from the soft contributions
(i.e., the ultrasoft and soft scales are set to be equal).
As an additional consequence, the uncertainties
in the extracted values of $\m_b$ in our approach are dominated
by the ultrasoft sector and are, especially in pQCD in
the $\MSbar$ scheme, larger than in the usual pQCD approaches. 
Further, the extracted values of $\m_b$ in
the 2$\delta$anQCD model, $\m_b \approx (4.35 \pm 0.08)$ GeV, 
are compatible with those of lattice calculations; cf.~Table \ref{bmbcomp}.
On the other hand, the 2$\delta$anQCD model indicates that the
ultrasoft regime in the $J/\Psi(1S)$ quarkonium ($c \bar c$)
is less important, principally because the leading renormalon cancellation
condition results in smaller ultrasoft coefficients in this system.
The extracted values, $\m_c \approx (1.27 \pm 0.02)$ GeV, are compatible
with those of pQCD (or pQCD+OPE) approaches, and those
of the lattice calculations. 
On the other hand, the (F)APT of Shirkov {\it et al.\/}, 
suppresses the infrared contributions
because the higher order couplings in (F)APT are more
strongly suppressed in the infrared than the 2$\delta$anQCD couplings.
The extracted values in (F)APT, $\m_b \approx (4.16 \pm 0.02)$ GeV and
$\m_c \approx (1.26 \pm 0.01)$ GeV, are compatible with those obtained from the 
sum rules and from the usual pQCD spectrum calculations.

\begin{acknowledgments}
\noindent
We thank A.~Pineda for useful comments.
C.A.~thanks Bogolyubov Laboratory of Theor.~Physics, of the
Joint Institute for Nuclear Research, Dubna, 
for warm hospitality during part of this work.
C.A.~further thanks J.~Ot\'alora for help in using calculational software.
This work was supported in part by 
MECESUP2 (Chile) Grant FSM 0605-D3021 (C.A.),
and by FONDECYT (Chile) Grant No.~1095196 and Anillos Project ACT119 (G.C.). 
\end{acknowledgments}

\appendix

\section{Analytic analogs of powers $a_{\rm pt}^{\nu}$ and of terms
$a_{\rm pt}^{\nu} \ln^k a_{\rm pt}$ in general analytic QCD models}
\label{app0}

We consider that a (general) anQCD model is defined via 
an analytic analog $\A_1(Q^2)$ of $a_{\rm pt}(Q^2)$
in ther complex plane, or, equivalently, by the discontinuity function
$\rho_1(\sigma)$ [$\equiv {\rm Im} \; \A_1(-\sigma - i \epsilon)$] on the positive
semiaxis $\sigma > 0$.

For such general anQCD models, the higher order couplings $A_{\nu}(Q^2)$ 
[analogs of $a_{\rm pt}^{\nu}(Q^2)$] were constructed
in Refs.~\cite{CV1,CV2} for integer $\nu$, 
and Ref.~\cite{GCAK} for general (noninteger) $\nu$.
Below we will summarize the basic aspects of such construction.

Since the general anQCD models, with the exception of APT, have
at low $\sigma$ ($\alt 1 \ {\rm GeV}^2$) 
different discontinuity function than the
pQCD coupling $a_{\rm pt}(Q^2)$, we cannot use the (F)APT method 
[Eq.~(\ref{MAangen})] for the construction of the analytic analogs of 
$a_{\rm pt}^{\nu} \ln^k a_{\rm pt}$. The analogs of the integer powers
$a_{\rm pt}^n$ in such general models were constructed in Refs.~\cite{CV1,CV2},
where it was shown that it is imperative to construct first the
analogs of the logarithmic derivatives of $a_{\rm pt}$ in 
the following way:\footnote{
If the analytization is performed in any other way, the renormalization scale and
scheme dependence of the resulting truncated analytic series of any observable
${\cal D}(Q^2)$ will in general increase (instead of decrease) when the number 
of terms in the series increases; cf.~\cite{CV1,CV2}.}
\be
\left( \frac{\partial^k a_{\rm pt}(Q^2)}{\partial (\ln Q^2)^k} \right)_{\rm an}
=
\frac{\partial^k \A_1(Q^2)}{\partial (\ln Q^2)^k} \quad (k=0,1,2,\ldots) \ .
\label{logder}
\ee
In pQCD, the logarithmic derivatives 
\be
{\ta}_{{\rm pt},k+1}(Q^2)
\equiv \frac{(-1)^{k}}{\beta_0^{k} k!}
\frac{ \partial^k a_{\rm pt}(Q^2)}{\partial (\ln Q^2)^k} \ , 
\qquad (k=0,1,2,\ldots) \ ,
\label{tan}
\ee
are related with the powers of $a_{\rm pt} \equiv \alpha_s/\pi$ 
in the following way (using RGEs in pQCD):
\begin{subequations}
\label{tatot}
\bea
{\widetilde a}_{{\rm pt},2} &=&
a_{\rm pt}^2 + c_1 a_{\rm pt}^3 + c_2 a_{\rm pt}^4 + 
c_3 a_{\rm pt}^5 + \cdots \ ,
\label{ta2}
\\
{\widetilde a}_{{\rm pt},3} &=&
a_{\rm pt}^3 + \frac{5}{2} c_1 a_{\rm pt}^4  
+ \left( 3 c_2 + \frac{3}{2} c_1^2 \right) a_{\rm pt}^5 + 
\cdots \ ,
\label{ta3}
\\
{\widetilde a}_{{\rm pt},4} &=& a_{\rm pt}^4 +  
\frac{13}{3} c_1 a_{\rm pt}^5 + \cdots \ , 
\qquad
{\widetilde a}_{{\rm pt},5} = a_{\rm pt}^5 + \cdots \ ,  
\qquad {\rm etc.}
\label{ta45}
\eea 
\end{subequations}
This means that the powers of $a_{\rm pt}$ are linear combinations
of logarithmic derivatives
\begin{subequations}
\label{atot}
\bea
a_{\rm pt}^2 & = & {\widetilde a}_{{\rm pt},2}
- c_1 {\widetilde a}_{{\rm pt},3} 
+ \left( \frac{5}{2} c_1^2 - c_2 \right) {\tilde a}_{{\rm pt},4} 
+ \left(-\frac{28}{3} c_1^3 + \frac{22}{3} c_1 c_2 - c_3 \right) 
{\tilde a}_{{\rm pt},5} + 
\cdots \ ,
\label{a2}
\\
a_{\rm pt}^3 & = & {\widetilde a}_{{\rm pt},3} 
- \frac{5}{2} c_1 {\widetilde a}_{{\rm pt},4} 
+ \left(\frac{28}{3} c_1^2 - 3 c_2 \right) {\widetilde a}_{{\rm pt},5} 
+ \cdots \ ,
\label{a3}
\\
 a_{\rm pt}^4  &=&  {\widetilde a}_{{\rm pt},4} 
- \frac{13}{3} c_1  {\widetilde a}_{{\rm pt},5} +  \cdots \ ,
\qquad 
 a_{\rm pt}^5  =  {\widetilde a}_{{\rm pt},5}  +  \cdots \ ,
\qquad {\rm etc.}
\label{a45}
\eea
\end{subequations}
These relations, in conjunction with the analytization Eq.~(\ref{logder}),
imply that the analytic analogs $\A_k$ of powers $a_{\rm pt}^k$, in general
anQCD models, can be expressed as linear combinations of the
logarithmic derivatives
\be
\tA_{k+1}(\mu^2)
= \frac{(-1)^k}{\beta_0^k k!}
\frac{ \partial^k \A_1(\mu^2)}{\partial (\ln \mu^2)^k} \ ,
\qquad (k=1,2,\ldots) \ .
\label{tAn}
\ee
in the following form:
\begin{subequations}
\label{Atot}
\bea
\A_2 & = & \tA_2
- c_1 \tA_{3} + \left( \frac{5}{2} c_1^2 - c_2 \right) \tA_4 
+ \left(-\frac{28}{3} c_1^3 + \frac{22}{3} c_1 c_2 - c_3 \right) \tA_5
+ \cdots \ ,
\label{A2}
\\
\A_3 & = & \tA_{3}
- \frac{5}{2} c_1 \tA_{4} 
+ \left(\frac{28}{3} c_1^2 - 3 c_2 \right) \tA_5 
+ \cdots \ ,
\label{A3}
\\
\qquad
\A_4  &=&  \tA_{4} 
- \frac{13}{3} c_1  \tA_5
+ \cdots \ , \qquad 
\A_5  =  \tA_5  +  \cdots \ ,
\qquad {\rm etc.}
\label{A45}
\eea
\end{subequations}
In Ref.~\cite{GCAK} this analytization was extended to the case when 
$k \mapsto \nu$ is noninteger\footnote{
\label{tAnu2}
This relation was also reformulated so as to be applicable in 
a larger $\nu$ interval: $-2 < \nu$; cf.Eq.~(22) of Ref.~\cite{GCAK}.}
\be
\tA_{\nu+1}(Q^2) = \frac{1}{\pi} \frac{(-1)}{\beta_0^{\nu} \Gamma(\nu+1)}
\int_{0}^{\infty} \ \frac{d \sigma}{\sigma} \rho_1(\sigma)  
{\rm Li}_{-\nu}\left( - \frac{\sigma}{Q^2} \right) \quad (-1 < \nu) \ ,
\label{tAnu1}
\ee
where, as always, $\rho_1(\sigma) \equiv {\rm Im} \; \A_1(Q^2=-\sigma - i \epsilon)$,
and ${\rm Li}_{-\nu}(z)$ is the polylogarithm function.\footnote{
In Mathematica \cite{Math8} 
it is implemented under the name PolyLog$[-\nu,z]$.}
The corresponding analogs of powers $a_{\rm pt}^{\nu}$ are then obtained 
by using the general relations
\be
a_{\rm pt}^{\nu} ~=~ {\ta}_{{\rm pt},\nu} + \sum_{m=1}^{\infty}
\tk_m(\nu) \; {\ta}_{{\rm pt},\nu + m} \ .
\label{anutanu}
\ee
and the linearity of analytization, i.e.
\be
\A_{\nu}(Q^2)  \equiv \left( a_{\rm pt}^{\nu}(Q^2) \right)_{\rm an}
= {\tA}_{\nu}(Q^2) + \sum_{m \geq 1}
\tk_m(\nu) \; {\tA}_{\nu + m}(Q^2) \quad (-1 < \nu) \ .
\label{AnutAnu}
\ee
The coefficients $\tk_m(\nu)$, for general real $\nu$ and positive
integer $m$, were calculated in Ref.~\cite{GCAK}, and are combinations
of gamma functions and their derivatives, with arguments involving $\nu + \ell$
($\ell$ being various integers).

Furthermore, since $a^{\nu} \ln^k a = \partial^k a^{\nu}/\partial \nu^k$,
the linearity of analytization then implies 
\bea
\A_{\nu,k}(Q^2) \equiv \left( a_{\rm pt}^{\nu}(Q^2) \ln^k a_{\rm pt}(Q^2) \right)_{\rm an} & = &
\frac{ \partial^k \A_{\nu}(Q^2)}{\partial \nu^k}
\nonumber\\
& = & 
\frac{ \partial^k}{\partial \nu^k}
\left[ {\tA}_{\nu}(Q^2) + \sum_{m \geq 1}
\tk_m(\nu) \; {\tA}_{\nu + m}(Q^2) \right]  \quad (-1 < \nu) \ , 
\label{Anulnk}
\eea
where in the terms on the right-hand side we use expressions for
${\tA}_{\nu + m}(Q^2)$ obtained by Eq.~(\ref{tAnu1}).
Comparing Eqs.~(\ref{Anulnk}) and (\ref{tAnu1}), we see that
the terms in the above sum represent integrals over the scale
$\sigma$ involving
the basic discontinuity function of the model ($\rho_1$) and
derivatives of the polylogarithm function ${\rm Li}_{-\nu - m +1}$ with
respect to its index $\nu$. In the evaluation of the binding
energy $E_{q \bar q}$ we will encounter the logarithmic terms
of the type (\ref{Anulnk}) with $\nu$ integer ($\nu=n$); however, 
the derivatives with respect to index $\nu$ in Eq.~(\ref{Anulnk})
imply that we must know the behavior of $\tA_{\nu}$ around the integer
value $\nu=n$, i.e., we need to use here the expression
(\ref{tAnu1}) for noninteger $\nu$, or a version of it with
improved integration convergence, Eq.~(22) of Ref.~\cite{GCAK}
(cf.~also the earlier footnote \ref{tAnu2}).

\section{Ground state quark-antiquark binding energy}
\label{app1}

According to Refs.~\cite{PinYnd, Chetyrkin:1999qi,Penin:2002zv}, the
perturbation expansion of the quark-antiquark binding energy $E_{q \bar q}$,
in terms of the quark pole mass $m_q \equiv m$ and $\alpha_s(\mu^2) \equiv \alpha_s$, is

\be
E_{q \bar q} = E_1^C + \delta E_1^{(1)} + \delta E_1^{(2)} + \delta E_1^{(3)} \ ,
\label{Eqqexp}
\ee
where
\begin{eqnarray}
E_1^C&=&-\frac{1}{4} \alpha _s^2 C_F^2 m ,
\end{eqnarray}
\begin{eqnarray}
\delta E_1^{(1)}&=&
E_1^C \; \frac{4 \alpha_s}{\pi} \left[ \beta_0 ( L_{\mu} + 1) + \frac{a_1}{8} \right] ,
\end{eqnarray}
\begin{eqnarray}
\delta E_1^{(2)}&=&
E_1^C \; \frac{\alpha_s^2}{\pi^2}
{\bigg \{}
12 \beta_0^2 L_{\mu}^2 + (16 \beta_0^2 + 3 a_1 \beta_0 + 4 \beta_1 ) L_{\mu}
+ \beta_0^2 \left( 4 + \frac{2 \pi^2}{3} + 8 \zeta(3) \right)
\nonumber\\
&&
+ \left( 2 a_1 \beta_0 + 4 \beta_1 + \frac{a_1^2}{16} + \frac{a_2}{8} \right)
+ C_A C_F \pi^2 + 
\pi^2 C_F^2 \left[ \frac{21}{16} - \frac{2}{3} S (1 + S) \right] 
{\bigg \}},
\end{eqnarray}

\begin{eqnarray}
\delta E_1^{(3)}&=&\delta E_1^{(3)}|_{\beta \left(\alpha _s\right)=0}+\delta E_1^{(3)}|_{\beta \left(\alpha _s\right)} \ .
\end{eqnarray}
The two contributions to $\delta E_1^{(3)}$ are
\begin{eqnarray}
\delta E_1^{(3)}|_{\beta \left(\alpha _s\right)=0}&=&
- E_1^C \; \frac{\alpha _s^3}{\pi }\left\{-\frac{a_1 a_2+a_3}{32 \pi ^2}+a_1 \left[-\frac{C_AC_F}{2}+C_F^2\left(-\frac{19}{16}+\frac{S (1+S)}{2} \right)\right]
\right.
\nonumber
\\
&&
\left.
+C_A^3 \left(-\frac{1}{36}+\frac{L_{\alpha _s}}{6}+\frac{\ln[2]}{6}\right)+C_A^2 C_F\left(-\frac{49}{36}+\frac{4}{3} \left(L_{\alpha _s}+\ln[2]\right)\right)
\right.
\nonumber
\\
&&
\left.
+C_AC_F^2\left[-\frac{5}{72}+\frac{37 L_{\alpha _s}}{6}+\left(\frac{85}{54}-\frac{7}{6}L_{\alpha _s}\right) S (1+S)+\frac{10 \ln[2]}{3}\right]
\right.
\nonumber
\\
&&
\left.
+C_F^3 \left(\frac{50}{9}+3 L_{\alpha _s}-\frac{S (1+S)}{3}+\frac{8 \ln[2]}{3}\right)+
C_F^2 T_F {\bigg (}
-\frac{32}{15}+S (1+S) (1-\ln[2])
\right.
\nonumber
\\
&&
\left.
+2 \ln[2]
{\bigg )}
+\frac{49}{36}C_A C_FN_fT_F+C_F^2 N_f T_F\left(\frac{11}{18}-\frac{10}{27} S (1+S)\right)+\frac{2}{3}C_F^3L_1^E\right\} ,
\end{eqnarray}

\begin{eqnarray}
\delta E_1^{(3)}|_{\beta \left(\alpha _s\right)}&=&
E_1^C \; \frac{\alpha _s^3}{\pi ^3} \left\{
32 \beta _0^3L_{\mu }^3+\left[
12 a_1\beta _0^2+40 \beta _0^3+28 \beta _0\beta _1
\right] L_{\mu }^2
\right.
\nonumber
\\
&&
\left.
+L_{\mu }\left[
10a_1\beta _0^2+3a_1\beta _1+4 \beta _2+ \beta _0
\left(
\frac{a_1^2}{2}+a_2+40 \beta _1+8 C_AC_F \pi ^2
\right.
\right.
\right.
\nonumber
\\
&&
\left.
\left.
\left.
+C_F^2 \left(
\frac{21 \pi ^2}{2}-\frac{16}{3} \pi ^2 S (1+S)\right)
\right)+
\beta _0^3
\left(
\frac{16 \pi ^2}{3}+64 \zeta (3)
\right)
\right]
+\beta _0^3 
{\bigg (}
-8+4 \pi ^2+\frac{2 \pi ^4}{45}
\right.
\nonumber
\\
&&
\left.
+64\zeta (3)-8 \pi ^2 \zeta (3)+96 \zeta (5) 
{\bigg )}
+a_1\beta_0^2 
\left(\frac{2 \pi ^2}{3}+8\zeta (3)
\right)+ 
\beta _0
\left[
-\frac{a_1^2}{8}+\frac{3 a_2}{4}
\right.
\right.
\nonumber
\\
&&
\left.
\left.
+C_A C_F\left(
6 \pi ^2-\frac{2 \pi^4}{3}
\right)
+C_F^2 
\left(
8 \pi ^2-\frac{4 \pi ^4}{3}+\left(-\frac{4 \pi ^2}{3}+\frac{4 \pi ^4}{9}\right) S (1+S)
\right)
\right.
\right.
\nonumber
\\
&&
\left.
\left.
+\beta _1
\left(
8+\frac{7 \pi ^2}{3}+16 \zeta (3)
\right)
\right]+2 a_1\beta _1+4 \beta _2
\right\} \ .
\end{eqnarray}
The following notations were used:
\begin{eqnarray}
L_{\mu }&=&\ln \left[\frac{\mu }{\alpha _s C_F m}\right]\ , \qquad 
L_{\alpha _s}=-\ln\left[C_F\alpha _s\right]\ , \qquad   
L_1^E=-81.5379 \ , \qquad
S= {\rm spin} (=1) \ , 
\\
C_A &=& 3  \ , \qquad    
C_F=4/3 \ , \qquad    
T_F=1/2 \ .
\end{eqnarray}
The RGE coefficients $\beta_j$ are in the $\MSbar$ scheme 
\begin{eqnarray}
\beta _0&=&\frac{1}{4}\left(11-\frac{2 }{3}N_f\right) \ ,
\qquad
\beta _1=\frac{1}{16}\left(102-\frac{38 }{3}N_f\right) \ ,
\qquad
\beta _2 = \frac{1}{64}\left(\frac{2857}{2}-\frac{5033}{18}N_f+\frac{325}{54}N_f{}^2\right) \ ,
\nonumber \\
\beta _3&=&\frac{1}{256}\left[\frac{149753}{6}+\frac{1093}{729}N_f{}^3+3564 \zeta (3)+N_f{}^2 \left(\frac{50065}{162}+\frac{6472 \zeta (3)}{81}\right)
-N_f\left(\frac{1078361}{162}+\frac{6508 \zeta (3)}{27}\right)\right] .
\end{eqnarray}
The constants $a_1$ and $a_2$ are
\begin{eqnarray}
a_1&=&\frac{1}{9} \left(31C_A-20N_fT_F\right) \ ,
\nonumber \\
a_2&=&\frac{400 N_f{}^2T_F{}^2}{81}-C_F N_f T_F \left(\frac{55}{3}-16 \zeta (3)\right)+C_A^2 \left(\frac{4343}{162}+\frac{1}{4} \left(16 \pi ^2-\pi ^4\right)+\frac{22\zeta (3)}{3}\right)
\nonumber\\
&&
-C_A N_f T_F \left(\frac{1798}{81}+\frac{56 \zeta (3)}{3}\right) \ .
\end{eqnarray}
The value for the constant $a_3$ associated with the three-loop soft 
contribution was obtained in Refs.~\cite{Smirnov:2009fh,Anzai:2009tm} 
and is
\begin{eqnarray}
a_3&=&a_3^{(0)}+a_3^{(1)} N_f+a_3^{(2)} N_f^2+a_3^{(3)} N_f^3 \ ,
\end{eqnarray}
where
\begin{eqnarray}
a_3^{(0)}&=&502.24 C_A^3-136.39\left(\frac{N_C\left(N_C^2+6\right)}{48}\right) \ ,
\nonumber \\
a_3^{(1)}&=&-709.717 C_A^2T_F+\left(\frac{-71281}{162}+264 \zeta (3)+80\zeta (5)\right)C_A C_FT_F
\nonumber
\\
&&
+\left(\frac{286}{9}+296\frac{\zeta (3)}{3}-160\zeta
(5)\right)C_F^2 T_F-56.83\left(\frac{18-6N_C^2+N_C^4}{96N_C^2}\right) \ ,
\nonumber \\
a_3^{(2)}&=&C_FT_F^2 \left(\frac{14002}{81}-\frac{416\zeta (3)}{3}\right)+C_AT_F^2 \left(\frac{12541}{243}+\frac{64 \pi ^4}{135}+\frac{368 \zeta
(3)}{3}\right) \ ,
\qquad
a_3^{(3)}=-\frac{8000 T_F^3}{729}\,.
\end{eqnarray}

\section{Renormalon-based estimate of $\sim a_{\rm pt}^4$ coefficient}
\label{app2}

The term $r_3$ in the expansion of $m_q/\m_q$ in Eq.~(\ref{mqbmq}) can be 
estimated by a method closely related with the approach presented in 
Sec.~II of Ref.~\cite{Contreras:2003zb}. The pQCD version of
the sum in Eq.~(\ref{mqbmq}) can be reexpressed in
terms of $a_{\rm pt}(\mu^2)$ at any other renormalization scale $\mu^2$
\begin{subequations}
\label{Sm}
\begin{eqnarray}
S \equiv \frac{m_q}{{\overline m}_q} - 1 & = & 
\frac{4}{3} a_{\rm pt}(\mu^2)
\left[ 1 + a_{\rm pt}(\mu^2) r_1(\mu^2) + a_{\rm pt}^2(\mu^2) r_2(\mu^2) + {\cal O}(a_{\rm pt}^3) \right] 
\ ,
\label{Smexp}
\\
r_1(\mu^2) & = & \kappa_1 + \beta_0 L_m(\mu^2) 
\ ,
\label{r1mu} 
\\
r_2(\mu^2) & = & \kappa_2 + ( 2 \kappa_1 \beta_0 + \beta_1) L_m(\mu^2) 
+ \beta_0^2 L_m^2(\mu^2) \ ,
\label{r2mu}
\\
(4/3) \kappa_1 & = & 6.248 \beta_0 - 3.739 \ ,
\label{k1}
\\
(4/3) \kappa_2 &= &   23.497 \beta_0^2 + 6.248 \beta_1 
+ 1.019 \beta_0 - 29.94 \ ,
\label{k2}
\end{eqnarray}
\end{subequations}
where $L_m(\mu^2) = \ln(\mu^2/{\overline m}_q^2)$, while
$\beta_0(N_f)$ and $\beta_1(N_f)$
are the renormalization scheme independent coefficients
[given just after Eq.~(\ref{RGE2l})].
Here, $N_f = N_{\ell}$ is the number of light active
flavors (quarks with masses lighter than $m_q$). 

Since $r_1$ and $r_2$ are explicitly known,
the Borel transform $B_S(b)$ is known to order $\sim\!b^2$
\begin{equation}
B_{S}(b; \mu) = \frac{4}{3} \left[ 1 + \frac{r_1(\mu^2)}{1! \; \beta_0} b +
\frac{r_2(\mu^2)}{2! \; \beta_0^2} b^2 + {\cal O}(b^3) \right] \ .
\label{BSm1}
\end{equation}
The function  $B_S(b)$ has renormalon singularities at 
$b = 1/2, 3/2, 2, \ldots, -1, -2, \ldots$ 
\cite{Bigi:1994em,Beneke:1994sw,Beneke:1999ui}.
The behavior of $B_S$ near the leading infrared (IR) renormalon
singularity $b=1/2$ is determined by the resulting
renormalon ambiguity of $m_q$. This ambiguity $\delta m_q$
is a (QCD) scale which, having the dimension 
of energy and being renormalization scale and
scheme independent, must be proportional to the QCD scale 
$\Lambda_{\rm QCD}$: $\delta m_q = const \times \Lambda_{\rm QCD}$ 
\cite{Beneke:1994rs}. This scale can be expressed in terms of
$a_{\rm pt}(\mu^2)$ and $\mu$ ($\mu$ being any renormalization scale)
in the form
\begin{equation}
\Lambda = const \times \mu \exp \left( - \frac{1}{2 \beta_0 a_{\rm pt}(\mu)}
\right) a_{\rm pt}(\mu)^{- \nu} c_1^{- \nu} \left[
1 + \sum_{k=1}^{\infty} \ (2 \beta_0)^k \nu (\nu-1) \cdots
(\nu-k+1){\widetilde c}_k a_{\rm pt}^k(\mu) \right] \ ,
\label{tL2}
\end{equation}
where
\begin{subequations}
\label{tr}
\begin{eqnarray}
\nu & = & \frac{c_1}{ 2 \beta_0} = \frac{\beta_1}{2 \beta_0^2} \ ,
\label{nu}
\\
{\widetilde c}_1 & = & \frac{ ( c_1^2 - c_2) }{(2 \beta_0)^2 \nu} \ ,
\qquad
{\widetilde c}_2 = \frac{1}{2 (2 \beta_0)^4 \nu (\nu\!-\!1)} \left[
( c_1^2 - c_2 )^2 - 2 \beta_0 ( c_1^3 - 2 c_1 c_2 + c_3 ) \right] \ ,
\label{tr1r2}
\\
{\widetilde c}_3 & = & \frac{1}{6 (2 \beta_0)^6 \nu (\nu\!-\!1)(\nu\!-\!2)} 
\left[
( c_1^2 - c_2 )^3 - 6 \beta_0 (c_1^2 - c_2 )(c_1^3 - 2 c_1 c_2 + c_3 )
+ 8 \beta_0^2 ( c_1^4 - 3 c_1^2 c_2 + c_2^2 + 2 c_1 c_3 - c_4 )
\right] .
\label{tr3}
\end{eqnarray}
\end{subequations}
The above constants, expressed in terms of $\beta_0$ and of $c_j = \beta_j/\beta_0$,
appear in the expansion of the residue of the Borel transform $B_S(b; \mu)$
at the pole $b=1/2$
\begin{eqnarray}
B_S(b; \mu) & = &  N_m \pi  \frac{\mu}{ {\overline m}_q }
 \frac{1}{ ( 1 - 2 b)^{1 + \nu} } \left[ 1 +
\sum_{k=1}^{\infty} {\widetilde c}_k ( 1 - 2 b)^k \right]
+ B_{S}^{\rm (an.)}(b; \mu) \ ,
\label{BSrenan}
\end{eqnarray}
where $B_{S}^{\rm (an.)}(b; \mu)$ is analytic on the disk $|b| < 1$ and
can be expanded in powers of $b$. The form of the 
representation (\ref{BSrenan}) is called bilocal and was proposed in 
Ref.~\cite{Lee:2003hh}.
We can assume that the coefficients ${\widetilde c}_k$ are known up
to $k=3$, because the coefficient $c_4 = \beta_4/\beta_0$ 
(in the ${\overline {\rm MS}}$ scheme)
is known to a large degree by Pad\'e-related methods of 
Ref.~\cite{Ellis:1997sb}
\be
\beta_4= \frac{1}{4^5} (A_4 + B_4 N_f + C_4 N_f^2 + D_4 N_f^3 + E_4 N_f^4)
\label{beta4}
\ee 
with $A_4= 7.59 \times 10^5$, $B_4= -2.19 \times 10^5$, $C_4= 2.05 \times 10^4$,
$D_4= -49.8$, and $E_4=-1.84$. This gives $c_4=123.7$ for $N_f=3$,
$c_4=97.2$ for $N_f=4$, and $c_4=86.2$ for $N_f=5$. 
The residue parameter $N_m$ can be determined with
high precision by using the idea of Refs.~\cite{Lee:1996yk},
i.e., by calculating
(cf.~Refs.~\cite{Pineda:2001zq,Lee:2003hh,Cvetic:2003wk}):
\begin{equation}
N_m = \frac{{\overline m}_q}{\mu} \frac{1}{\pi} R_S(b=1/2) \ ,
\label{Nmform}
\end{equation}
where
\begin{equation}
R_S(b; \mu) \equiv  (1 - 2 b)^{1 + \nu} B_{S}(b; \mu) \ ,
\label{RSm}
\end{equation} 
and the first coefficients in the expansion in powers of $b$
of this quantity are known from the known coefficients $r_1$ and $r_2$.
We can use a combination of truncated perturbation series
and Pad\'e approximants [1/1] for $R_S(b)$, as presented in
 Ref.~\cite{Cvetic:2003wk}, and obtain
\begin{eqnarray}
 N_m & \approx & 0.575 (N_f=3) \ , \quad \approx  0.555 (N_f=4) \ ,
\quad \approx 0.533 (N_f=5) \ .
\label{Nmnf}
\end{eqnarray}
with the uncertainties in these values of roughly $\pm 0.020$.

In the bilocal expansion (\ref{BSrenan}), the analytic part
$B_{S}^{\rm (an.)}(b; \mu)$ can be taken as a polynomial in $b$, i.e.,
a truncated expansion in powers of $b$. The coefficients of
the latter expansion can be related with $r_j(\mu^2)$'s
by equating the expansion of Eq.~(\ref{BSrenan}) in powers of $b$
with the expansion (\ref{BSm1}), resulting in
\begin{subequations}
\label{BSan}
\begin{eqnarray}
B_{S}^{\rm (an.)}(b; \mu) & = & h^{(m)}_0 + 
\sum_{k \geq 1} \frac{h^{(m)}_k}{k! \; \beta_0^k} b^k \ ,
\label{BSanexp}
\\
h^{(m)}_k & = & \frac{4}{3} r_k  - \pi N_m \frac{\mu}{{\overline m}_q}
(2 \beta_0)^k \sum_{n \geq 0} {\widetilde c}_n 
\frac{ \Gamma ( \nu + k + 1 - n) }{ \Gamma(\nu + 1 - n) } \ ,
\label{hms}
\end{eqnarray}
\end{subequations}
where, by convention, $r_0 = {\widetilde c}_0 = 1$. The numbers ${\widetilde c}_n$
of Eqs.~(\ref{tr}),
which enter the sum in Eq.~(\ref{hms}), are known only up to $n=3$,
because, in ${\overline {\rm MS}}$, only $c_k$ up to $k=4$ are reasonably known
($c_4$ approximately, as mentioned). For $N_f=3$, these values are:
${\widetilde c}_1= -0.1638$, ${\widetilde c}_2= 0.2372$, ${\widetilde c}_3= -0.1205$ 
(and $\nu= 0.3951$). For $N_f=4$, they are:
${\widetilde c}_1= -0.1054$, ${\widetilde c}_2= 0.2736$, ${\widetilde c}_3= -0.1610$ 
(and $\nu= 0.3696$). And for $N_f=5$ they are: 
${\widetilde c}_1= 0.0238$, ${\widetilde c}_2= 0.3265$, ${\widetilde c}_3= -0.2681$ 
(and $\nu= 0.3289$).
Therefore, the sums in (\ref{hms}) are truncated at $n=3$.

Theoretically, the pole closest to the origin in $B_{S}^{\rm (an.)}(b; \mu)$
is at $b=-1$, at least in the large-$\beta_0$ approximation.\footnote{
Nonetheless, there is a possibility that at two-loop order 
the kinetic term contributes to an IR renormalon at $b=+1$ 
in $B_S(b)$, cf.~Ref.~\cite{Neubert:1996zy}.}
Since in  $B_{S}^{\rm (an.)}(b; \mu)$ the coefficients
$h^{(m)}_k(\mu^2)$ for $k=0,1,2$ are known [because $r_1(\mu^2)$ and $r_2(\mu^2)$
are known], we can construct the Pad\'e ${\rm [1/1]}_{B_{S}^{\rm (an.)}}(b)$
and check the pole of it. It turns out that this 
Pad\'e, at the natural scale $\mu=\m_q$, has the pole 
at $b=-1.09, -0.96, -1.12$, for $N_f=3, 4, 5$,  
respectively, reflecting correctly the theoretical expectation.\footnote{
However, the $\mu$ dependence of this position is rather strong. For example, when
$\mu^2$ varies by 10 \% around $\m_q^2$, the pole position in [1/1]
varies between $-1.6$ and $-0.7$ in the $N_f=3$ case, between $-1.2$ and $-0.7$
in the $N_f=4$ case, and between $-1.26$ and $-1.00$ in the $N_f=5$ case.} 
We can extend further this reasoning and obtain
the next coefficient $h^{(m)}_3$ (at $\mu=\m_q$) by requiring that
the Pad\'e ${\rm [2/1]}_{B_{S}^{\rm (an.)}}(b)$ has the pole at
$b=-1$. This gives us
\be
h^{(m)}_3(\m_q^2) = -25.18 (N_f=3) \ , \ -28.28 (N_f=4) \ ,
\ -35.62 (N_f=5) \ .
\label{hm3}
\ee
If constructing with these values of $h^{(m)}_3$ the other possible
Pad\'e approximant of index 3, 
namely ${\rm [1/2]}_{B_{S}^{\rm (an.)}}(b)$, it turns out that the
nearest to origin pole of such Pad\'e is then at $b=-1.003,-1.001,-1.008,$ for 
$N_f=3, 4, 5$, respectively. This indicates that the
obtained values of $h^{(m)}_3$, Eq.~(\ref{hm3}), are consistent.\footnote{
We used for the ${\overline {\rm MS}}$ scheme coefficient $c_4$ the estimated
values (\ref{beta4}), with $c_4=123.7, 97.2, 86.2$, for $N_f=3,4,5$, respectively,
from Ref.~\cite{Ellis:1997sb}.
Simpler Pad\'e-based estimates of $c_4$ were obtained in
Ref.~\cite{Elias:1998bi}: $c_4 =40,70$ for $N_f=4,5$, respectively 
(and a large negative and uncertain value $c_4=-850$ for $N_f=3$).
The $c_4=40$ value (for $N_f=4$) in this case differs substantially 
from the value $c_4=97.2$.
If we repeat for the $c_4=40$ value ($N_f=4$)
the same procedure described above, we obtain 
${\widetilde c}_4 \approx 0.0055$ (for $c_4=97.2$ we got: ${\widetilde c}_4= -0.1610$);
hence the expressions of $h^{(m)}_k(\m_q^2)$ of Eq.~(\ref{hms}) change,
and the pole of ${\rm [1/1]}_{B_{S}^{\rm (an.)}}(b)$ becomes $b \approx -5.9$
(before: $b \approx -0.96$), not close to the theoretical pole $b=-1$.
Furthermore, from the requirement that ${\rm [2/1]}_{B_{S}^{\rm (an.)}}(b)$ 
has the pole at $b=-1$ we now get $h^{(m)}_3(\m_q^2)=4.10$ (before: $-28.28$),
and using this value of $h^{(m)}_3$ in the Pad\'e
${\rm [1/2]}_{B_{S}^{\rm (an.)}}(b)$ we obtain the pole nearest to the origin
$b = 1.94$ (before: $b=-1.001$). This indicates that the estimate
 $c_4 =40$ (for $N_f=4$) is not giving results consistent with the
theoretical expectations of the renormalon structure of  $B_{S}^{\rm (an.)}$.}
Using these values, we obtain from the relation (\ref{hms}) (with
the natural choice $\mu^2=\m_q^2$) at $k=3$ an estimate for $r_3$
\begin{eqnarray}
\frac{4}{3} r_3(\m_q^2) &=&  h^{(m)}_3(\m_q^2) + 
\pi N_m (2 \beta_0)^3 \sum_{n=0}^3 {\widetilde c}_n 
\frac{ \Gamma ( \nu + 4 - n) }{ \Gamma(\nu + 1 - n) } \ .
\label{r3mq2}
\end{eqnarray}
This gives us numerically the following estimates (we recall that
$N_f \equiv N_{\ell}=3,4,5$ for $c$, $b$, $t$ quark, respectively):
\be
\frac{4}{3} r_3 = 1785.9  (N_f=3) \ , \ 1316.4 (N_f=4) \ , \ 920.1
 (N_f=5) \ .
\label{r3}
\ee
The principal origin of the uncertainties in these expressions 
is the uncertainty in the residue parameter 
$N_m$ (roughly $\pm 0.020$, i.e., less than 4 \%), 
implying an uncertainty in $r_3$ of a few percent (below 4 \%).

An analysis similar to this one has been performed in 
Ref.~\cite{Pineda:2001zq}. There, however, the term ${\widetilde c}_3$
and the coefficients $h^{(m)}_k$ were not included in the analysis.
The results of Ref.~\cite{Pineda:2001zq} are: 
$(4/3) r_3 = 1818.6, 1346.7, 947.9.,$ for $N_f \equiv N_{\ell}=3,4,5$, 
respectively.
These results are by about $2$-$3 \%$ higher than 
ours [Eq.~(\ref{r3})].
In another approach, applying the effective charge method (ECH)
of Refs.~\cite{ECH} to a Euclidean analog of the quantity $m_q$,
an approach using the idea of Ref.~\cite{KSt} extended in Ref.~\cite{CKSir}
to the mass-dependent Minkowskian quantities,
the authors of Ref.~\cite{KK} obtained for these coefficients
the estimates $1281.05, 986.097, 719.339$, respectively. These
quantities are by about $22$-$28 \%$ lower than ours.
On the other hand, the corresponding estimates in Ref.~\cite{CKSir} are
$1544.1, 1091.0, 718.74$, respectively.\footnote{
At the time, the coefficient $r_2$ was not known, and the authors of
Ref.~\cite{CKSir} used in the estimates of $(4/3)r_3$ the analogously
ECH-estimated values of $(4/3)r_2=124.1, 97.729, 73.616$, respectively
(the exact values are $116.30, 94.21, 73.43$).}

\section{Variation of pQCD coupling with scales and schemes}
\label{app3}

In this appendix we give the relation between 
$a_0 \equiv a_{\rm pt}(Q_0^2;c_2^{(0)},c_3^{(0)},\ldots)$
and $a \equiv a_{\rm pt}(Q^2;c_2,c_3,\ldots)$, where the latter
is expressed as power expansion of the former
(cf.~Appendix A of Ref.~\cite{Cvetic:2000mz} for details)
\begin{eqnarray}
a & = & a_0 + a_0^2 (-x) + a_0^3 ( x^2 - c_1 x + \delta c_2)
\nonumber\\
&& + a_0^4 \left( - x^3 + \frac{5}{2} c_1 x^2 - c_2^{(0)} x
- 3 x \delta c_2 + \frac{1}{2} \delta c_3 \right) 
\nonumber\\
&& + a_0^5 {\Bigg [} x^4 - \frac{13}{3} c_1 x^3 + 
\left( \frac{3}{2} c_1^2 + 3 c_2^{(0)} + 6 \delta c_2 \right) x^2
\nonumber\\
&&+ ( - c_3^{(0)} - 3 c_1 \delta c_2 - 2 \delta c_3 ) x
+ \left( \frac{1}{3} c_2^{(0)} \delta c_2 + 
\frac{5}{3} (\delta c_2)^2 - \frac{1}{6} c_1 \delta c_3 +
\frac{1}{3} \delta c_4  \right) {\Bigg ]} + {\cal O}(a_0^6) \ ,
\label{aexpina0}
\end{eqnarray}
where we denote
\begin{subequations}
\label{aa0tot}
\begin{eqnarray}
a & \equiv & a_{\rm pt}(Q^2;c_2,c_3,\ldots) \ , 
\quad
a_0 \equiv  a_{\rm pt}(Q_0^2;c_2^{(0)},c_3^{(0)},\ldots) \ , 
\label{aa0}
\\
x &\equiv& \beta_0 \ln \frac{Q^2}{Q_0^2}, \quad
\delta c_k \equiv c_k - c_k^{(0)} \ .
\label{xdcj}
\end{eqnarray}
\end{subequations}
For the purposes of our paper, it is sufficient to consider
in the above relation (\ref{aexpina0})
terms up to (including) terms $\sim a_0^4$.
 
The three-loop threshold connection of $a_{\rm pt}$ in the $\MSbar$ scheme
at the threshold scale $\mu_{\rm thr}^2 = ({\cal K} \m_c)^2$ 
(where ${\cal K} \sim 1$; usually ${\cal K}=2$) can
be written as the following relation between
$a_{\rm pt}(\mu_{\rm thr}^2+0;N_f=4) \equiv a_{+}$ and
$a_{\rm pt}(\mu_{\rm thr}^2-0;N_f=3) \equiv a_{-}$:
\bea
a_{+} & = & a_{-} \left[ 1 + x_1 a_{-} + x_2 a_{-}^2 + x_3 a_{-}^3
+ {\cal O}(a_{-}^4) \right] \ ,
\label{apm}
\eea
where 
\bea
x_1 & = & - k_1 \ , \qquad x_2 = - k_2 + 2 k_1^2 \ ,
\qquad
x_3  =  - k_3 + 5 k_1 k_2 - 5 k_1^3 \ ,
\label{xjs}
\eea
and the coefficients $k_j$ were calculated in Ref.~\cite{CKS}
\bea
k_1 & = & -\frac{1}{6} \ell_h \ , \qquad
k_2 = \frac{1}{36} \ell_h^2 - \frac{19}{24} \ell_h + \frac{11}{72} \ ,
\nonumber\\
k_3 & = & - \frac{1}{216} \ell_h^3 - \frac{131}{576} \ell_h^2
+ \frac{(-6793 + 281 N_{\ell})}{1728} \ell_h +
\left( - \frac{82043}{27648} \zeta(3) + \frac{564731}{124416} 
- \frac{2633}{31104} N_{\ell} \right) \ ,
\label{kjs}
\eea
where $\ell_h = \ln(\mu^2_{\rm thr}/\m_c^2) = \ln {\cal K}^2$,
and $N_{\ell}$ in Eq.~(\ref{kjs}) is the number of light quark flavors,
i.e., $N_{\ell}=3$ in the considered case of transition from $N_f=4$ to $N_f=3$.

\end{document}